\documentclass[11pt]{article}
\usepackage[utf8]{inputenc}
\usepackage[a4paper, margin=2.5cm]{geometry}
\usepackage{graphicx}
\usepackage{xcolor}
 \usepackage{float} 
\usepackage{latexsym}
\usepackage{comment}
\usepackage{multicol}
\usepackage{dsfont}
\usepackage{braket}
\usepackage{newunicodechar}
\newunicodechar{⁻}{$^{-}$}
\usepackage{amssymb,amsthm, amsmath}
\usepackage{quantikz}
\usepackage{hyperref}
\usepackage{subcaption}
\usepackage{caption}
\usepackage{mathtools} 
\usepackage{dsfont}
\usepackage{booktabs}
\usepackage{array}
\usepackage[backend=bibtex,sorting=none,giveninits=true,maxnames=99,maxcitenames=99]{biblatex}
\usepackage{soul}
\usepackage{authblk}
\usepackage{enumitem}
\usepackage{makecell}
\usepackage{algorithm}
\usepackage{algpseudocode}
\usepackage[page]{appendix}

\title{
Universal linear manipulation\\ via routing and projective measurements}

\author[1,*]{Alessio Baldazzi}
\author[2,3]{ Sonia Mazzucchi }
\author[1]{ Lorenzo Pavesi }

\affil[1]{Department of Physics, University of Trento, Trento, 38123, Italy}
\affil[2]{Department of Mathematics, University of Trento, Trento, 38123, Italy}
\affil[3]{Istituto Nazionale di Fisica Nucleare - Trento Institute for Fundamental Physics and Applications (INFN-TIFPA), Trento, 38123, Italy}

\affil[*]{ alessio.baldazzi@unitn.it }

\date{}

\begin{document}

\maketitle

\begin{abstract}

Multiport interferometers with $N$ ports are basic devices in both classical and quantum photonics. 
Ideally, they implement a linear unitary transformation between the input and output electric field vectors with $N$ components, each associated with a spatial mode of classical coherent light or a single photon. Standard designs for a fully reconfigurable universal multiport interferometer are given by the Reck or the Clements schemes.
In this work, we introduce routing schemes to implement a generic unitary transformation on classical coherent light or single photons using linear or tree geometries via multiple projective measurements on a single detector with the minimum number of components. 
Then, we generalize this result to the case of any multi-photon state for scattershot boson sampling experiments with multi-routing schemes.
Finally, we test the robustness of routing schemes compared to universal schemes with respect to losses and phase noise. 

\end{abstract}

\section{Introduction}
\label{sec:intro}

In photonics, unitary matrices are usually realized through multiport interferometers \cite{reck_experimental_1994,clements_optimal_2016,mi10100646,Lin_24}.
More precisely, a reconfigurable multiport interferometer can execute different unitary linear transformations of an input electric vector with $N$ spatial components associated with a classical coherent light or a single-photon state. Therefore, they are fundamental elements in both classical and quantum photonics \cite{daldosso2009nanosilicon,fox2006quantum,bogaerts2020programmable}, since they can in principle implement any unitary operation in a $N$-dimensional complex Hilbert space \cite{Wang_2020}. On the one hand, they are used for optical signal processing and communications \cite{eldada2004optical,minzioni2019roadmap}, photonic matrix and tensor computation \cite{Peserico_23,nano11071683}, and optical neural networks in machine learning applications \cite{shen2017deep,de2019photonic,sui2020review}. On the other hand, they are employed to implement universal quantum computation with photons \cite{knill_scheme_2001,obrien_demonstration_2003,Flamini_2018,baldazzi2024}, quantum photonic simulators \cite{Wang_2018,Baldazzi2025}, and boson sampling \cite{boson_sampling,boson_sampling_review,wang2019boson,metcalf2013multiphoton}.
As shown by Reck et al. \cite{reck_experimental_1994}, multiport interferometers are obtained using precise arrangements of Mach-Zehnder interferometers (MZIs) \cite{zetie2000does,mzi_21}, which can be understood as reconfigurable beam-splitters made of phase shifters \cite{ellinger2010integrated,sun2022silicon} and 50:50 beam-splitters. Their MZI meshes are based on particular mathematical decompositions of the unitary matrices in terms of $2\times2$ sub-matrices. 
However, even if all designs ideally produce the same transformation, real implementations \cite{Miller_15} will always be characterized by different performances for different schemes, as demonstrated by Clements et al. \cite{clements_optimal_2016,burgwal2017using} for non-ideal MZIs.
Following Clements decomposition algorithm, other MZI designs have been found and studied with respect to different non-idealities \cite{exagone_17,Fldzhyan_20,Marchesin_25,pereira2025minimum,61zbk52n}.

\begin{figure}[t]
    \centering
    \includegraphics[width=0.83\textwidth]{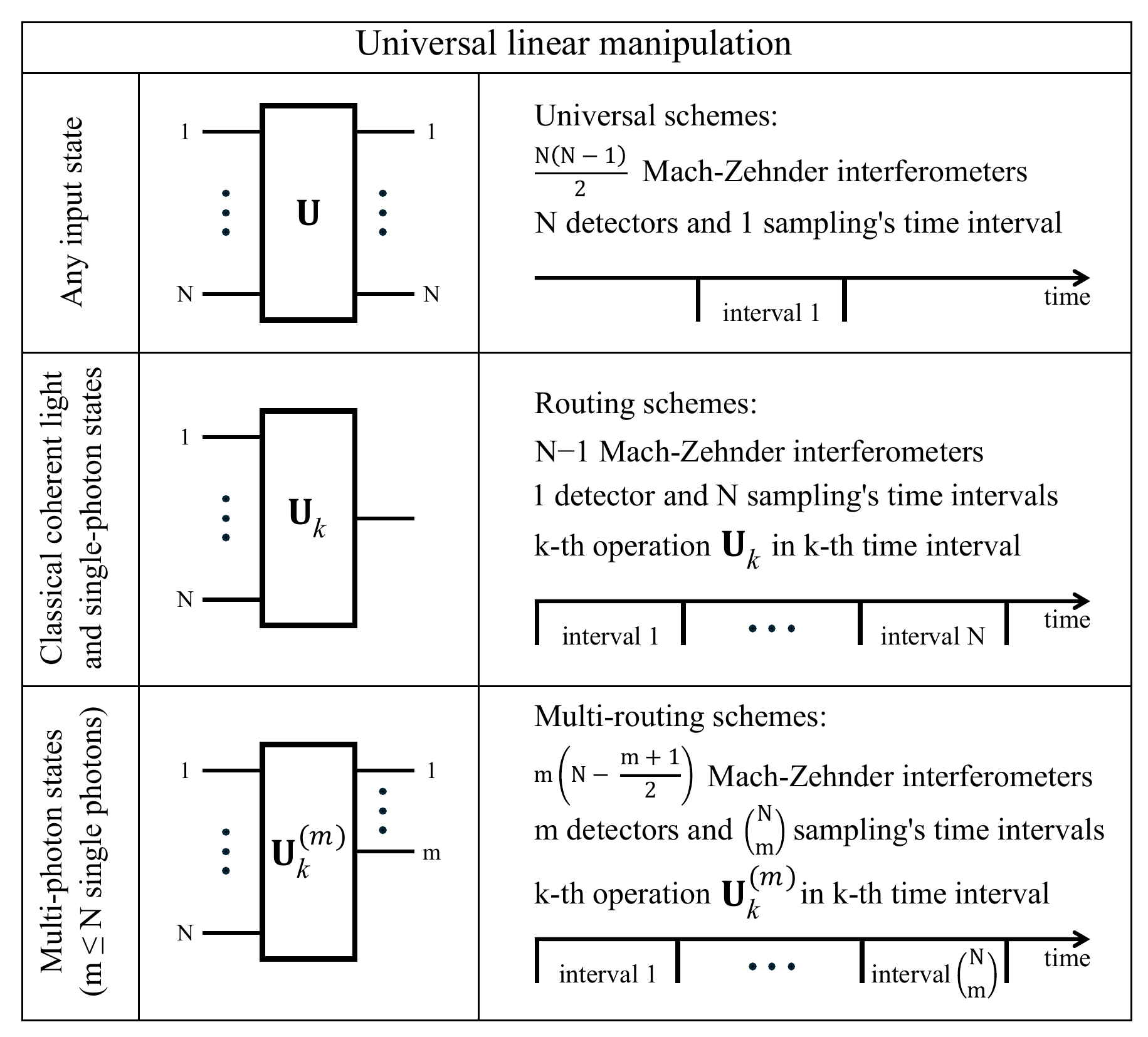}
    \caption{\textbf{Different techniques to achieve universal linear manipulation.}
    Each row of the table shows a method to implement the generic unitary transformation $\mathbf{U}$. Each technique is tailored with respect to a class of input states, shown in the left column. In the first row, the universal schemes are based on MZI meshes with $N(N+1)/2$ MZIs, and the operation associated with $\mathbf{U}$ is performed with $N$ detectors in a single temporal interval. In the second row, routing schemes use $N-1$ MZIs and only one detector, but $N$ temporal intervals are needed. In each time interval, a different operation $\mathbf{U}_k$ is executed to implement $\mathbf{U}$. In the third row, multi-routing schemes for $m$ single photons use $m\,(N-(m+1)/2)$ MZIs and $m$ detectors. Like the routing case, each time interval is associated with an operation $\mathbf{U}_k^{(m)}$, and $\binom{N}{m}=\frac{N!}{m!(N-m)!}$ temporal intervals are needed to implement $\mathbf{U}$. }
    \label{fig:ToC}
\end{figure}

The design of photonic circuits based on multiport interferometers always requires the quantification and management of spatial and temporal resources. The former are characterized by the number of phase shifters and detectors, while the latter are determined by the amount of time required to perform the experiments. Of course, the two categories of resources are related.
More phase shifters make the circuit more trainable with respect to a specific task, while more detectors imply faster sampling/tomography of the processed signals.
However, in many tasks, the resources are redundant, since they are not used simultaneously. For example, consider the case of single-photon states propagating in a multiport interferometer with $N$ modes: the detectors are clicking one at a time, and so only one detector is actually needed, if proper photon routing is used. This is exactly what is happening in \cite{Wang_2018} and in \cite{Baldazzi2025,baldazzi2025resourceefficientvariationalquantumsolver} for a $16$- and a four-dimensional photonic system, respectively. In both cases, a tree MZI scheme allows the use of the minimum number of phase shifters and detectors through multiple projective measurements.
The use of the minimum number of detectors implies insensitivity to possible unbalanced coupling losses at the output of the interferometers, and the use of the minimum number of MZIs reduces the circuit footprint, power consumption, control complexity and the intrinsic losses during the manipulation. 
In this work, we generalize the use of multiple projective measurements for classical coherent light and single-photon states to the general $N$-dimensional case for two different schemes: the tree scheme, introduced in \cite{Wang_2018}, and the linear schemes, introduced here.
Then, we present multi-routing schemes able to obtain the general scattershot boson sampling results with any multi-photon state through the minimum number of resources, such as detectors and MZIs.
In particular, we demonstrate that universal schemes employing $N$ detectors are equivalent to routing schemes with a single detector in the case of classical coherent light and single-photon states, and to multi-routing schemes employing $m$ detectors in the case of $m$-photon states.
This concept is summarized in Figure \ref{fig:ToC}: in each row, we show the input state and the spatial and temporal resources for the different classes of schemes.
The proposed schemes together with multiple projective measurements allow to reduce the overhead of universal schemes in terms of spatial resources (MZIs and detectors) through the use of more temporal resources. 
Depending on the problem at hand, this trade-off can be advantageous.

This paper is structured as follows. 
In Section \ref{sec:design}, we review the theory behind multiport interferometers.
In Section \ref{sec:routers}, we explain how architectures with linear and tree meshes can be used as routers and splitters exploiting the minimum amount of spatial resources.
Section \ref{sec:genuniman} shows the equivalence of the routing and multi-routing schemes with multiple projective measurements and universal interferometers for universal linear unitary manipulation for classical coherent light and any number of coincident single photons.
Section \ref{sec:disc} presents the simulation results for lossy MZIs and for random noise in the phase setting.
Section \ref{sec:conclu} reports the comparison of the routing and multi-routing designs with universal multiport interferometers.
Finally, the Appendices discuss the decomposition algorithms for the Reck, the Clements, the routing and the multi-routing schemes, together with additional simulations on non-idealities.

\section{Multiport interferometers}
\label{sec:design}

A multiport interferometer with $N$ modes is composed of MZIs arranged in a specific mesh. 
The associated unitary operator reads as follows:
\begin{equation}
    \mathbf{U}_{\rm MZI\,scheme}^{(N)}
    =
    \prod_{k\in\,\Xi} \mathbf{U}_{\rm MZI}^{(m_k,n_k)}[\theta_{k},\phi_{k}] \,,
    \label{eq:MZIscheme_dec_princ}
\end{equation}
where $\Xi$ is the ordered string of indexes related to the MZI scheme and $\mathbf{U}_{\rm MZI}^{(m_k,n_k)}[\theta_{k},\phi_{k}]$ is the unitary operator associated with an MZI acting on the pair of modes $(m_k,n_k)$. Different phase settings $\{\theta_{k},\phi_{k}\}_{k\in\Xi}$ give rise to different unitary transformations. The phases $\{\theta_{k},\phi_{k}\}_{k\in\Xi}$ are set on the phase shifters (PSs) contained in each MZI of the interferometer.
Essentially, the multiport interferometer generalizes the MZI to the case of $N$ modes. Appendix \ref{app:design} provides further details about the operators associated with MZIs and its linear optical components, such as beam-splitter (BS) and PS. \\
A multiport interferometer scheme is defined to be {\em universal} if, for any unitary matrix $\mathbf{U} \in U(N)$, there exists a specific mesh $\Xi$ of MZIs and a set of phases $\{\theta_k, \phi_k\}_{k}$ such that the scheme implements $\mathbf{U}$ up to an $N$-dimensional diagonal phase matrix $\mathbf{d}$, that is 
\begin{equation}
    \mathbf{U}_{\rm MZI\,scheme}^{(N)}=\mathbf{d}\,\mathbf{U} \,. 
    \label{eq:uNdecu2_princ}
\end{equation}
This definition of universality is based on an equivalence relation between the unitary transformations associated with a universal multiport interferometer and its phase settings. This choice is justified by the fact that the phases encoded in $\mathbf{d}$ are not directly measurable at the output of the interferometer.
In view of Eq. \eqref{eq:MZIscheme_dec_princ}, one of the first questions to address concerns the number of transformations $\mathbf{U}_{\rm MZI}^{(m_k,n_k)}$, or equivalently the number of MZIs, required for the interferometer scheme to be universal. Note that the unitary group $U(N)$ has dimension $N^2$, whereas the space of diagonal phase matrices $\mathbf{d}$, which is isomorphic to $U(1)^N$, has dimension $N$. Therefore, the space of equivalence classes relevant to universal schemes is the quotient $U(N)/U(1)^N$, whose dimension is $N\,(N-1)$. To match the dimensions of the two sides of Eq. \eqref{eq:MZIscheme_dec_princ}, one requires $N\,(N-1)/2$ operators of the form $\mathbf{U}_{\rm MZI}^{(m_k,n_k)}[\theta_{k},\phi_{k}]$. Indeed, each such operator is associated with the two-dimensional quotient space $U(2)/U(1)^2$, consisting of $U(2)$ matrices modulo multiplication by a two-dimensional diagonal phase matrix. Consequently, a universal $N$-port interferometer scheme requires $N\,(N-1)/2$ MZIs \cite{reck_experimental_1994,clements_optimal_2016}. \\
The Reck and Clements schemes are the two principal examples of universal interferometer schemes. The former is based on a triangular mesh, whereas the latter employs a rectangular mesh. Both meshes are illustrated for an eight-dimensional multiport interferometer in the left column of Fig. \ref{fig:scheme_8dim}.
Further details on the Reck and Clements decomposition algorithms are provided in Appendices \ref{app:reck} and \ref{app:clements}, respectively.
The right column of Fig. \ref{fig:scheme_8dim} shows the linear V-shaped and tree schemes for the eight-dimensional case. These schemes are not universal, as can be established simply by counting their MZIs and comparing the resulting number of free parameters with the dimensionality requirement in Eq.\eqref{eq:MZIscheme_dec_princ}.

\begin{figure}[t]
\centering
        \includegraphics[width=\textwidth]{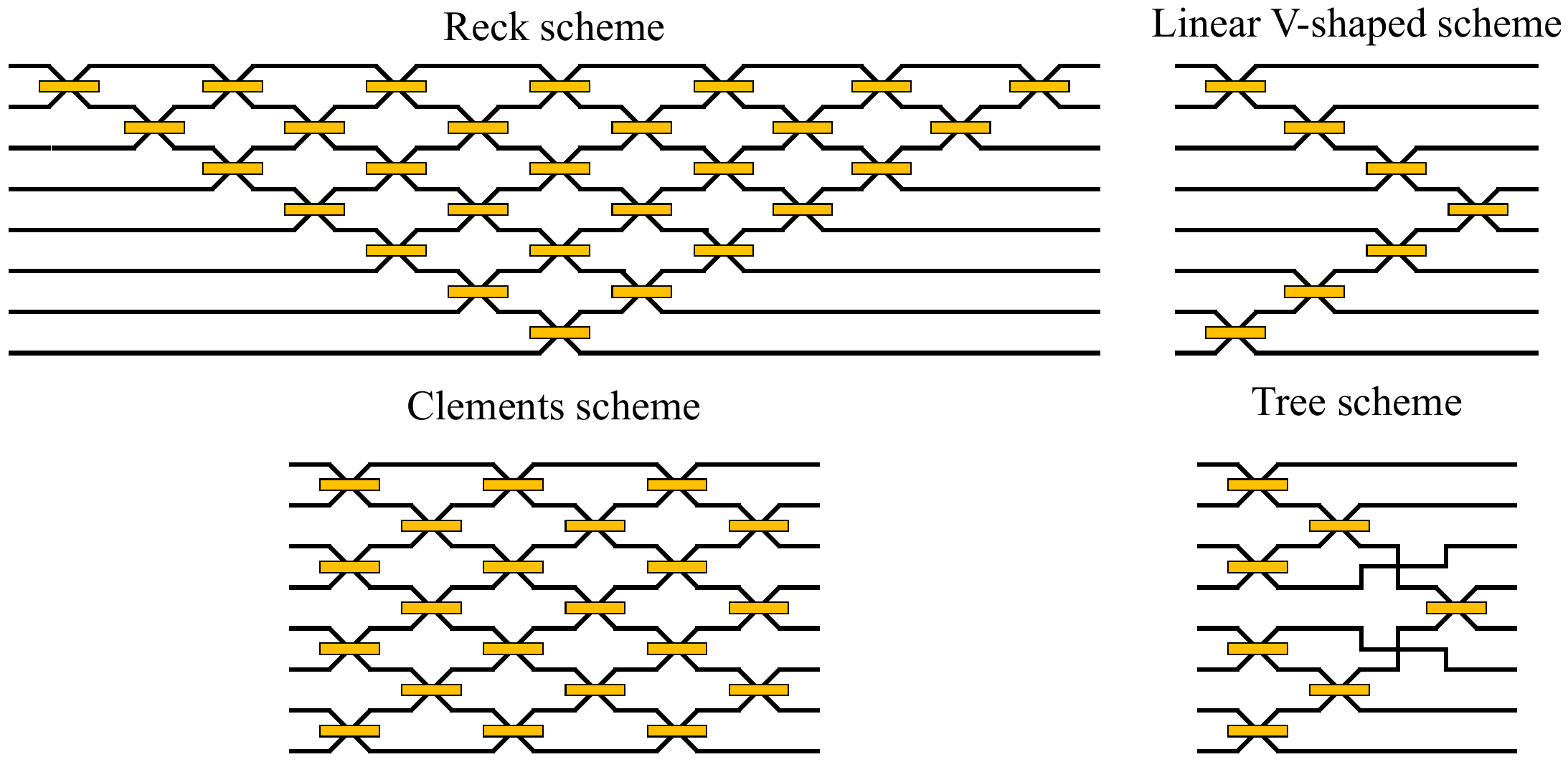}
    \caption{\textbf{Different schemes of multiport interferometers with eight modes.}
    The MZI is represented by the crossing lines with the yellow rectangle. 
    \textbf{(left column)} Universal schemes: the Reck scheme \cite{reck_experimental_1994} has a triangular mesh, while the Clements scheme \cite{clements_optimal_2016} a rectangular mesh. For both universal meshes, the number of MZIs grows quadratically with the mode number: for eight modes, there are 28 MZIs.
    \textbf{(right column)} Non-universal schemes: the linear V-shaped and the tree meshes have the same number of MZIs, but they differ in the arrangement of the MZIs when the number of modes is greater than four. For both non-universal meshes, the number of MZIs grows linearly with the mode number: for eight modes, there are 7 MZIs.
    }
    \label{fig:scheme_8dim}
\end{figure}

We conclude this section by discussing the typical input states of multiport interferometers.
First, we consider the quantum state most-closely resembling the ideal output of a laser \cite{Glauber1963} and the simplest quantum state made with one photon: classical coherent light and a single photon, respectively. The action of the multiport interferometer preserves the statistics of these states in both cases. If $\{|\alpha_m^{in/out}|^2 \}_{m\in1\ldots N}$ denote the input/output averaged numbers of photons in the $m$-th mode for coherent states and the input/output probabilities of one photon in the $m$-th mode for single-photon states, the amplitudes $\alpha$ transform as $\vec \alpha^{out} = \mathbf{U} \,\cdot \,\vec\alpha^{in}$, with $\mathbf{U}$ the unitary transformation associated with the interferometer and $\vec \alpha^{in/out}$ the vector containing the input/output mode amplitudes.
Therefore, in the cases of classical coherent light and a single photon, the evolution through a multiport interferometer is the same, modulo swapping the words {\em intensity} and {\em probability}. Interestingly, the results for the single-photon state can be brought to the classical coherent regime through a simple change of words. Splitting, routing, and most generally manipulating a single photon and a classical coherent state involve the same MZI scheme with the same phase setting.
This result shows that, within linear multiport interferometers, the propagation of a single-photon state entails no greater computational complexity than that of an attenuated coherent state \cite{Barnett_2022}.\\
Finally, we comment on the input states made of a multi-photon state and a multi-modal squeezed coherent state.
After a multiport interferometer associated with $\mathbf{U}$, for multi-photon states, the statistics of the detector clicks is proportional to the permanent of the submatrix of $\mathbf{U}$ with the row and columns determined by the input modes and the output modes associated with the detector clicks \cite{Scheel_2004vsh}. On the contrary, for multi-mode squeezed states, the statistics of the detector clicks is proportional to the Hafnian of a specific matrix built from $\mathbf{U}$ and from the covariance matrix of the squeezed input state, with the row and columns determined by the output modes associated with the detector clicks \cite{hamilton2017gaussian}.
Essentially, the permanent is the sum of the element products over all the permutations, while the Hafnian is the sum of the element products over all the pair partitions. These two cases are denoted scattershot and Gaussian boson sampling, respectively \cite{boson_sampling,boson_sampling_review,wang2019boson}. 
Since the permanent and the Hafnian are classically hard to evaluate for random matrices, boson sampling offers a natural way to compute them by sampling the photon statistics with linear multiport interferometers.

\section{Multiport routers and splitters}
\label{sec:routers}

In this section, we explore new configurations of MZI schemes: in particular, we are interested in multiport interferometers able to perform fully reconfigurable routing and splitting of classical coherent light or a single-photon state by exploiting the minimum number of components.

The unitary operations will be explicitly written using the bra-ket notation, which provides a compact representation of the standard vector formalism by identifying $N$-dimensional vectors with ket-vectors $|\ldots\rangle$ as follows: $(1,0,\ldots,0)^{\rm T}\equiv |1\rangle$, $(0,1,\ldots,0)^{\rm T} \equiv |2\rangle$, \ldots $(0,0,\ldots,1)^{\rm T} \equiv |N\rangle$. The bra-vectors $\langle\ldots|$ are the complex conjugates of ket-vectors. The scalar products are associated with the {\em bra-ket} multiplication, i.e. $\langle\ldots||\ldots\rangle$, while matrices $N\times N$ are given by the {\em ket-bra} multiplication, i.e. $|\ldots\rangle\langle\ldots|$.

First, we consider the unitary transformation $\mathbf{U}=\sum_{n,m=1}^Nu_{n,m}|n\rangle\langle m|$, with components $\{u_{n,m}\}_{n,m\in 1\ldots N}$, and a fixed reference output mode $|r\rangle$, with $r\in 1\ldots N$. Then, for any $k=1,\dots, N$ we define the {\em routers} $\mathbf{R}^{\mathbf{u}}_{k,r}$ associated with $\mathbf{U}$ and $|r\rangle$ as follows:
\begin{equation}
    \mathbf{R}^{\mathbf{u}}_{k,r} \equiv 
    \sum_{n=1}^N u_{k,n} |r\rangle\langle n| 
    \,.
    \label{eq:genericrouter}
\end{equation}
Each transformation $\mathbf{R}^{\mathbf{u}}_{k,r}$ belongs to the coset of rank-one operators mapping the normalized state
\begin{equation}
    |\psi_k^{\mathbf{u}}\rangle \equiv \mathbf{U}^\dagger\, |k\rangle =\sum_{n=1}^N \bar{u}_{k,n} | n\rangle 
    \label{eq:psikroutedvector}
\end{equation}
to the reference output mode $|r\rangle$, with the bar denoting the complex conjugation. Note that each operator $\mathbf{R}^{\mathbf{u}}_{k,r}$ is associated with one row of $\mathbf{U}$.
Since a router $\mathbf{R}^{\mathbf{u}}_{k,r}$ is uniquely determined by the normalized mode defined in Eq. \eqref{eq:psikroutedvector}, it is specified by $2(N-1)$ real parameters. This coincides with the number of real parameters required to characterize a normalized complex $N$-dimensional state up to an overall phase.
Note that the states $\{|\psi_k^{\mathbf{u}}\rangle \}_{k\in 1\ldots N}$ form an orthonormal basis, since $\mathbf{U}$ is a unitary transformation. Clearly, $\mathbf{R}^{\mathbf{u}}_{k,r}$ is not unitary since it is a rank-one operator. We define the {\em unitary router} as the set of unitary maps $\mathbf{U}_{k,r}$ such that
\begin{equation}
    \mathbf{P}_{r}\, \mathbf{U}_{k,r} = {\rm e}^{ {\rm i} \Gamma_k}\,\mathbf{R}^{\mathbf{u}}_{k,r} 
    \,,
    \label{eq:router_to_uni_router}
\end{equation}
where $k\in1\ldots N$, $\mathbf{P}_{r}=|r\rangle\langle r|$ is the projection operator on the subspace generated by the reference mode $|r\rangle$, while $\Gamma_k$ is a global phase depending on $\mathbf{U}$ and $|k\rangle$.
Thus, the action of the router $\mathbf{R}^{\mathbf{u}}_{k,r}$ is equivalent to the action of the unitary router $\mathbf{U}_{k,r}$ followed by $\mathbf{P}_{r}$, which describes the detection of a photon on the reference output mode $| r \rangle$. In particular, Eq. \eqref{eq:router_to_uni_router} implies that $\mathbf{U}_{k,r}$ admits the following decomposition:
\begin{equation}
    \mathbf{U}_{k,r} = {\rm e}^{ {\rm i} \Gamma_k}\,\mathbf{R}^{\mathbf{u}}_{k,r}+\mathbf{V}_k
    \label{eq:unitaryU_as_UV}
\end{equation} 
for some operator $\mathbf{V}_k$ satisfying the conditions: $\mathbf{V}_k=(\mathbf{1}-\mathbf{P}_{r})\mathbf{V}_k(\mathbf{1}-|\psi_k^{\mathbf{u}}\rangle\langle \psi_k^{\mathbf{u}}| )$, $\mathbf{V}_k\mathbf{V}_k^\dagger=(\mathbf{1}-\mathbf{P}_{r})$ and $\mathbf{V}_k^\dagger\mathbf{V}_k=(\mathbf{1}-|\psi_k^{\mathbf{u}}\rangle\langle \psi_k^{\mathbf{u}}| )$. 
Thus, $\mathbf{U}_{k,r}$ maps $|\psi_k^{\mathbf{u}}\rangle$ into a vector proportional to $|r\rangle$, and any other vector orthogonal to $|\psi_k^{\mathbf{u}}\rangle$ into a vector orthogonal to $|r\rangle$. The freedom in the choice of $\mathbf{U}_{k,r}$ is encoded in the phase $\Gamma_k$ and in the rank-$(N-1)$ operator $\mathbf{V}_k$, which maps the subspace orthogonal to $|\psi_k^{\mathbf{u}}\rangle$ onto the subspace orthogonal to $|r\rangle$.
Multiplying both sides of Eq. \eqref{eq:router_to_uni_router} on the left by  $\langle r|$, we obtain $\langle r |\mathbf{U}_{k,r}=e^{{\rm i} \Gamma_k}\langle k |\mathbf{U}$, or equivalently, the identity $\langle r|\mathbf{U}_{k,r}|v\rangle=e^{{\rm i} \Gamma_k}\langle k|\mathbf{U}|v\rangle$, valid for any state $|v\rangle \in \mathbb{C}^N$. This means that for every initial state $|v\rangle$, the probability distribution $|\langle k|\mathbf{U}|v\rangle |^2$ associated with the measurement of the observable $\{|k\rangle\langle k|\}_{k=1,\dots, N}$ on the transformed state $\mathbf{U}|v\rangle$ can be equivalently estimated in terms of the transition probability $|\langle r|\mathbf{U}_{k,r}|v\rangle|^2 $ of the state $\mathbf{U}_{k,r}|v\rangle$ and the reference output state $|r\rangle$. 

The unitary routers $\{\mathbf{U}_{k,r} \}_{k\in1\ldots N}$ are precisely the transformations we want to implement with new MZI meshes.
Even if Eq. \eqref{eq:genericrouter} uniquely specifies each router $\mathbf{R}^{\mathbf{u}}_{k,r}$ given the associated column of the unitary map $\mathbf{U}$, the situation for a unitary router is completely different. Indeed, Eq. \eqref{eq:router_to_uni_router} does not uniquely identify the unitary router $\mathbf{U}_{k,r}$, as it is satisfied by infinitely many unitary matrices.
The idea behind routing schemes consists of exploiting these degrees of freedom in realizing the unitary router $\mathbf{U}_{k,r}$ with the minimum number of resources. From the mathematical point of view, this amounts to decomposing the matrix $\mathbf{U}_{k,r}$ as the product of a minimum number of maps in  $U(2)/U(1)^2$, i.e. $U(2)$ matrices modulo a two-dimensional diagonal phase matrix. From the physical point of view, this consists of looking for meshes made of the minimum number of MZIs. 
Since the number of free parameters of a router $\mathbf{R}^{\mathbf{u}}_{k,r}$ is $2(N-1)$ and two is the dimension of the set $U(2)/U(1)^2$, we are looking for decompositions of unitary routers made of $N-1$ elements. Consequently, the desired multiport interferometers able to implement a unitary router $\mathbf{U}_{k,r}$, associated with the unitary map $\mathbf{U}$, consist of $N-1$ MZIs. Moreover, since the final measurement is performed on the mode $|r\rangle$, the structure of $\mathbf{V}_k$, Eq. \eqref{eq:unitaryU_as_UV}, does not play any role and its action can be neglected. 

\begin{figure}[t]
\centering
\begin{subfigure}[b]{\textwidth}
       \centering
        \includegraphics[width=0.97\textwidth]{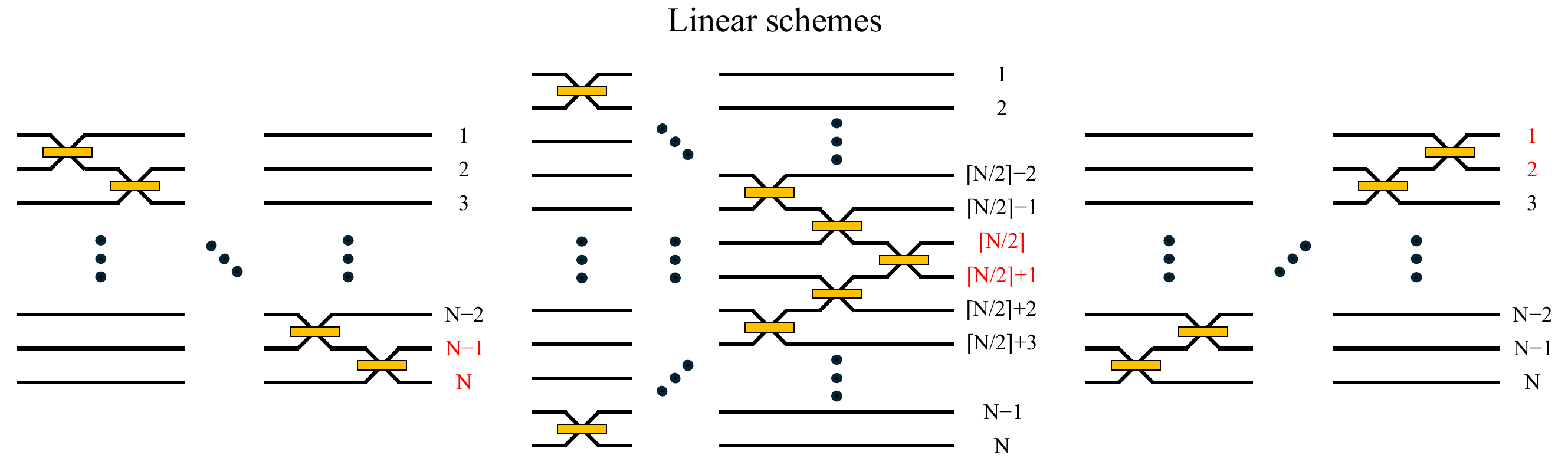}
        \subcaption*{(a)}
    \end{subfigure}
    \hfill
    \begin{subfigure}[b]{\textwidth}
        \centering
        \includegraphics[width=0.97\textwidth]{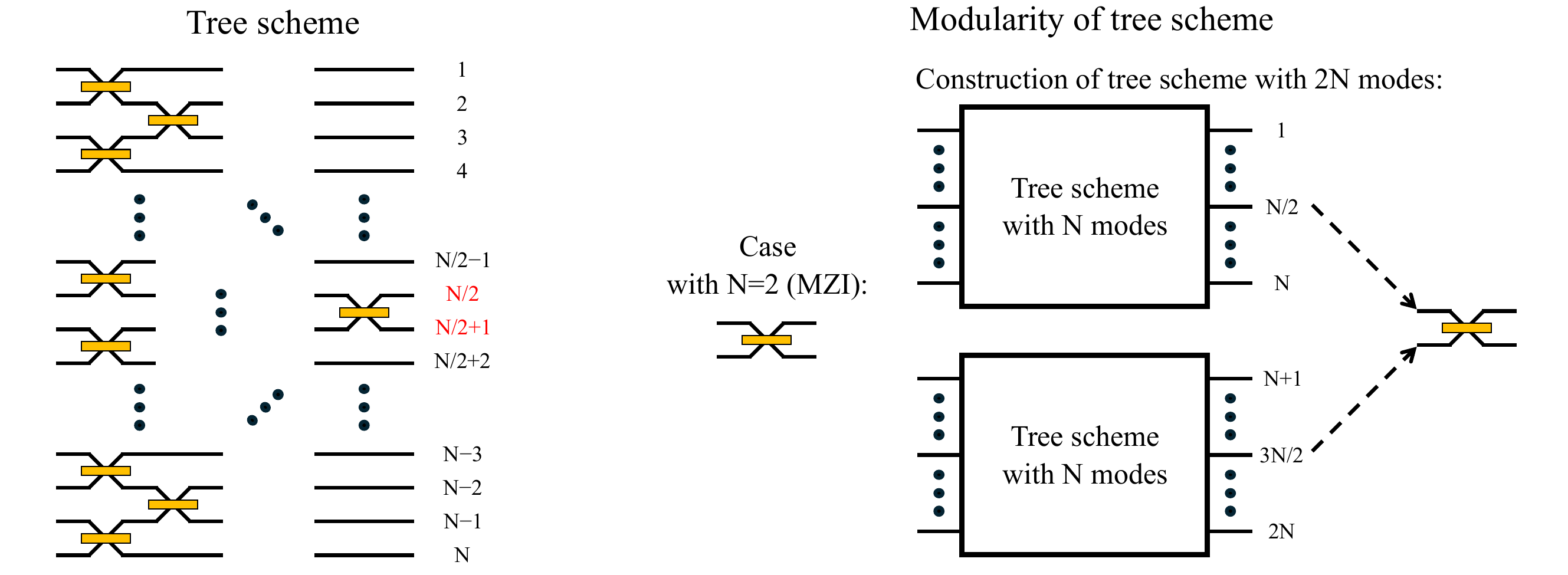}
        \subcaption*{(b)}
    \end{subfigure}
    \caption{\textbf{Multiport routers with $N$ modes.}
    The MZI is represented by the crossing lines with the yellow rectangle. The input enters from the left to the right, and only one particular input superposition is routed to the reference output by choosing the associated phase setting of the MZIs. Both schemes have $N-1$ MZIs for $N$ modes.
    The red labels in the outputs are the possible reference output modes for each scheme.
    \textbf{(a)} Three examples of linear schemes. Among the linear multiport interferometers, the central configuration is denoted as the linear V-shaped scheme. Besides the 'V' shape, there are diagonal shapes, but any other intermediate configuration is allowed.
    \textbf{(b-left)} The tree scheme \cite{Wang_2018} is characterized by a number of modes equal to a power of two and by $N/2$ MZIs in the first layer, $N/4$ MZIs in the second, and so on. 
    \textbf{(b-right)} The recursive procedure to build a tree scheme with $2N$ modes given a tree scheme with $N$ modes, with $N$ equal to power of two.
    }
    \label{fig:router_scheme}
\end{figure}

As said in Section \ref{sec:design}, the linear schemes and the tree scheme, shown in Fig. \ref{fig:scheme_8dim}(right column) for eight modes and in Fig. \ref{fig:router_scheme} for $N$ modes, are not universal schemes.
However, they can execute the unitary routing of any input state, such as one state in the set $\{|\psi_k^{\mathbf{u}}\rangle \}_{k\in 1\ldots N}$, to the reference output state $|r\rangle$, associated with the specific mesh.
The choice of the reference output state for the routing is arbitrary, and it assesses the {\em flexibility} of linear schemes. As shown in Fig. \ref{fig:router_scheme}(a), we can move from a 'V' shape to diagonal shapes or any other intermediate configuration. Each choice of linear scheme is associated with a reference output where the detection is performed. In particular, we denote the central mesh of Fig. \ref{fig:router_scheme}(a) as the linear V-shaped scheme.
Note that for the linear V-shaped scheme and the tree scheme, there are only two options for the reference output mode: either $|\lceil N/2\rceil\rangle$ or $|\lceil N/2\rceil+1\rangle$.

To establish the universal routing capability of both linear schemes and the tree scheme, we begin with the two-dimensional case, corresponding to a single MZI. Given the two complex components of a generic state $|v\rangle \in \mathbb{C}^2$, it is always possible to find a pair of phases $(\theta,\phi)$ such that 
\begin{equation}
    \mathbf{U}_{\rm MZI}[\theta,\phi]\, |v\rangle = \quad {\rm e}^{{\rm i} \gamma_{1} } |1\rangle \quad \mbox{or} \quad {\rm e}^{{\rm i} \gamma_{2} }|2\rangle
    \, ,
    \label{eq:mziasrouter}
\end{equation}
with $\gamma_{1/2}$ being global phases depending on $ |v\rangle$ and  $(\theta,\phi)$.
Indeed, since the MZI can prepare any superposition at the output for both computational basis inputs, it is also true that it can route a specific superposition to one of the two possible reference output modes. 
Then, to demonstrate that the generic unitary router $\mathbf{U}_{k,r}$ of $N$-mode states can be implemented via either the linear schemes or the tree scheme, we present a constructive algorithm similar to the cases of Reck or Clements schemes and based on the routing capability of the single MZI.
Indeed, by concatenating $N-1$ MZIs, it is possible to execute $N-1$ times the single MZI routing action, Eq. \eqref{eq:mziasrouter}. 
The difference between the linear and the tree meshes reveals the different reduction algorithm for the $N-1$ components of the input state.
In the case of the linear schemes, the reduction of the input state is linear, while in the case of the tree scheme, the input state is binary reduced. Specifically, each layer of linear schemes removes one or two components of the input state, whereas each layer of the tree scheme reduces half of the components.
Among the linear schemes, we focus on the linear V-shaped scheme and reference output mode $|\lceil N/2\rceil\rangle$. However, the same procedure with a different reference output mode can be applied to obtain the generic routing action on any desired reference output through the associated linear scheme shape.
For the linear V-shaped scheme, the two MZIs of the first layer \footnote{To be precise, if $N$ is even, all layers of the linear V-shaped scheme contain two MZIs, except the first one, while if $N$ is odd, both the first and the last layers contain one MZI.} route the components of the mode pairs $(1,2)$ and $(N-1,N)$ to the second and $(N-1)$-th modes, respectively. Then, the previous operation is iterated on the resulting state seen as a $(N-2)$-dimensional input state.
In the case of the tree scheme, the first layer consists of $N/2$ MZIs, one acting on each pair of modes. By exploiting their routing action, the output state after the first layer has at most $N/2$ non-zero components. The same binary reduction can then be applied recursively to the resulting $(N/2)$-dimensional state.
At the end of this procedure for both schemes, the output state contains a non-zero amplitude only for the middle mode $|\lceil N/2\rceil\rangle$.
Appendices \ref{app:vshape} and \ref{app:triangular} contain further details on the summarized procedures and decomposition algorithms for the linear V-shaped scheme and the tree scheme, respectively.

\begin{figure}[t]
\centering
        \includegraphics[width=0.75\textwidth]{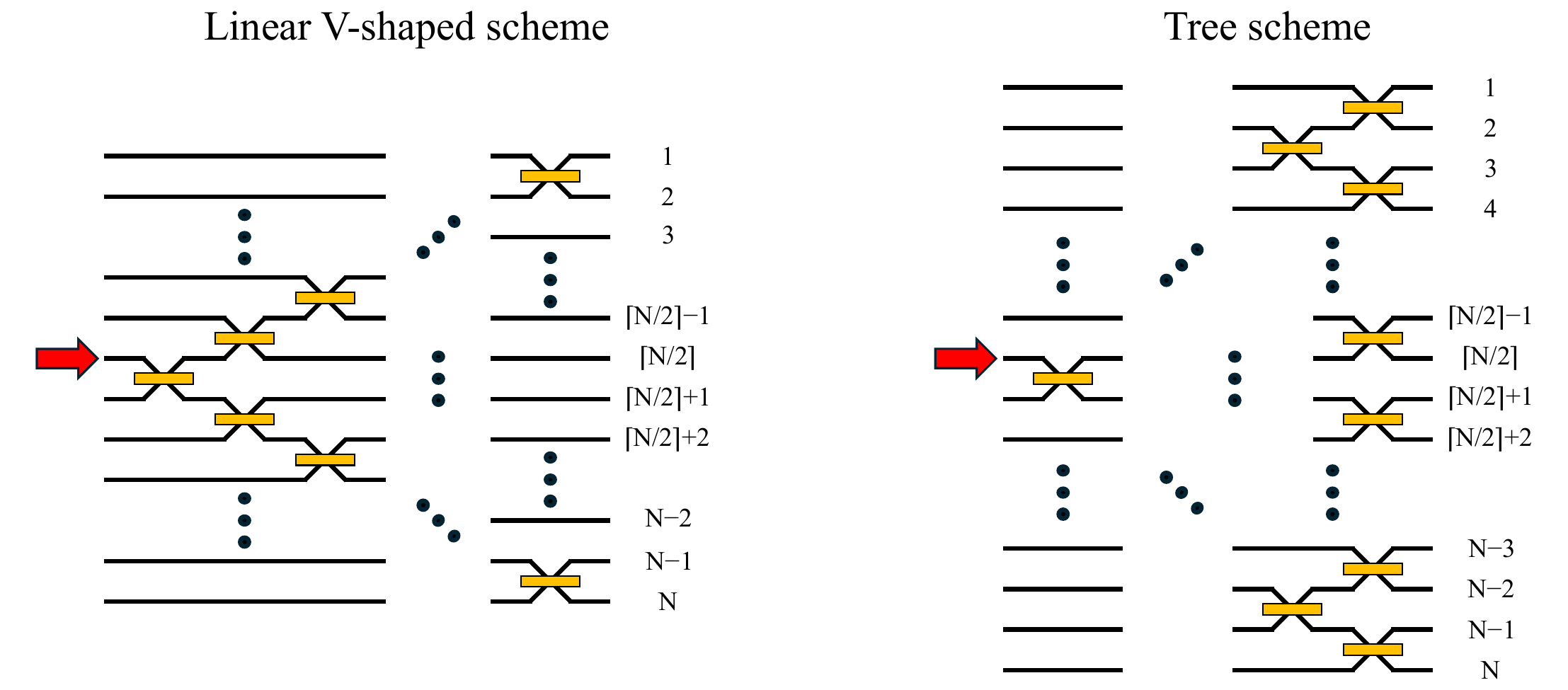}
    \caption{\textbf{Multiport splitters with $N$ modes.}
    The linear V-shaped scheme, on the left, and the tree scheme, on the right, are possible configurations for generic splitters.
    The MZI is represented by the crossing lines with the yellow rectangle. The input is highlighted by the red arrows on the left of the schemes, i.e. on the middle mode $|\lceil N/2\rceil\rangle$. By choosing the precise phase setting of the MZIs, it is possible to coherently and arbitrarily split classical coherent states and a single-photon state. 
    Note that these schemes are just the mirrored version of the routing schemes reported in Fig. \ref{fig:router_scheme}.
    Among linear schemes, we choose the linear V-shaped scheme because it is the most convenient for practical uses.
    }
    \label{fig:splitter_scheme}
\end{figure}

Regarding the multiport splitters, let's consider the schemes in Fig. \ref{fig:splitter_scheme}, which are the mirrored versions of the multiport routers.
The generic splitting means that we inject the input state in one reference mode, and we want to prepare the most generic $N$-mode output state. Without loss of generality, we can choose the input in the middle mode $|\lceil N/2\rceil\rangle$. In the case the state to be prepared has the form 
$|\psi_k^{\mathbf{u}}\rangle \equiv \mathbf{U}^\dagger\, |k\rangle$ defined in  Eq. \eqref{eq:psikroutedvector},
the splitting operation $\mathbf{S}^{\mathbf{u}}_{k,\lceil N/2\rceil}$ is exactly the conjugate transpose of the generic router $\mathbf{R}^{\mathbf{u}}_{k,r}$:
\begin{equation}
    \mathbf{S}^{\mathbf{u}}_{k,\lceil N/2\rceil} = (\mathbf{R}^{\mathbf{u}}_{k,r})^\dagger=|\psi_k^{\mathbf{u}}\rangle\langle \lceil N/2\rceil| 
    \label{eq:genericsplitter}
\end{equation}
while the unitary transformations $ \mathbf{U}^{\rm split}_{k,\lceil N/2\rceil}$ implementing the splitting operations are the adjoint of the unitary routers $\mathbf{U}_{k,\lceil N/2\rceil}$ and by Eq. \eqref{eq:unitaryU_as_UV} have the form
$\mathbf{U}^{\rm split}_{k,\lceil N/2\rceil}= \mathbf{U}_{k,r}^\dagger = {\rm e}^{- {\rm i} \Gamma_k}\,\mathbf{S}^{\mathbf{u}}_{k,\lceil N/2\rceil} +\mathbf{V}_k^\dagger$. The conjugate transposition is captured by the mirroring action on the schemes in Fig. \ref{fig:router_scheme}, keeping the MZI phase settings of the associated routing. 
Practically, the conjugate transposition means that the routing schemes are flipped (inputs$\iff$outputs), the input is the middle state $|\lceil N/2\rceil \rangle$ and we can prepare the generic superposition of a single-photon state or a classical coherent state at the $N$ outputs. Due to the direct relation between routers and splitters, what has been said about the resources needed for the routers also holds for the splitters. Thus, the minimum number of MZIs for $N$-mode splitters is $N-1$.

Although the Reck and Clements architectures can implement arbitrary routing and splitting operations, they require many redundant MZIs for these tasks. In contrast, the linear and tree schemes achieve generic routing and splitting with the minimum number of MZIs.
In the case of linear schemes, the number of modes $N$ can be any odd or even number. There are $N-1$ possible configurations, each associated with two possible choices of the reference output mode, which can be one of the output modes of the final MZI. For the linear V-shaped scheme with odd mode numbers, the first layer is simply composed of one MZI acting on the first and second modes.
Generally, as in the two-dimensional case, the choice of the output mode in which we want to route the input is arbitrary. It is enough to modify the procedure and the 'vertex' of the linear scheme can be moved to any position, moving from a 'V' shape to diagonal shapes. However, the linear V-shaped mesh with routing to mode $|\lceil N/2\rceil\rangle$ or $|\lceil N/2\rceil+1\rangle$ has a better behaviour for lossy MZIs, since the maximum loss for an $N$-mode multiport router is $\alpha_{\rm IL}^{\lceil N/2\rceil}$, assuming the same loss $\alpha_{\rm IL}$ for all MZIs, contrary to the diagonal cases with maximum loss equal to $\alpha_{\rm IL}^{ N-1}$. Indeed, for the linear schemes the MZI layers span from $N-1$ to $\lceil N/2\rceil$, whose minimum value is achieved by the linear V-shaped scheme.
Thus, the choice of the middle modes $|\lceil N/2\rceil\rangle$ or $|\lceil N/2\rceil+1\rangle$ is not necessary, but convenient for practical uses.
In the case of the tree scheme, the number of modes must be a power of two and the number of MZI layers is $\log_2 N$. Practically, to extend the tree scheme to a generic number of modes, it is enough to add auxiliary modes.
The advantage of having a number of modes equal to a power of two is based on the modularity of the tree scheme, shown in Fig. \ref{fig:router_scheme}(b-right). Starting from $N$ modes, in order to double its dimensionality, reaching $2N$ modes, it is enough to make two copies with $N$ modes and connect the outputs of the middle modes associated with each copy to the final MZI. This implies an increasing number of crossings or interruptions of paths as the dimension increases, contrary to linear schemes that do not require any crossings. Without crossings, one can simply truncate the unused paths, but the calibration of the MZIs is more complicated, since it is not possible to address each MZI individually.
Despite the rigidity with respect to the dimensionality and the need for crossings, the tree scheme is very resilient to loss effects, since all paths encounter the same number of MZIs, contrary to the linear V-shaped scheme, where the paths involve from one to $\lceil N/2\rceil$ MZIs. In particular, all paths have losses equal to $\alpha_{\rm IL}^{\log_2 N}$, assuming the same loss $\alpha_{\rm IL}$ for all MZIs.\\
Finally, for a balanced splitting on an output with dimension equal to a power of two, it is preferable to use passive 50:50 devices with the tree scheme, but for a tunable splitting one can choose between the linear V-shaped and tree schemes. 
The tree mesh generally has a smaller footprint, but it requires a number of modes equal to a power of two. The linear V-shaped mesh can be used for a generic number of modes, it does not require path crossings, and it is generally easier to characterize its components.

\section{Universality through multiple projective measurements }
\label{sec:genuniman}

After establishing the routing capability of routing schemes for coherent states and single-photon states, the next step to demonstrate our concept is to show that the linear and the tree schemes are able to implement the generic unitary transformation $\mathbf{U}$ through multiple projective measurements. This means that these schemes perform the $N$ generic unitary routers $\mathbf{U}_{k,r}$, associated with $\mathbf{U}$ through Eq. \eqref{eq:genericrouter}, and each router is followed by a measurement on the reference output mode. The universality of these multiport interferometers is recovered by dividing the measurement process on the chosen reference output mode in $N$ temporal slots. 
Since the actions of multiport interferometers on classical coherent states and a single-photon state are equivalent, it is enough to prove a result for just one of the two cases. We choose to consider the single-photon case, and we use the bra-ket notation summarized at the beginning of Section \ref{sec:design}. 

We consider a general initial state $\rho_{in}$ of the single photon on the $N$ input modes and the measurement set at the outputs of the interferometer associated with the projection operators $\{\mathbf{P}_k = |k\rangle\langle k|\}_{k=1\ldots N}$.
If the interferometer is universal, a generic unitary matrix $\mathbf{U}\in U(N)$ can be implemented, while if the interferometer has a routing mesh, the unitary matrices $\{\mathbf{U}_{k,r}\}_{k=1\ldots N}$, related to $\mathbf{U}$ through Eq. \eqref{eq:router_to_uni_router}, are available.
In both cases, the detection probabilities $\{p_k\}_{k\in 1\ldots N}$ can be estimated:
\begin{equation}\label{condition-Uk-1}
     p_k={\rm Tr}\,\left[ \mathbf{P}_k \,\mathbf{U} \,\rho_{in} \,\mathbf{U}^\dagger\right]={\rm Tr}\,\left[ \mathbf{P}_{r} \,\mathbf{U}_{k,r} \,\rho_{in} \,\mathbf{U}_{k,r}^\dagger\right]  \,,
\end{equation}
where $|r\rangle$ is the reference output mode of the unitary routers $\{\mathbf{U}_{k,r}\}_{k=1\ldots N}$.
The first equality is the usual result, while the second equality shows the new scenario where only a single detector is used, e.g. the one corresponding to the mode $|r\rangle$, and the detection probabilities $\{p_k\}_{k\in 1\ldots N}$ occur with $N$ MZI phase settings, each associated with a unitary $\mathbf{U}_{k,r}\in U(N)$.\\
By linearity, without loss of generality, we consider pure states $ \rho_{in}=|\psi\rangle\langle\psi|$ and Eq. \eqref{condition-Uk-1} becomes:
  \begin{equation}\label{condition-Uk-2}
  |\langle k | \,\mathbf{U}\, |\psi\rangle |^2 =|\langle r|\,\mathbf{U}_{k,r} \,|\psi\rangle |^2\qquad ,\quad\forall |\psi\rangle \in \mathbb{C}^N\,.
\end{equation}
The previous identity is equivalent to
\begin{equation}\label{condition-Uk-2bis}
  \langle \psi|\,\mathbf{U}^\dagger\, |k\rangle  =e^{i\alpha (\psi)}\langle \psi|\,\mathbf{U}_{k,r}^\dagger\,|r\rangle \qquad,\quad \forall |\psi\rangle \in \mathbb{C}^N\,,
\end{equation}
where $\alpha$ is a phase that can depend on $|\psi\rangle$. 
A sufficient condition for \eqref{condition-Uk-2} and \eqref{condition-Uk-2bis} is:
\begin{equation}\label{condition-Uk-3}
  |r\rangle=e^{i\Gamma_k}\,\mathbf{U}_{k,r}\cdot\mathbf{U}^\dagger\, |k\rangle\ \,,
\end{equation}
with $k\in1\ldots N$ and $\Gamma_{k}$ a global phase depending on $\mathbf{U}$ and $|k\rangle$.
In fact, according to condition \eqref{condition-Uk-3}, $\mathbf{U}_{k,r}$ is required to be an element of $U(N)$ mapping the unitary vector $\mathbf{U}^\dagger \, |k\rangle=(\bar u_{k1},\dots,\bar u_{kN} )$ to the vector $|r\rangle$ up to an additional phase. This peculiar degree of freedom is fundamental in the reduction of the number of resources necessary to practically implement $\mathbf{U}_{k,r}$. Indeed, the possibility to tune the additional phase $\Gamma$ allows to realize $\mathbf{U}_{k,r}$ as a product of block matrices, each acting only on a couple of modes \emph{in a non-universal way} through the linear schemes or the tree scheme.
More specifically, the interferometric configuration implemented relies on MZIs acting on mode pairs as described in Eq. \eqref{eq:mziasrouter}.

Eq. \eqref{condition-Uk-1} captures the initial statement on universality, based on the fact that each unitary router $\mathbf{U}_{k,r}$ followed by the projection to the reference output mode executes the generic manipulation $\mathbf{U}$ followed by the projection to the mode $|k\rangle$.
In the first equality of Eq. \eqref{condition-Uk-1}, the measurement process takes place at different locations $k$ simultaneously, as usually done for multiport interferometers. In the last equality of Eq. \eqref{condition-Uk-1}, the measurement process is performed at the same location at different times associated with different unitary routers $\mathbf{U}_{k,r}$. 
Therefore, the probability distribution $\{p_k\}_k$ associated with the sampling at the $N$ outputs of a generic unitary transformation $\mathbf{U}$ for classical coherent light and one single-photon state can be equivalently estimated by sampling at one output of the $N$ unitary transformations implementing all routers associated with $\mathbf{U}$, i.e.
\begin{equation}
    \mathbf{U} + N \;\mbox{detectors at modes}\,\{ |k\rangle\}_{k\in1\ldots N} \iff \{ \mathbf{U}_{k,r}+ 1 \;\mbox{detector at mode}\,|r\rangle \}_{k\in1\ldots N} \,,
\end{equation}
where unitary routers are given in Eq. \eqref{eq:unitaryU_as_UV}.

To conclude, we stress the fact that the previous result holds for classical coherent states and a single-photon state.
For boson sampling with more than one photon, we cannot work with only one detector, since it is not enough to recover the measurement statistics.
Thus, for an input state with $m$ photons, we consider $m$ detectors, placed in the mode set $\{r_j\}_{j\in 1\ldots m}$. Then, in order to understand the structure of the multiport interferometer able to reproduce the standard result of scattershot boson sampling, we modify Eq. \eqref{eq:unitaryU_as_UV} into the following form: 
\begin{equation}
\mathbf{U}_{\mathbf{k},\mathbf{r}}^{(m)} = \sum_{j =1 \ldots m}{\rm e}^{i\gamma_{j,k}}\mathbf{R}^{\mathbf{u}}_{k_j,r_j}+\mathbf{V}_{\mathbf{k}}^{(m)} \qquad\mbox{with}\,\,\
    \mathbf{R}^{\mathbf{u}}_{k,r_j} \equiv \sum_{n=1}^N u_{k,n} |r_j\rangle\langle n| = | r_j\rangle \langle \psi_k^{\mathbf{u}}|  \,,
    \label{eq:unitaryU_as_UV_morephotons}
\end{equation} 
where $\mathbf{k}=(k_1,\ldots,k_m)$, $\gamma_{j,k}\in [0,2\pi]$, $\mathbf{R}^{\mathbf{u}}_{k,r_j}$ is a routing operator where the middle mode is substituted with the output mode $|r_j\rangle$, Eq. \eqref{eq:genericrouter}, and $\mathbf{V}_{\mathbf{k}}^{(m)}$ is an operator with rank $(N-m)$, kernel equal to $\mbox{span}\{|\psi_{k_1}^{\mathbf{u}}\rangle, \ldots, |\psi_{k_m}^{\mathbf{u}}\rangle \}$, Eq. \eqref{eq:psikroutedvector}, and image orthogonal to $\mbox{span}\{|r_1\rangle,\ldots, |r_m\rangle \}$. 
We denote the maps $\mathbf{U}_{\mathbf{k},\mathbf{r}}^{(m)}$ as {\em unitary multi-routers}.
There are $N!/(m!(N-m)!)$ combinations of output modes $\{k_1,\ldots,k_m\}$ and consequently $N!/(m!(N-m)!)$ unitary multi-routers, since the probability in the output does not depend on the order for indistinguishable photons.
As shown in Appendix \ref{app:multilinear}, the sampling at the $N$ outputs of a generic unitary transformation $\mathbf{U}$ for an input state composed of $m$ photons is equivalent to the sampling at $m$ outputs of the $N!/(m!(N-m)!)$ unitary transformations implementing all multi-routers associated with $\mathbf{U}$, i.e.
\begin{equation}
    \mathbf{U} + N \;\mbox{detectors} \iff \{ \mathbf{U}_{\mathbf{k},\mathbf{r}}^{(m)}+ m \;\mbox{detectors at modes}\,\{|r_1\rangle,\ldots, |r_m\rangle \} \}_{\mathbf{k}\in [1\ldots N]^m} \,,
\end{equation}
where unitary multi-routers are given in Eq. \eqref{eq:unitaryU_as_UV_morephotons}. 
More specifically, if the occupation number vector of the input state is  $\mathbf{s}=(s_1,...,s_N)$, with $s_i\geq 0$ and $\sum_{i=1}^Ns_i=N$, then the probability distribution of the occupation number vector of the output state $\mathbf{n}=(n_1,...,n_N)$ is given by 
\begin{equation}\label{eq-boson-sampling-ideal} P(\mathbf{n}|\mathbf{s})=\frac{|\mathrm{Perm}(\tilde U_{\mathbf{n},\mathbf{s}})|^2}{\prod_{i=1}^Ns_i!\prod_{j=1}^Nn_j!}=\frac{|\sum_{\sigma\in S_m}\prod_{\alpha=1}^m\langle k_\alpha|\mathbf{U}| j_{\sigma(\alpha)}\rangle|^2}{\prod_{i=1}^Ns_i!\prod_{j=1}^Nn_j!}\,, \end{equation}
where $\tilde U_{\mathbf{n},\mathbf{s}}$ is the $m\times m$ matrix obtained from $\mathbf{U}$ by repeating the rows according to the output occupation numbers and the columns according to the input ones, while $\mathrm{Perm}(\tilde U_{\mathbf{n},\mathbf{s}})$ denotes its permanent (see Appendix \ref{app:multilinear_BS}). On the right hand side,  $\{j_1,...,j_m\}$ are the input modes associated with the non vanishing occupation numbers appearing in  $\mathbf{s}=(s_1,...,s_N)$, and $\{k_1,\ldots, k_l\}$, with $l\leq m$, are the output modes associated with the non vanishing occupation numbers appearing in  $\mathbf{n}=(n_1,...,n_N)$. The sum  is taken over the symmetric group $S_m$, that is, over all permutations $\sigma(\ldots)$ of $\{1,\dots, m\}$.  Thanks to the routing properties of the unitaries $\mathbf{U}_{\mathbf{k},\mathbf{r}}^{(m)}$, the probabilities \eqref{eq-boson-sampling-ideal} can be equivalently computed as:
\begin{equation}\label{eq-boson-sampling-multirouters}  P(\mathbf{n}|\mathbf{s})=\frac{|\sum_{\sigma\in S_m}\prod_{\alpha=1}^m\langle r_\alpha|\mathbf{U}_{\mathbf{k},\mathbf{r}}^{(m)}| j_{\sigma(\alpha)}\rangle|^2}{\prod_{i=1}^Ns_i!\prod_{j=1}^Nn_j!} \,.
\end{equation}
The previous equivalence implies that, for any initial state with $m$ photons, the statistics of detector clicks associated with the photon detection at the outputs $\mathbf{k}$ can be estimated through the detection at the reference outputs $\{r_j\}_{j\in 1\ldots m}$.

\begin{figure}[t]
\centering
        \includegraphics[width=0.45\textwidth]{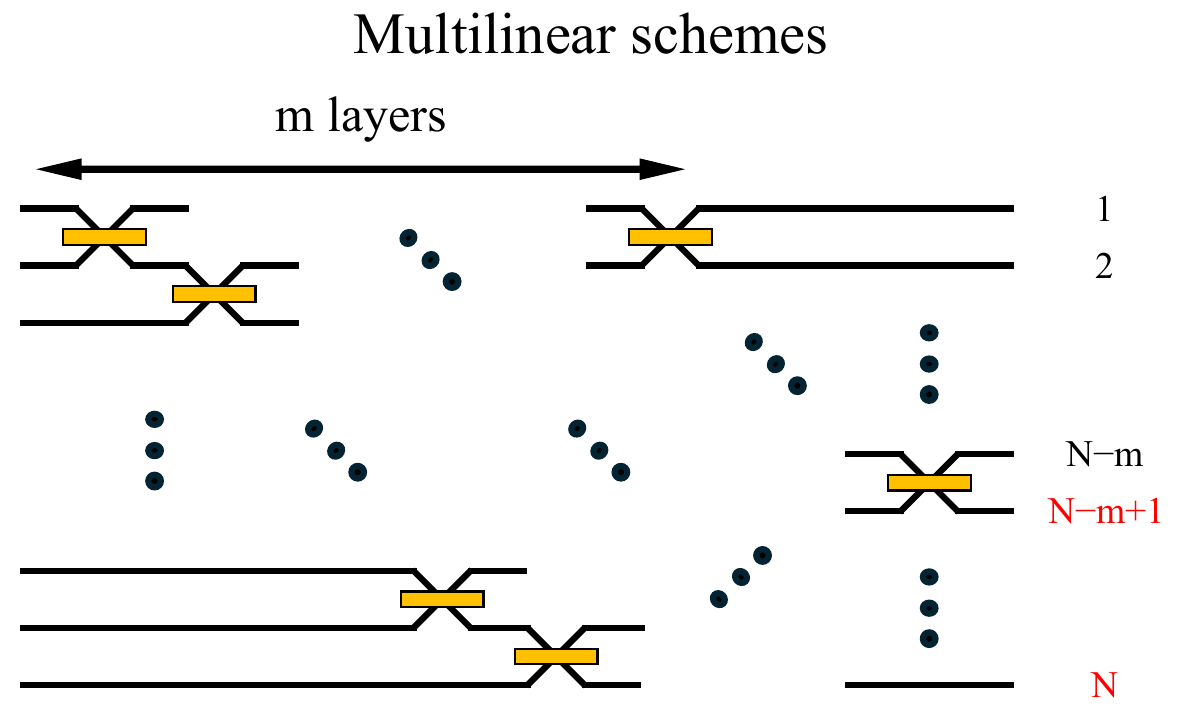}
    \caption{\textbf{Multi-router with $N$ modes for $m$ input single photons.}
    The MZI is represented by the crossing lines with the yellow rectangle. The input enters from the left to the right, and through $N!/(m!(N-m!)$ phase settings of the MZIs it is possible to find the correct statistics of clicks by looking only at the reference output modes. The scheme has $m\,(N-(m+1)/2)$ MZIs for $N$ modes.
    The red labels in the outputs give a possible set of reference output modes.
    Note that for $m=1$ we recover the first linear scheme of Fig. \ref{fig:router_scheme}(a).
    The multilinear scheme can be seen as a combination of $m$ linear schemes, where the first has dimension $N$, the second $N-1$ and so on.
    Appendix \ref{app:multilinear} contains further details about how these multilinear schemes work.
    }
    \label{fig:multirouter_scheme}
\end{figure}

Analogously to the unitary routers, unitary multi-routers are not uniquely identified by Eq. \eqref{eq:unitaryU_as_UV_morephotons}. 
Also in this case, we exploit these degrees of freedom to realize the unitary multi-router $\mathbf{U}_{\mathbf{k},\mathbf{r}}^{(m)}$ with the minimum number of resources. From the mathematical point of view, this means to decompose the unitary $\mathbf{U}_{\mathbf{k},\mathbf{r}}^{(m)}$ as the product of a minimum number of  $U(2)/U(1)^2$ maps, while from the physical point of view this consists of looking for meshes made of the minimum number of MZIs. 
The number of free parameters of a multi-router $\mathbf{U}_{\mathbf{k},\mathbf{r}}^{(m)}$ is $2m \, (N-(m+1)/2)$ and two is the dimension of the set $U(2)/U(1)^2$, hence we aim at decompositions of unitary routers made of $m \, (N-(m+1)/2)$ elements. This result can be found from the dimension of the $m$ vectors $\{|\psi_{k_1}^{\mathbf{u}}\rangle, \ldots, |\psi_{k_m}^{\mathbf{u}}\rangle \}$ , i.e. $2m\,(N-1)$, and taking into account the orthogonality constraints, which are $m\,(m-1)/2$. This implies that the desired multiport interferometers able to implement a generic unitary multi-router $\mathbf{U}_{\mathbf{k},\mathbf{r}}^{(m)}$, associated with the unitary map $\mathbf{U}$, consist of $m \, (N-(m+1)/2)$ MZIs. Moreover, since the measurement is done on the reference modes $\{r_j\}_{j\in 1\ldots m}$, the structure of $\mathbf{V}_{\mathbf{k}}^{(m)}$, Eq. \eqref{eq:unitaryU_as_UV_morephotons}, does not require any constraints, and its action can be neglected.
In this way, the standard result with the permanent of $m$-dimensional sub-matrices of $\mathbf{U}$ with rows $\mathbf{k}=(k_1,\ldots,k_m)$ can be obtained with any scheme associated with the transformation in Eq. \eqref{eq:unitaryU_as_UV_morephotons}. Such a scheme uses the minimum number of MZIs and detectors, but it requires $N!/(m!(N-m)!)$ manipulations, each associated with a corresponding time slot to acquire the statistics.
We point out that for $N$ input single photons, one for each input mode, the number of MZI matches exactly the case of universal schemes, and we need only one phase setting and one time slot for the sampling. From this perspective, the multi-routing schemes provide a bridge between the routing scheme and the universal schemes. The maximum number of needed multi-routers to perform the same sampling is achieved for $m= N/2$ for $N$ even, and $m= (N\pm 1)/2$ for $N$ odd.

Figure \ref{fig:multirouter_scheme} shows an example of a multi-router interferometer belonging to a class of meshes that we refer to as multilinear schemes, which are capable of implementing generic unitary multi-routing transformations. These schemes use the minimum number of MZIs and can be constructed by starting from an $N$-mode Reck scheme and retaining only $m$ diagonals. One representative of this family is the incomplete Reck scheme shown in Fig. \ref{fig:multirouter_scheme}. By contrast, finding the analogue of the tree mesh for multi-routing does not appear to be straightforward. Appendix \ref{app:multilinear} provides further details on the construction, decomposition algorithms, and routing procedures for multilinear schemes.

\section{Tolerance to losses and phase noise} 
\label{sec:disc}

In any realistic implementation, there will always be discrepancies between the ideal unitary $\mathbf{U}$ one aims to achieve and the one actually realized by the physical interferometer, due to losses and random fluctuations in the phase parameters associated with the MZI. In the following, we denote the transformation associated with the actual interferometer with $\mathbf{U}^{(\rm exp)}$ for universal schemes and with $\{\mathbf{U}_{k,r}^{(\rm exp)}\}_{k\in 1\ldots N}$ for routing schemes.
As in \cite{clements_optimal_2016}, we consider the fidelity of universal and routing multiport interferometers for lossy MZIs. 
To quantify the fidelity of the transformation implemented by a non-ideal $N \times N$ multiport interferometer, we use the following metrics:
\begin{equation}
    \begin{split}
        F = 
        \begin{cases}
            \frac{\Big|{\rm Tr}\left[ \mathbf{U}^{(\rm exp)} \cdot \mathbf{U}^\dagger\right]\Big|^2}{ N \,{\rm Tr}\left[ \mathbf{U}^{(\rm exp)} \cdot \left(\mathbf{U}^{(\rm exp)}\right)^\dagger\right]} &\qquad \mbox{for universal schemes}\,,
        \\
        \frac{1}{N} \sum_{k=1}^N \frac{| \langle r | \,\mathbf{U}_{k,r}^{(\rm exp)} \cdot \mathbf{U}^\dagger | k\rangle|^2}{ \,\langle r | \,\mathbf{U}_{k,r}^{(\rm exp)} \cdot \left(\mathbf{U}_{k,r}^{(\rm exp)}\right)^\dagger | r\rangle} & \qquad \mbox{for routing schemes}\,,
        \end{cases}
    \end{split}
    \label{eq:fidelity}
\end{equation}
where the fidelity for universal schemes is the one used in \cite{clements_optimal_2016}, and the fidelity for routing schemes such as linear and tree schemes is based on the property of routers associated with a specific unitary transformation given in Eq. \eqref{condition-Uk-3}.
In the case of the tree scheme, auxiliary modes are added to make the number of modes equal to a power of two. 

We point out that this definition of $F$ is ambiguous. Indeed, for universal schemes it is not an invariant of the set $U(N)/U(1)^N$; in other words it depends on the pair of diagonal phase matrices $\mathbf{d}$, Eq. \eqref{eq:uNdecu2_princ}, used for $\mathbf{U}$ and $\mathbf{U}^{(\rm exp)}$. Because of this, $F$ must be evaluated not only using the same MZI mesh and phase-shifter assignment, but also after factoring out the same global phases from the different output modes.
Moreover, $F$ for routing schemes is not sensitive to different global losses associated with the different unitary routers $\mathbf{U}_{k,r}$. This condition can be experimentally observed when different settings of the multiport interferometer produce different insertion losses. In this case, we can have $\mathbf{U}_{k,r}^{\rm(exp)} =10^{-\alpha_k/10}\,\mathbf{U}_{k,r}$, with $\alpha_k$ a positive number quantifying the insertion loss for the configuration $\mathbf{U}_{k,r}$, but $F$ does not change even if all the parameters $\{\alpha_k\}_k$ are different, thus affecting the detection statistics. 
Indeed, since the matrix $\mathbf{U}^{(\mathrm{exp})}$ is no longer unitary due to losses, the detection probabilities associated with the $N$ runs of the experiment, one for each mode $k$, given by
\begin{equation}\label{realistic-probabilities}
\tilde p_k=
\begin{cases}
        |\langle k|\,\mathbf{U}^{(\mathrm{exp})}\,| \psi\rangle |^2 &\qquad \mbox{for universal schemes}\,,
        \\
        |\langle r|\,\mathbf{U}^{(\mathrm{exp})}_{k,r}\,| \psi\rangle |^2 & \qquad \mbox{for routing schemes}\,,
\end{cases}
\end{equation}
will not only differ from the ideal probabilities $p_k$ shown in Eq. \eqref{condition-Uk-1}, but will also be non-normalized:
$\sum_{k=1}^N\tilde p_k<1$.
This is a consequence of the particular experimental configuration: when the detector associated with the measurement of $p_k$ does not click, the event is interpreted as the photon potentially having been detected in one of the other modes, rather than as a loss.\\
Via post-selection, the estimators $\hat p_k$ for the theoretical probabilities $p_k$ are given by
\begin{equation}
    \label{estimator-probabilities}
    \hat p_k \equiv\frac{\tilde p_k}{\sum_{j=1}^N\tilde p_j} \,.
\end{equation}
Contrary to the fidelity, if different $\mathbf{U}_{k,r}^{\rm(exp)}$ have different insertion losses $\alpha_k$, i.e. $\mathbf{U}_{k,r}^{\rm(exp)} =10^{-\alpha_k/10}\,\mathbf{U}_{k,r}$, the probabilities $\{\hat p_k\}_k$, and consequently any related quantity, change.
Figures of merit quantifying the distance between the ideal probability distribution $P=\{p_k\}_k$ and the estimated one $\hat P=\{\hat p_k\}_k$ are provided by the \emph{Kullbach-Leibner (KL) divergence} $KLD(\hat P|P)$ \cite{kullback1997information} and the {\em total variation (TV) distance $TV(P,\hat P)$} \cite{cohn2013measure} defined as:
\begin{equation}
    \label{KLdivergence}
    KLD(\hat P|P)=\sum_{k=1}^N\hat p_k\log\left(\frac{\hat p_k}{p_k}\right)\,,
    \quad \mbox{and} \qquad
    TV(P,\hat P)=\frac{1}{2}\sum_{k=1}^N|p_k-\hat p_k|\,.
\end{equation}
Note that KL divergence and TV distance are zero if $\hat P \sim P$. 
Contrary to the KL divergence, which in principle can attain any non negative value, the TV distance is bounded, as $0\leq TV(P,\hat P)\leq 1$.
We choose to report only the TV distance, since the two figures of merit show essentially the same behavior.

By looking at $F$ and the TV distance, we simulate two different scenarios: a constant loss on MZIs and a random phase noise on the phase shifters setting. For both figures of merit, given mode number $N$, 1000 random unitary matrices are generated using the polar decomposition of complex random matrices \cite{polar_dec}, and are decomposed with respect to the different schemes. In the first scenario, a constant loss to both outputs of all MZIs in the different multiport schemes is added: thus, we simply introduce a constant factor $10^{-\alpha_{\rm IL}/10}$ in front of the MZI matrices, assuming equal insertion loss for every MZI. In the second scenario, the phase setting of the real transformation is modified by adding Gaussian random phase noise with mean equal to zero and standard deviation $\sigma_{\rm phase\,noise}$ and by sampling over 100 random phase noise for each unitary matrix. For TV distance, the ideal and non-ideal probabilities are calculated by considering 100 random normalized complex states $|\psi\rangle$ for each random unitary matrix.

\begin{figure}[!t]
    \centering
    \begin{subfigure}[b]{0.45\textwidth}
       \centering
        \includegraphics[width=\textwidth]{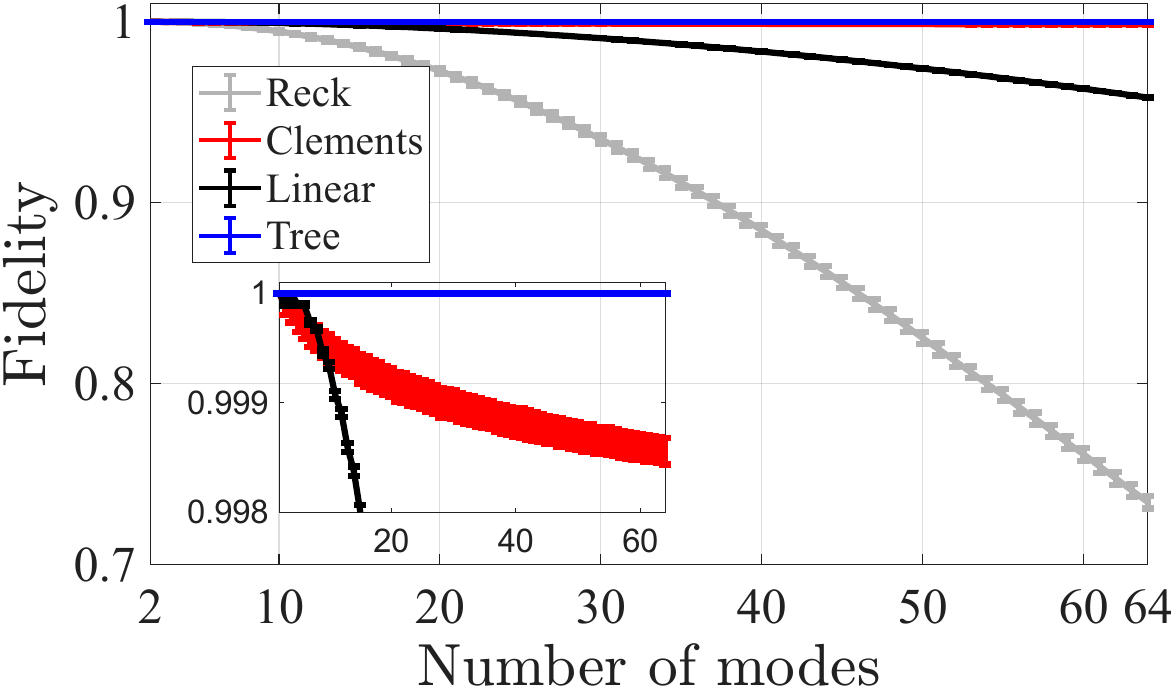}
        \subcaption*{(a)}
    \end{subfigure}
    \hfill
    \begin{subfigure}[b]{0.45\textwidth}
        \centering
        \includegraphics[width=\textwidth]{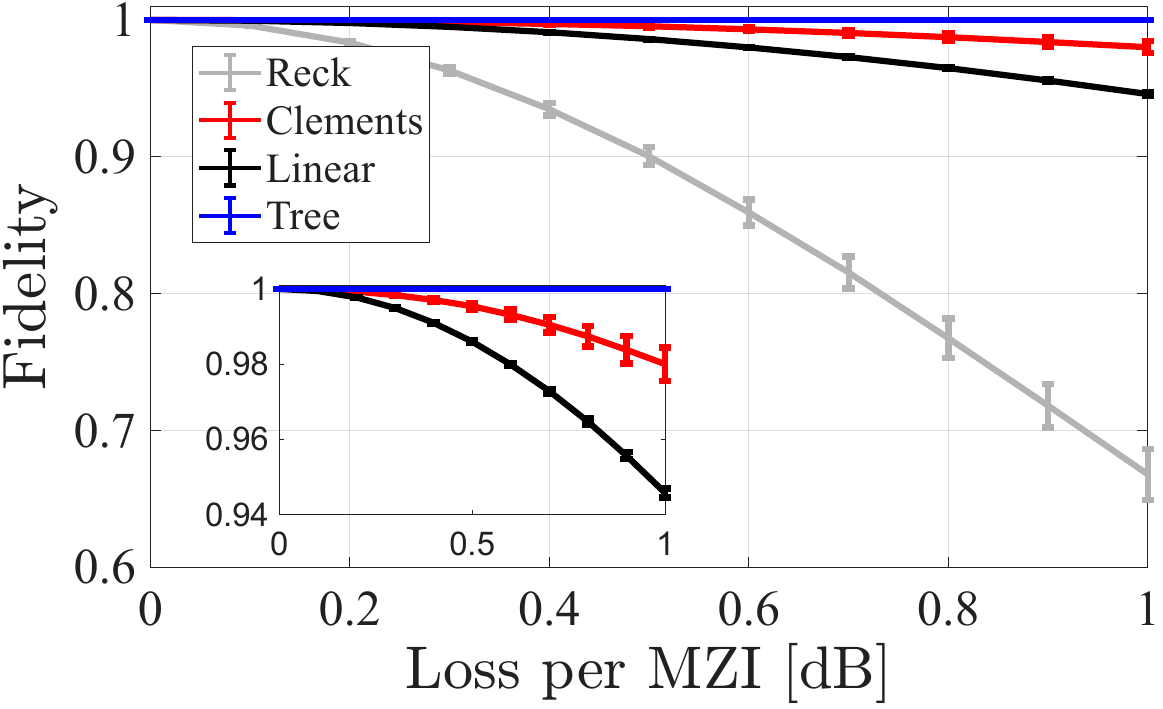}
        \subcaption*{(b)}
    \end{subfigure}
    \begin{subfigure}[b]{0.45\textwidth}
       \centering
        \includegraphics[width=\textwidth]{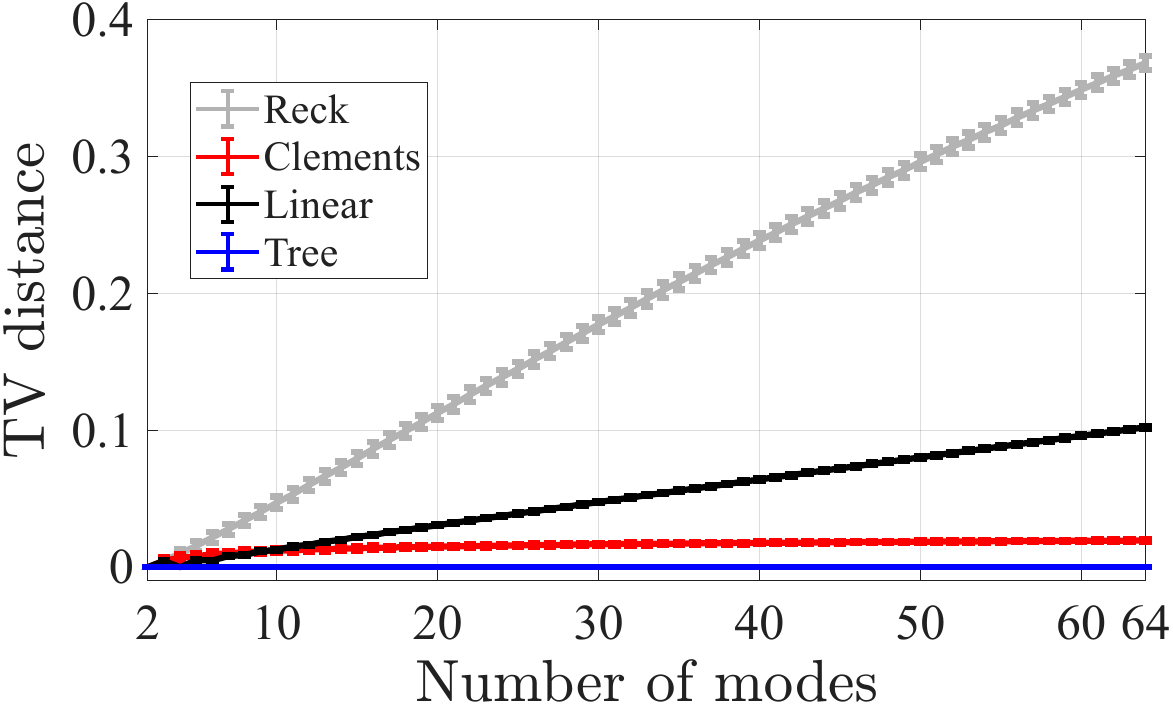}
        \subcaption*{(c)}
    \end{subfigure}
    \hfill
    \begin{subfigure}[b]{0.45\textwidth}
        \centering
        \includegraphics[width=\textwidth]{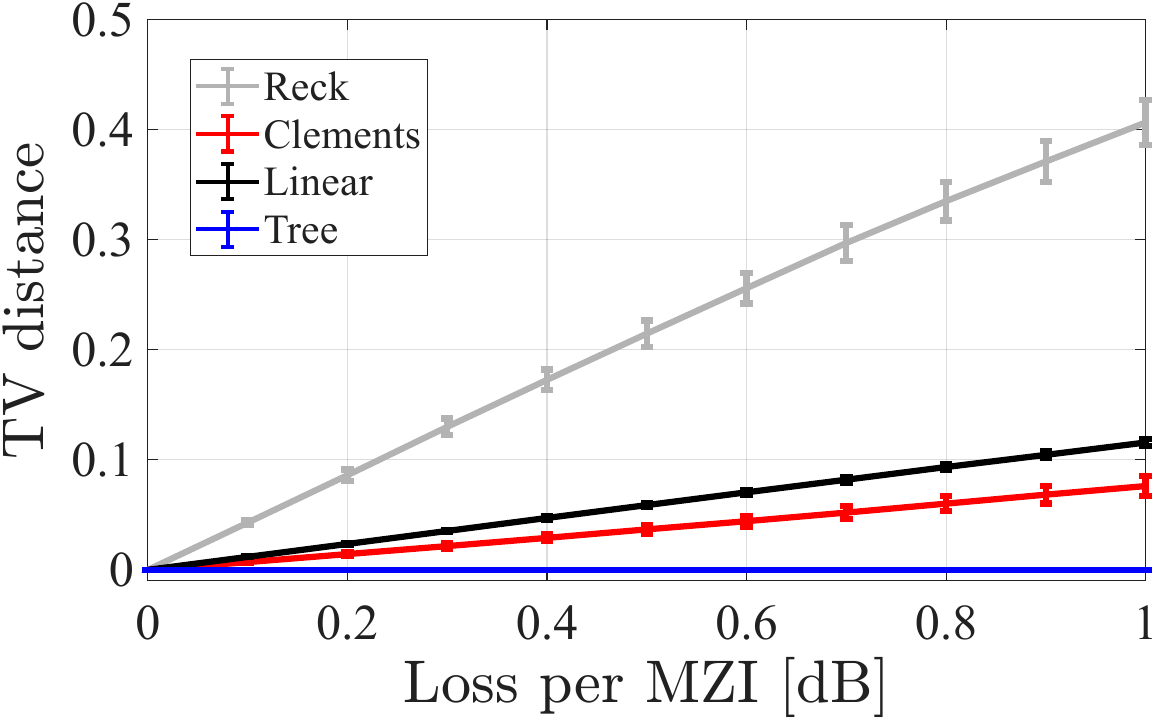}
        \subcaption*{(d)}
    \end{subfigure}
    \caption{ \textbf{Fidelity, Eq. \eqref{eq:fidelity}, and TV distance, Eq. \eqref{KLdivergence}, of the universal schemes and routing schemes for lossy MZIs.} 
    Average \textbf{(a)} fidelity and \textbf{(c)} TV distance for different multiport interferometers with a constant loss of 0.2 dB per MZI for interferometers built according to the Reck scheme (grey), Clements scheme (red), linear V-shaped scheme (black) and tree scheme (blue), for different interferometer sizes.
    Average \textbf{(b)} fidelity and \textbf{(d)} TV distance as a function of loss for $16\times 16$ multiport interferometers. 
     Insets: zoom of the fidelity/TV distance for the Clements, the linear V-shaped and the tree schemes.
    In every plot, the fidelity/TV distance of the tree scheme is always equal to one/zero, since all the paths have the same optical depth. 
    Among the linear schemes, we choose the linear V-shaped scheme since it is the most balanced. Therefore, for both routing schemes (linear and tree) the reference output mode $|r\rangle$ is the middle mode $|\lceil N/2\rceil\rangle$.
    }
    \label{fig:simu_MZIloss}
\end{figure}

Fig. \ref{fig:simu_MZIloss} shows the average fidelity and TV distance over 1000 random unitary transformations implemented with the different schemes as a function of the mode number with fixed constant MZI loss and as a function of constant MZI loss for a fixed mode number.
The result of $F$ for universal schemes is the same as in \cite{clements_optimal_2016}. The linear V-shaped and tree schemes achieve higher fidelity and lower TV distance than the Reck scheme. However, only the tree scheme outperforms the Clements scheme. This can be attributed to the fact that the optical paths in the Clements mesh are more balanced than those in the linear schemes, despite the larger number of components in the former. In particular, the tree scheme always achieves unit fidelity and zero TV distance, since all possible optical paths have the same depth. The linear V-shaped scheme does not share this property. 
From this perspective, the linear V-shaped scheme is the router analogue of the Reck scheme, and the tree scheme is the router analogue of the Clements scheme.
We conclude that the tree scheme is a better choice with respect to the linear V-shaped scheme for lossy MZIs. The drawback of the tree scheme is the need for crossings and auxiliary modes for a generic mode number.
For fewer than 20 modes, nearly unit fidelity and nearly zero TV distance can be achieved either with the Clements scheme or through multiple projective measurements using the linear V-shaped and tree schemes.

\begin{figure}[H]
    \centering
    \begin{subfigure}[b]{0.45\textwidth}
       \centering
        \includegraphics[width=\textwidth]{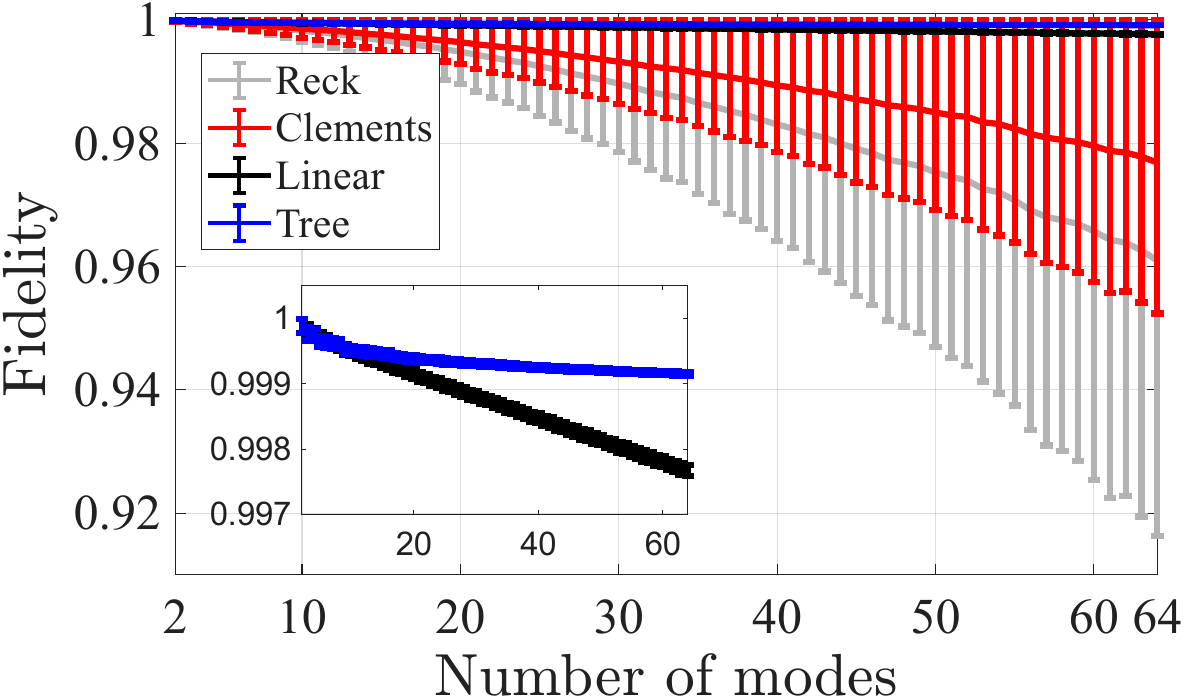}
        \subcaption*{(a)}
    \end{subfigure}
    \hfill
    \begin{subfigure}[b]{0.45\textwidth}
        \centering
        \includegraphics[width=\textwidth]{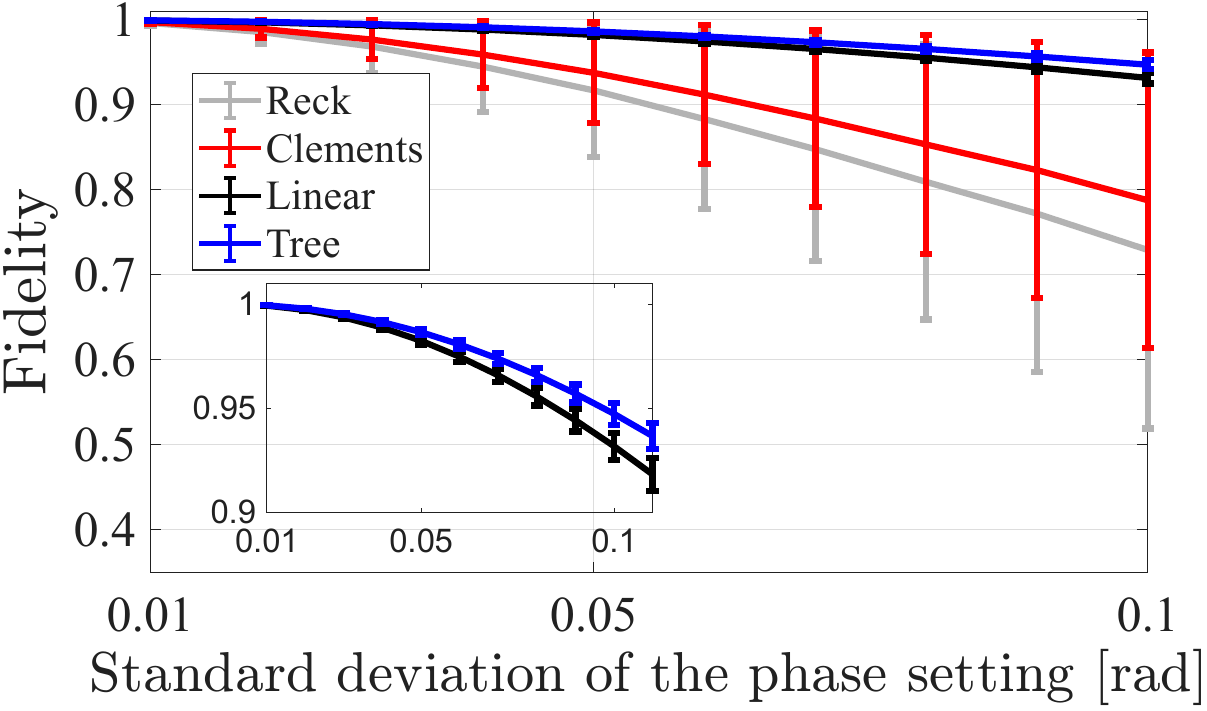}
        \subcaption*{(b)}
    \end{subfigure}
    \begin{subfigure}[b]{0.45\textwidth}
       \centering
        \includegraphics[width=\textwidth]{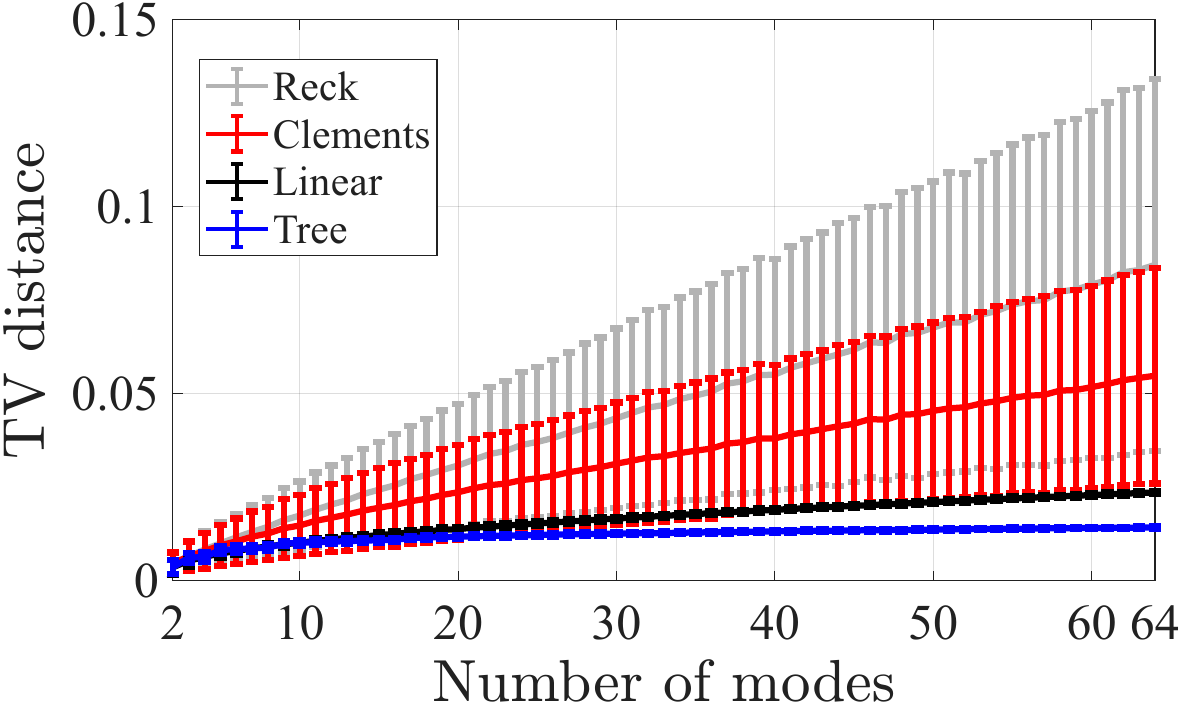}
        \subcaption*{(c)}
    \end{subfigure}
    \hfill
    \begin{subfigure}[b]{0.45\textwidth}
        \centering
        \includegraphics[width=\textwidth]{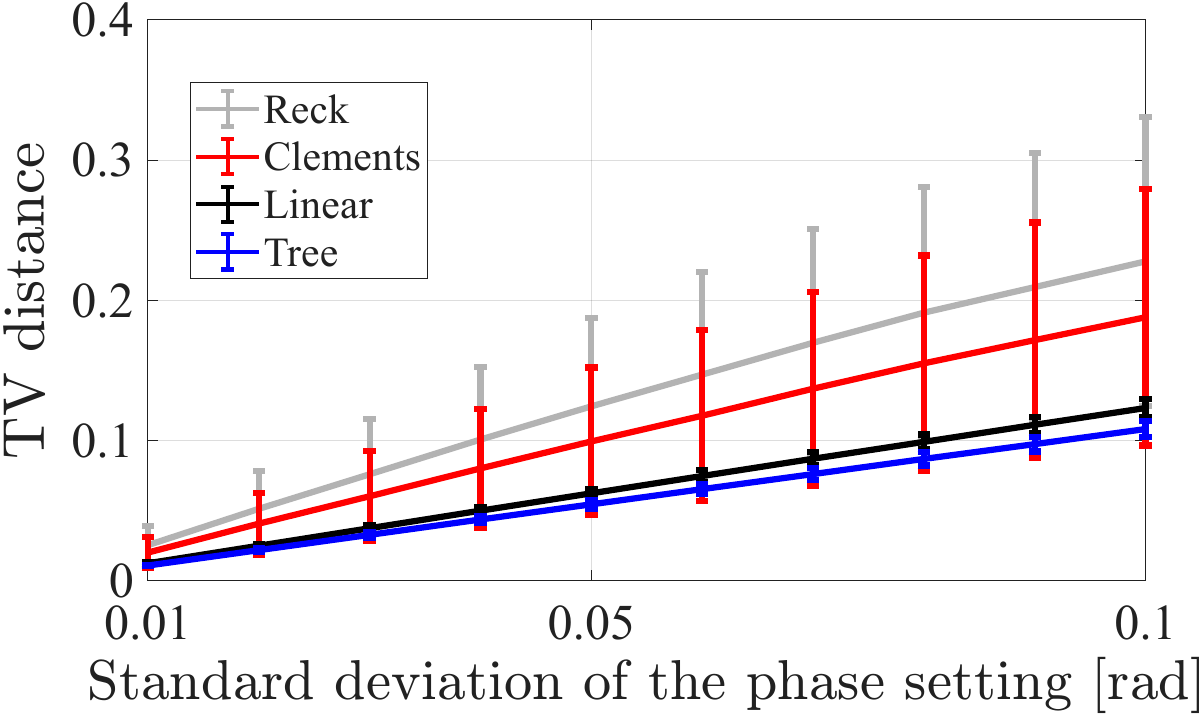}
        \subcaption*{(d)}
    \end{subfigure}
    \caption{ 
    \textbf{Fidelity, Eq. \eqref{eq:fidelity}, and TV distance, Eq. \eqref{KLdivergence}, of the universal schemes and routing schemes for random phase noise.} 
    Average \textbf{(a)} fidelity and \textbf{(c)} TV distance for different multiport interferometers with Gaussian-distributed noise on the phase setting with mean equal to zero and standard deviation equal to 0.01 rad for interferometers built according to the Reck scheme (grey), Clements scheme (red), linear V-shaped scheme (black) and tree scheme (blue), for different interferometer sizes. 
    Average \textbf{(b)} fidelity and \textbf{(d)} TV distance as a function of Gaussian-distributed noise standard deviation for $16\times 16$ multiport interferometers. 
    Insets: zoom of the fidelity for the linear V-shaped and the tree schemes. 
    Among the linear schemes, we choose the linear V-shaped scheme since it is the most balanced. Therefore, for both routing schemes (linear and tree) the reference output mode $|r\rangle$ is the middle mode $|\lceil N/2\rceil\rangle$.}
    \label{fig:simu_phasenoise}
\end{figure}

Fig. \ref{fig:simu_phasenoise} shows the average fidelity and TV distance over 1000 random unitary transformations implemented with the different schemes as a function of the mode number with random phase noise with mean equal to zero and standard deviation equal to 0.01 rad and as a function of phase noise's standard deviation for a fixed mode number.
We can note that the fidelities and the TV distances for Reck and Clements scheme are characterized by wider standard deviations, while for the routing schemes the fidelity/TV distance is decreasing/increasing very slowly and with very small standard deviations.
For universal schemes, the bigger optical depth is the main cause for the decreased fidelity and the increased TV distance of the transformation, since the phase noise of different MZIs can constructively interfere, and thus this effect can be enhanced. For routing schemes, this does not happen, since non-correctly-routed signals are simply lost.
Therefore, even if a unitary transformation requires $N$ temporal slots, the routing schemes are more robust against random noise in the phase settings.
Appendix \ref{app:nonideal} reports the results for random MZI losses and random coupling losses on the outputs of the interferometer. 

We remind the reader that the linear and the tree schemes achieve universality through multiple projective measurements only for classical coherent light and a single-photon state.
A similar analysis can be performed for multi-routing schemes with more than one photon.
In particular, for $m$ single photons, the fidelity can be defined as:
\begin{equation}
        F = 
        \frac{m!(N-m)!}{N!}  \sum_{\mathbf{k}}\frac{\sum_{j=1}^m| \langle r_j | \,\mathbf{U}_{\mathbf{k},\mathbf{r}}^{(\rm exp)} \cdot \mathbf{U}^\dagger | k_j\rangle|^2}{ \sum_{j=1}^m\,\langle r_j | \,\mathbf{U}_{\mathbf{k},\mathbf{r}}^{(\rm exp)} \cdot \left(\mathbf{U}_{\mathbf{k},\mathbf{r}}^{(\rm exp)}\right)^\dagger | r_j\rangle}  \qquad \mbox{for multi-routing schemes}\,,
    \label{eq:fidelity_multi}
\end{equation}
where the sum is over all possible subsets $\mathbf{k}\equiv\{k_1,\dots, k_m\}$ of $m$ modes out of the $N$ possible ones, $\{|r_1\rangle,\ldots, |r_m\rangle \}$ are the reference output modes, and we have dropped the upper $(m)$ of $\mathbf{U}_{\mathbf{k},\mathbf{r}}$. 
As done for the other schemes, other figures of merit are the KL divergence and the TV distance, defined in Eq. \eqref{KLdivergence}, where the probabilities $\tilde p_k$ are replaced by the measurement distribution of occupation number $\mathbf{n}=(n_1,...,n_N)$, with $n_i\geq 0$ denoting the number of photons detected at the mode $i$, fulfilling the constraint $\sum_{i=1}^Nn_i=m$ (see Eq. \eqref{eq-boson-sampling-ideal}).  In the case of universal schemes, the detection probabilities $\tilde P(\mathbf{n}|\mathbf{s})$ that are experimentally accessible are  given by Eq. \eqref{eq-boson-sampling-ideal} with $\tilde U_{\mathbf{n},\mathbf{s}}$ replaced by $\tilde U^{(\mathrm{exp})}_{\mathbf{n},\mathbf{s}}$.
In the case of multi-routing schemes  with reference modes $r_1,\ldots,r_m$, the detection probabilities are given by Eq. \eqref{eq-boson-sampling-multirouters} with $\mathbf{U}_{\mathbf{k},\mathbf{r}}^{(m)}$ replaced by $\mathbf{U}_{\mathbf{k},\mathbf{r}}^{(\rm exp)}$.
Again, as in the cases of one photon and classical coherent light, the probability distribution $\tilde P(\,\cdot\,|\mathbf{s})$ is not normalized due to losses, and the estimators for the theoretical probabilities $P(\,\cdot\,|\mathbf{s})$ are given by:
\begin{equation}
    \hat P(\mathbf{n}|\mathbf{s})=\frac{\tilde P(\mathbf{n}|\mathbf{s})}{\sum_{\mathbf{n}'}\tilde P(\mathbf{n}'|\mathbf{s})} \,,
\end{equation}
where the sum appearing in the denominator is taken over the set of all possible occupation-number vectors $\mathbf{n}$. We remark that when some of the occupation numbers $n_i$ are strictly greater than 1 and, consequently, the output modes $\{k_1,\ldots, k_l\}$ associated with the non-vanishing occupation number are strictly less than $m$, there are $ \binom{N-l}{m-l}$ experimental settings allowing the estimation of the detection probability $\tilde P(\mathbf{n}'|\mathbf{s})$, each associated with a different $m$-tuple of output modes  containing the set $\{k_1,\ldots, k_l\}$. This redundancy can actually be exploited to enhance the detection statistics. Indeed, in this case an experimental estimator for the probability $\tilde P(\mathbf{n}'|\mathbf{s})$ is provided by the ratio $\sum_{\mathbf{k}}N_{\mathbf{k},\mathbf{n}}/\sum_{\mathbf{k}}N_{\mathbf{k}}$, where the sum is over all $m-$tuples $\mathbf{k}$ containing the modes $\{k_1,\ldots, k_l\}$, $N_{\mathbf{k}}$ is the number of runs of the experiment in the setting associated with $\mathbf{k}$ and $N_{\mathbf{k},\mathbf{n}}$ is the corresponding number of counts for the occupation number $\mathbf{n}$.
Numerical results are left for future investigations.

\section{Conclusion}
\label{sec:conclu}

In conclusion, we have demonstrated how the linear and the tree designs for multiport interferometers can perform routing and splitting with the minimum number of MZIs. 
Moreover, we have shown how these multiport routers via multiple projective measurements can perform the universal manipulation of a classical coherent state and a single-photon state. 
Finally, we extend the result to the general case of scattershot boson sampling experiments where the input is composed of multi-photon states and the manipulation is achieved through multiport multi-routers and multiple projective measurements.
Thus, we expect that the presented designs for fully programmable routing and splitting multiport interferometers will play an important role for photonic processors for both classical and quantum applications.

\begin{table}[ht]
\centering
\scriptsize
\begin{tabular}{c c c c c c c c}
\toprule
 Design & MZI & MZI layers & Detectors & Input & Runs & Calibration & Crossings \\ 
\midrule
\midrule
Reck \cite{reck_experimental_1994}  & $\frac{N(N-1)}{2}$ & $2N -3$ & $N$ & any & $1$ & easy & no \\
\midrule
Clements \cite{clements_optimal_2016}  & $\frac{N(N-1)}{2}$ & $N-\delta_{2,N}$ & $N$ & any & $1$ & hard & no \\
\midrule
Braided \cite{Marchesin_25}   & $\frac{N(N-1)}{2}$ & $N-1$ & $N$ & any & $1$ & hard & yes \\
\midrule
Fldzhyan \cite{Fldzhyan_20}   & $N(N-1)^*$ & $2N^*$ & $N$ & any & $1$ & hard & no \\
\midrule
Tree \cite{Wang_2018}  & $N-1$ & $\lceil \log_2 N\rceil$ & $1$ & one photon & $N$ & hard/easy & no/yes \\
\midrule
Linear  & $N-1$ & $(\lceil N/2\rceil,\ldots, N-1)$ & $1$ & one photon & $N$ & easy & no \\
\midrule
Multilinear  & $m \, \left(N-\frac{m+1}{2}\right)$ & $N+m-2-\delta_{m,N}$ & $m$ & $m$ photons & $\frac{N!}{m!(N-m)!}$ & easy & no \\
\bottomrule
\end{tabular}
\caption{ \textbf{Comparison among different multiport interferometer schemes.}
The first four schemes are universal, while the last three are two multiport routers and one multi-router. 
The phase shifters in the tree scheme are hard/easy to be characterized if there are not/are crossings.
For the MZI layers, $\delta_{i,j}$ stands for the Kronecker delta.
The Fldzhyan scheme has the same mesh of the Clements scheme, but the MZIs are substituted by one PS and a balanced BS: '*' in the table means that the word 'MZI' must be substituted with 'PS+BS' in corresponding column. This implies that the number of total PSs of the Fldzhyan scheme is the same as the other universal schemes, and its footprint is equal to the Clements one.
The number of MZI layers of linear schemes spans from $N-1$ to $\lceil N/2\rceil$, which is achieved with the linear V-shaped scheme.
Note that the universal schemes can be used for classical coherent light and any Fock state, the routing scheme (tree and linear), for classical coherent light and single-photon state, and the multi-routing scheme (multilinear) with $m$ reference outputs for Fock states with $m$ photons.}
\label{tab:comparison}
\end{table}

Moreover, our matrix decomposition methods can be utilized for other architectures that use structures analogous to beam-splitters and phase shifters, such as photonics with frequency-bin \cite{frequency_bin}, orbital angular momentum \cite{OAM_bin}, time-bin \cite{time_bin}, etc., or other platforms like ion traps \cite{ion_boson_sampling} and superconducting circuits \cite{peropadre2016proposal,Peropadre2017}.
In all of these scenarios, it is important to decompose complex operations in terms of elementary transformations via the available amount of resources.

We have compared the two routing schemes with universal schemes provided by the Reck and Clements decompositions for lossy experimental realizations of different unitary transformations.
We have not included the braided \cite{Marchesin_25} and Fldzhyan \cite{Fldzhyan_20} schemes in the simulations, because these designs do not have an algorithm to uniquely identify the phase setting associated with a unitary matrix. Unlike the other schemes, where the decomposition procedure is based on an analytical method, these schemes do not have a decomposition algorithm and their phase setting can only be found with holistic optimization algorithms, which are computationally difficult for mode numbers bigger than 10.

Table \ref{tab:comparison} summarizes different features of universal schemes, i.e. Reck, Clements, braided and Fldzhyan, multiport routing schemes. i.e. linear and tree, and multilinear scheme.
The number of PSs in the universal schemes is always $N(N-1)$ and the sampling is done with $N$ detectors working simultaneously at the different outputs. The Clements, braided and Fldzhyan require half the optical depth of the Reck design, and they are significantly more
robust to optical losses. Contrary to the Reck design, their main drawback is given by the PS calibration, since it is not possible to address each PS individually. Without crossings, the tree scheme faces the same difficulty.
The linear V-shaped scheme has half the optical depth of the Clements scheme, while the tree scheme has a logarithmic scaling of the optical depth with respect to the number of modes.
Considering the scaling of the layers and the possible difficulty for PSs calibration, the linear V-shaped scheme is the router analogue of the Reck scheme, and the tree scheme is the router analogue of the Clements scheme.
In real experiments, even if the Clements scheme is theoretically better, the Reck scheme for a small number of modes has the same performance for real lossy MZI with simple PSs calibration. The same holds for the linear V-shaped scheme in comparison with the tree scheme.
Finally, multilinear and more generally multi-routing schemes can be used to perform scattershot boson sampling by using less MZIs than universal schemes and without calibration problems.
The price of using fewer spatial resources is paid in terms of temporal resources: we need $N!/(m!(N-m)!)$ temporal slots, each associated with a different phase setting of the interferometer.

\newpage

\newpage
\begin{appendices}

\section{Linear optical devices and multiport interferometers}
\label{app:design}

\begin{figure}[t]
    \centering
    \includegraphics[width=0.55\textwidth]{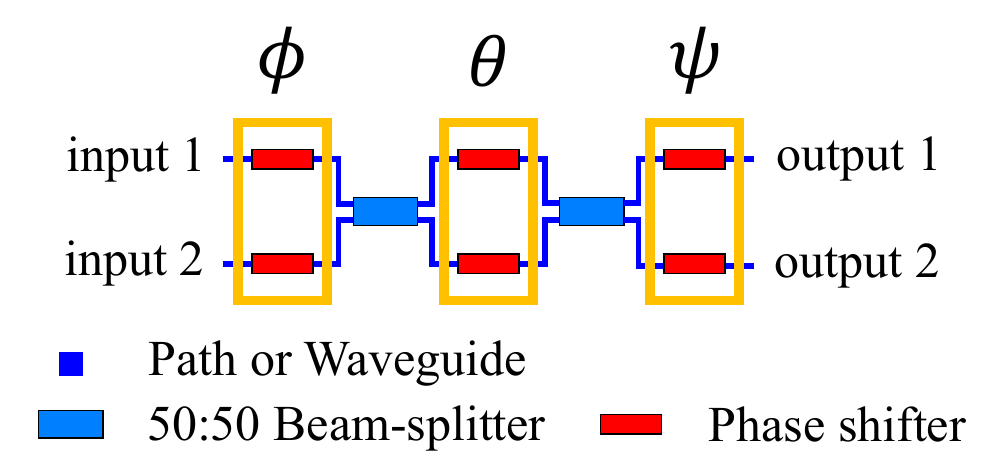}
    \caption{\textbf{Graphical representation of a Mach Zehnder interferometer.}
    A Mach Zehnder interferometer is a reconfigurable $2\times2$ device made of two balanced beam-splitters and three pairs of phase shifters. A generic input made of a classical coherent state or a single-photon state can be linearly transformed in a generic output made of a classical coherent state or a single-photon state. In classical photonics, it can route, switch, split and filter the light \cite{marchisio2025comprehensive,shamy_22}. In quantum photonics, it is used as the fundamental component for the qubit manipulation in dual-rail encoding \cite{Kok_2007}.
    }
    \label{fig:mzi}
\end{figure}

We consider $N$ spatial modes and the ladder operators $\{a_k,a_k^\dagger\}_{k\in 1\ldots N}$ associated with each mode. These pairs of operators are associated with the annihilation and creation of one photon in one mode.
Under a unitary map $\mathbf{U}$, an operator $A$ is transformed as usual, i.e. $A \to \mathbf{U}^\dagger\, A \, \mathbf{U}$. 
In linear optics \cite{Kok_2007}, linear unitary transformations are made of two elements: the phase shifter (PS) and the beam-splitter (BS).
The unitary operator associated with a PS acting on the $m$-th mode reads as follows
\begin{equation}
    \mathbf{U}_{\rm PS}^{(m)}[\phi] = \exp\left[ -{\rm i}\,\phi\,a^\dagger_m\,a_m\right] \,,
    \label{eq:PSoperator}
\end{equation}
and the unitary operator associated with a BS with transmittance $t$ acting on the mode pair $(m,n)$ reads as follows
\begin{equation}
    \mathbf{U}_{\rm BS}^{(m,n)}[\theta] = \exp\left[ -{\rm i}\,\theta\left(a^\dagger_m\,a_n+a^\dagger_n\,a_m\right)\right] \,,
    \label{eq:BSoperator}
\end{equation}
where $t=\cos \theta$ and reflectance $r=\sqrt{1-t^2}$. A BS is balanced for $\theta=\pm\pi/4$. Since any composition of these operators commutes with the number operator, each sector of the Fock space associated with a specific value of the number operator is invariant under such a composition.
The action of previous transformations on the ladder operators is linear, and they read as follows:
\begin{equation}
\begin{split}
    \left(\mathbf{U}_{\rm PS}^{(m)}[\phi] \right)^\dagger\, a_{m}^\dagger \,  \mathbf{U}_{\rm PS}^{(m)}[\phi] &= {\rm e}^{{\rm i} \phi} \, a_{m}^\dagger \,,
    \\
    \left( \mathbf{U}_{\rm BS}^{(m,n)}[\theta] \right)^\dagger\, a_{m/n}^\dagger\, \mathbf{U}_{\rm BS}^{(m,n)}[\theta]  &= \cos\theta \, a_{m/n}^\dagger+ {\rm i}\,\sin\theta \, a_{n/m}^\dagger \,.
    \\
\end{split}
\end{equation}
These outcomes make clear why the operators in Eq. \eqref{eq:PSoperator} and Eq. \eqref{eq:BSoperator} are associated with real PSs and BSs, respectively.
Then, an MZI is a $2\times2$ device composed of PSs and two balanced BSs as shown in Fig. \ref{fig:mzi}. The unitary operator associated with an MZI acting on the mode pair $(m,n)$ reads as follows
\begin{equation}
    \mathbf{U}_{\rm MZI}^{(m,n)}[\theta,\phi,\psi,\Theta,\Phi,\Psi] = 
    \mathbf{U}_{\rm PS}^{(m,n)}[\psi_1,\psi_2]\mathbf{U}_{\rm BS}^{(m,n)}[\pm\pi/4] \mathbf{U}_{\rm PS}^{(m,n)}[\theta_1,\theta_2] \mathbf{U}_{\rm BS}^{(m,n)}[\pm\pi/4] \mathbf{U}_{\rm PS}^{(m,n)}[\phi_1,\phi_2] \,,
    \label{eq:mzioperator}
\end{equation}
where $\mathbf{U}_{\rm PS}^{(m,n)}\equiv \mathbf{U}_{\rm PS}^{(m)}\mathbf{U}_{\rm PS}^{(n)}$, $\theta \equiv (\theta_1-\theta_2)/2$, $\phi \equiv (\phi_1-\phi_2)/2$, $\psi \equiv (\psi_1-\psi_2)/2$, $\Theta \equiv (\theta_1+\theta_2)/2$, $\Phi \equiv (\phi_1+\phi_2)/2$ and $\Psi \equiv (\psi_1+\psi_2)/2$.
If the state is an eigenstate of the total number operator associated with the mode pair $(m,n)$, the phase differences $(\theta,\phi,\psi)$ determine the interference between the mode pair $(m,n)$, while the phase sums $(\Theta,\Phi,\Psi)$ contribute as a global phase with respect to the mode pair $(m,n)$. The simplest case is provided by the single-photon states with a mode pair, which are not affected by a global phase with respect to such a mode pair.
In particular, for dual rail encoding (one photon propagating in two paths), the MZI can implement any single-qubit transformation \cite{Kok_2007}, which can be tuned by using the PSs associated with the phase differences $(\theta,\phi,\psi)$.
Therefore, for states of a single photon with two modes, MZIs implement $U(2)$ matrices, modulo a global phase. Hence, an MZI can be naturally associated as an element of the quotient space $U(2)/U(1)$, i.e. the set of equivalent classes of $2\times 2$ unitary matrices endowed with the equivalence relation $M\sim M'$ if $M'=e^{i\phi}M$ for some $\phi\in [0,2\pi]$.\footnote{We stress the fact that the MZI action to a dual-rail state of one photon does not generally belong to $SU(2)$, since the global phase cannot be tuned. Note that the number of parameters is three, which exactly matches the dimension of $U(2)/U(1)$ and $SU(2)$.}.
Appendix \ref{app:mzi} shows further details about the MZI action on classical coherent states and single-photon states.
In the following, we will neglect the dependence on the global phases $\{\Theta,\Phi,\Psi\}$, introduced in Eq. \eqref{eq:mzioperator}, since they cannot be operationally used and measured at the output of the MZI. Moreover, experimentally, only one phase of each pair of PSs contained in an MZI is activated, and thus, the MZI's action depends only on phases $\{\theta,\phi,\psi\}$.

A multiport interferometer with $N$ modes is composed of MZIs arranged in a specific mesh. 
The associated linear unitary operator is shown in Eq. \eqref{eq:MZIscheme_dec_princ}.
Note that the MZI operator in Eq. \eqref{eq:MZIscheme_dec_princ} depends only on $\{\theta_k,\phi_k\}_k$: this is understood as the phases $\{\psi_k\}_k$ set to zero. Indeed, the PSs associated with $\{\psi_k\}_k$ are typically not present in real implementations, and this choice is justified by the fact that their absence can be compensated with the PSs belonging to the following MZIs. For the rest of the manuscript, $\mathbf{U}_{\rm MZI}^{(m_k,n_k)}[\theta_{k},\phi_{k}]$ stands for $\mathbf{U}_{\rm MZI}^{(m_k,n_k)}[\theta_{k},\phi_{k},0]$ in order to simplify the notation.\\
Starting from a generic ladder operator, any linear unitary combination of the ladder operators \cite{Kok_2007} can be reached through the unitary map in Eq. \eqref{eq:MZIscheme_dec_princ} \cite{reck_experimental_1994}:
\begin{equation}
        \left( \mathbf{U}_{\rm MZI\,scheme}^{(N)}\right)^\dagger\, a_m^\dagger \,  \mathbf{U}_{\rm MZI\,scheme}^{(N)}  = \sum_{n=1}^N u_{m,n} \, a_n^\dagger 
        \,,
\end{equation}
where $\{u_{m,n}\}_{m,n\in \{1\ldots N\}}$ are the components of a unitary matrix $\mathbf{U}\in U(N)$.
Essentially, the multiport interferometer generalizes the MZI to the case of $N$ modes. 
A multiport interferometer scheme is said to be {\em universal} if for every unitary matrix $\mathbf{U} \in U(N)$ there exists a specific mesh $\Xi$ of MZIs and a corresponding set of phase settings $\{\theta_k, \phi_k\}_{k}$, such that the interferometer implements $\mathbf{U}$ up to a diagonal phase matrix $\mathbf{d}$. Therefore, in analogy to Eq. \eqref{eq:MZIscheme_dec_princ}, the unitary matrix associated with $\mathbf{U}_{\rm MZI\,scheme}^{(N)}$ has the following form:
\begin{equation}
    \mathbf{U}
    = \mathbf{d} \cdot
    \prod_{k\in\,\Xi} \mathbf{U}^{(k)}_2 \,.
    \label{eq:uNdecu2}
\end{equation}
The matrix $\mathbf{U}^{(k)}_2$ represents $\mathbf{U}_{\rm MZI}^{(m_k,n_k)}[\theta_{k},\phi_{k}]$, and it is an $N\times N$ block matrix containing in positions $(m_k,n_k)$ a $2\times 2 $ block associated with an element of the quotient space $U(2)/U(1)^2$ and identity components otherwise. The set $U(N)/U(1)^N$ is made of equivalence classes of $N\times N$ unitary matrices, endowed with the equivalence relation $M\sim M'$ if $M'=D\,M$, with $D$ a $N\times N$ diagonal phase matrix. The set $U(N)/U(1)^N$ with $N\ge 2$ is not a group, since the subgroup $U(1)^N$ is not normal to $U(N)$, i.e. $g\cdot h\cdot g^{-1} \notin U(1)^N$ given $h\in U(1)^N$ and $g\in U(N)$.
We remark that $U(2)/U(1)^2$ is a 2-dimensional manifold; indeed, the number of parameters in $\mathbf{U}_{\rm MZI}^{(m_k,n_k)}[\theta_{k},\phi_{k}]$ is two, while the dimensions of $U(1)$ and $U(2)$ are one and four, respectively.
Since an MZI with $\psi=0$ is the physical creation of $U(2)/U(1)^2$ matrices, multiport interferometer schemes are the physical creation of the set $U(N)/U(1)^N$ decomposed in terms of $U(2)/U(1)^2$ transformations.

\section{MZI as a unitary transformation}
\label{app:mzi}

Fig. \ref{fig:mzi} shows the MZI, which is the physical realization of matrices belonging to $U(2)/U(1)$ for classical coherent light and a single photon.
The generic unitary transformation achieved through an MZI is described by the following matrix
\begin{equation}
\mathbf{U}_{\rm MZI}[\theta,\phi,\psi]=
{\rm e}^{{\rm i}\xi}
\begin{pmatrix}
{\rm e}^{{\rm i}(\phi+\psi)}\,\sin\theta & {\rm e}^{-{\rm i}(\phi-\psi)}\,\cos\theta \\
{\rm e}^{{\rm i}(\phi-\psi)}\,\cos\theta & -{\rm e}^{-{\rm i}(\phi+\psi)}\,\sin\theta
\end{pmatrix} \,,
\end{equation}
where 
\begin{equation}
\begin{split}
    \theta &\equiv \frac{\theta_1-\theta_2}{2}\quad, \quad
    \phi \equiv \frac{\phi_1-\phi_2}{2} \quad , \quad
    \psi \equiv \frac{\psi_1-\psi_2}{2}  \\
    \xi &\equiv
     \frac{\pi+\theta_1+\theta_2+\phi_1+\phi_2+\psi_1+\psi_2}{2} 
\end{split}
\end{equation}
and we have used the convention $|1\rangle \iff (1,0)^{\rm T}$ for the light/photon in the upper path and $|2\rangle \iff (0,1)^{\rm T}$ for the light/photon in the lower path. 
Note that the global phase $\xi$ cannot be tuned arbitrarily. However, for a two-mode system it is physically irrelevant, as it is not observable. Likewise, the phase $\psi$ is not observable when the output state is measured in the computational basis following the application of $\mathbf{U}_{\rm MZI}$.
A generic unitary map $\mathbf{U}$ associated with a $2\times2$ matrix composed of elements $\{u_{i,j}\}_{i,j\in[1,2]}$ can be implemented by choosing
\begin{equation}
\begin{split}
    \phi &= \frac{\arg\left[u_{11}\right]-\arg\left[u_{12}\right]}{2} = \frac{\arg\left[u_{21}\right]-\arg\left[u_{22}\right]}{2} \,,\\
    \theta &= \mbox{atan}\left[\frac{|u_{11}|}{|u_{12}|}\right] =-\mbox{atan}\left[\frac{|u_{22}|}{|u_{21}|}\right]  \,,\\
    \psi &= \frac{\arg\left[u_{11}\right]+\arg\left[u_{12}\right]}{2} =-\frac{\arg\left[u_{21}\right]+\arg\left[u_{22}\right]}{2}
\end{split}
\end{equation}
modulo an overall global phase.

Since our goal is to decompose a generic element of the quotient set $U(N)/U(1)^N$  as the product of elements of   $U(2)/U(1)^2$, we first introduce a change of variables that simplifies the decomposition procedure.\\
As in Ref. \cite{clements_optimal_2016}, we parametrize the cosets $U(2)/U(1)^2$ space of matrices as follows
\begin{equation}
\mathbf{u}_2[\theta,\phi]\equiv
\begin{pmatrix}
{\rm e}^{{\rm i}\phi}\,\cos\theta & -\sin\theta \\
{\rm e}^{{\rm i}\phi}\,\sin\theta & \cos\theta
\end{pmatrix} \,,
\label{eq:u2matrix}
\end{equation}
which is related to $\mathbf{U}_{\rm MZI}$ through the following changes
\begin{equation}
    \theta \to  \pi/2-\theta \quad,\quad \phi\to(\phi+\pi)/2 \quad,\quad \psi\to0 \,,
\end{equation}
or equivalently
\begin{equation}
\mathbf{u}_2[\theta,\phi]\equiv
{\rm e}^{-{\rm i}\tilde\xi}\,\mathbf{U}_{\rm MZI}[\pi/2-\theta,(\phi+\pi)/2,0]
\,.
\end{equation}

Since the phase $\psi$ s fixed to zero in multiport interferometers, the matrix $\mathbf{u}_2$ is equivalent, up to unobservable phases, to the transformation implemented by an MZI. \\
In the following sections of the appendix we present four different way to decompose a matrix belonging to $U(N)/U(1)^N$ in terms of $\mathbf{u}_2$ matrices. The four decompositions are associated with four different types of meshes.

The two-dimensional case of all four meshes coincides with the MZI with $\psi=0$, and it is described by $\mathbf{u}_2$, Eq. \eqref{eq:u2matrix}.

The first interesting property of $\mathbf{u}_2$ is the ability to route any input state of the two modes to the upper/lower state, i.e. $(1,0)^{\rm T}/ (0,1)^{\rm T}$. Indeed, the routing action reads as follows
\begin{equation}
\begin{split}
\mathbf{u}_2[\theta,\phi]\,
\begin{pmatrix}
a  \\
b 
\end{pmatrix} 
&= \begin{pmatrix}
{\rm e}^{{\rm i}\phi}\,\cos\theta \,a-\sin\theta\,b \\
{\rm e}^{{\rm i}\phi}\,\sin\theta \,a+\cos\theta\,b
\end{pmatrix}\\
&=
\begin{cases}
    \begin{pmatrix}
    {\rm e}^{{\rm i} \arg(b)}\sqrt{|a|^2+|b|^2}  \\
    0
    \end{pmatrix} \quad \mbox{if}\,\, \theta = -\mbox{atan}\frac{|b|}{|a|} \,,\, \phi= \arg(b)-\arg(a)\,,\\
    \begin{pmatrix}
    0  \\
    {\rm e}^{{\rm i} \arg(b)}\sqrt{|a|^2+|b|^2}
    \end{pmatrix} \quad \mbox{if} \,\, \theta =\mbox{atan}\frac{|a|}{|b|} \,,\, \phi= \arg(b)-\arg(a) \,.\\
\end{cases}
\end{split}
\label{eq:eliminationvector}
\end{equation}
This property is exploited in both  the linear and tree schemes, and it is summarized in the following algorithm.

\begin{algorithm}[H]
\caption{The algorithm for element elimination for a vector}\label{alg:vectelementelimination}
\begin{algorithmic}
\Function{vector\_element\_elimination}{$a;b;c=\pm1$}\\
$r \gets \sqrt{|a|^2 + |b|^2}$
\If{$r = 0$} \Comment{If zero amplitude, trivial action.}
    \State $\theta \gets 0$ and $\phi \gets 0$ 
\Else 
    \If {$c=+1$} \Comment{Route to the upper mode.}
        \State $\theta \gets -\text{atan}(|b|/ |a|)$ and $\phi\gets \arg(b)-\arg(a) $
    \ElsIf {$c=-1$}  \Comment{Route to the lower mode.}
        \State $\theta \gets \text{atan}(|a|/ |b|)$ and $\phi\gets \arg(b)-\arg(a) $
    \EndIf
\EndIf
\State return $\theta, \phi$
\EndFunction
\end{algorithmic}
\end{algorithm}

The second property of $\mathbf{u}_2$ plays a key role in the Reck decomposition. Specifically, right multiplication of a unitary matrix by $\mathbf{u}_2^{-1}$ eliminates a prescribed off-diagonal entry:
\begin{equation}
\begin{split}
\begin{pmatrix}
* & * \\
b & a
\end{pmatrix}\cdot\mathbf{u}_2[\theta,\phi]^{-1}
& =
\begin{pmatrix}
* & * \\
{\rm e}^{-{\rm i}\phi}\,\cos\theta \,b-\sin\theta\, a & {\rm e}^{-{\rm i}\phi}\,\sin\theta \,b+\cos\theta\, a
\end{pmatrix} \\
& =
\begin{pmatrix}
* & * \\
0 & {\rm e}^{{\rm i}\arg(a)}\,\sqrt{|a|^2+|b|^2}
\end{pmatrix}
\,,
\quad \mbox{if} \,\, \theta =\mbox{atan}\frac{|b|}{|a|} \,,\, \phi= \arg(b)-\arg(a) \,.
\end{split}
\label{eq:eliminationfromright}
\end{equation}

The third property of $\mathbf{u}_2$ is required for the Clements decomposition. Indeed, by left-multiplying a unitary matrix by $\mathbf{u}_2$, one can eliminate a selected off-diagonal entry:
\begin{equation}
\begin{split}
\mathbf{u}_2[\theta,\phi]\cdot
\begin{pmatrix}
a & * \\
b & *
\end{pmatrix} &=
\begin{pmatrix}
{\rm e}^{{\rm i}\phi}\,\cos\theta \,a-\sin\theta\, b & * \\
{\rm e}^{{\rm i}\phi}\,\sin\theta \,a+\cos\theta\, b & *
\end{pmatrix} 
\\
& =
\begin{pmatrix}
{\rm e}^{{\rm i}\arg(b)}\,\sqrt{|a|^2+|b|^2} & * \\
0 & *
\end{pmatrix}
\,,
\quad \mbox{if} \,\, \theta =-\mbox{atan}\frac{|b|}{|a|} \,,\, \phi= \arg(b)-\arg(a) \,.
\end{split}
\label{eq:eliminationfromleft}
\end{equation}

Finally, the fourth property of $\mathbf{u}_2$ is needed for the Clements decomposition, and it reads as follows:
\begin{equation}
\begin{split}
\mathbf{u}_2[\theta,\phi]^{-1}\cdot
\begin{pmatrix}
{\rm e}^{{\rm i}\alpha} & 0 \\
0 & {\rm e}^{{\rm i}\beta}
\end{pmatrix} &=
\begin{pmatrix}
{\rm e}^{{\rm i}\gamma} & 0 \\
0 & {\rm e}^{{\rm i}\delta}
\end{pmatrix} \cdot \mathbf{u}_2[\tilde\theta,\tilde\phi] 
\hspace{1cm}\mbox{if}\,
\begin{cases}
    \tilde\theta =-\theta \quad\mbox{and}\quad \tilde\phi=\alpha-\beta \\
    \gamma = \beta-\phi \quad \mbox{and} \quad \delta = \beta
\end{cases}\,.
\end{split}
\label{eq:propu2withdiag}
\end{equation}

In the $N$-dimensional case, if we multiply a generic $N\times N$ unitary matrix $\mathbf{u}$ from the right/left with a block matrix equal to the identity except for one $2\times2$ block in positions $(i,i+1)$ equal to $\mathbf{u}_2^{-1}$/$\mathbf{u}_2$, we can eliminate a component of $\mathbf{u}$. More details are given in Sections \ref{app:reck} and \ref{app:clements}.
The last two properties are summarized in the following algorithm.
\begin{algorithm}[H]
\caption{The algorithm for element elimination decomposition}\label{alg:elementelimination}
\begin{algorithmic}
\Function{matrix\_element\_elimination}{$\mathbf{u};i;j;c=\pm1$}
\If {$c=+1$} \Comment{Elimination from the right, Eq. \eqref{eq:eliminationfromright}.}
    \State $a \gets \mathbf{u}(i, j+1)$ , $b \gets \mathbf{u}(i, j)$ and $r \gets \sqrt{|a|^2 + |b|^2}$
        \If{$r = 0$}
            \State $\theta \gets 0$ and $\phi \gets 0$
        \Else
            \State $\theta \gets \text{atan}(|b|, |a|)$
            and $\phi\gets \arg(b)-\arg(a) $
        \EndIf
\ElsIf {$c=-1$} \Comment{Elimination from the left, Eq. \eqref{eq:eliminationfromleft}.}
    \State $a \gets \mathbf{u}(i-1, j)$ , $b \gets \mathbf{u}(i, j)$ and $r \gets \sqrt{|a|^2 + |b|^2}$
        \If{$r = 0$}
            \State $\theta \gets 0$ and $\phi \gets 0$
        \Else
            \State $\theta \gets -\text{atan}(|b|, |a|)$
            and $\phi\gets \arg(b)-\arg(a) $
        \EndIf
\EndIf
\State return $\theta, \phi$
\EndFunction
\end{algorithmic}
\end{algorithm}

We conclude this subsection by defining the embedding of $\mathbf{u}_2$ in $N\times N$ matrix. This matrix describes the action of a single MZI on the mode pair $(k,k+1)$, and it reads as follows 
\begin{equation}
\mathbf{U}_2^{(k)} \equiv
    \begin{pmatrix}
    1 & 0 &  \ldots & \ldots & \ldots & 0 \\
    0 & \ddots & & & &\vdots \\
    \vdots & & {\rm e}^{{\rm i}\phi}\,\cos\theta & -\sin\theta &  & \vdots \\
    \vdots & & {\rm e}^{{\rm i}\phi}\,\sin\theta & \cos\theta & & \vdots \\
    \vdots & & & & \ddots & 0 \\
    0 & \ldots & \ldots & \ldots & 0 & 1 \\
    \end{pmatrix} \,.
    \label{Ukemb}
\end{equation}

\section{Reck scheme}
\label{app:reck}

\begin{figure}[t!]
    \centering
    \includegraphics[width=0.9\textwidth]{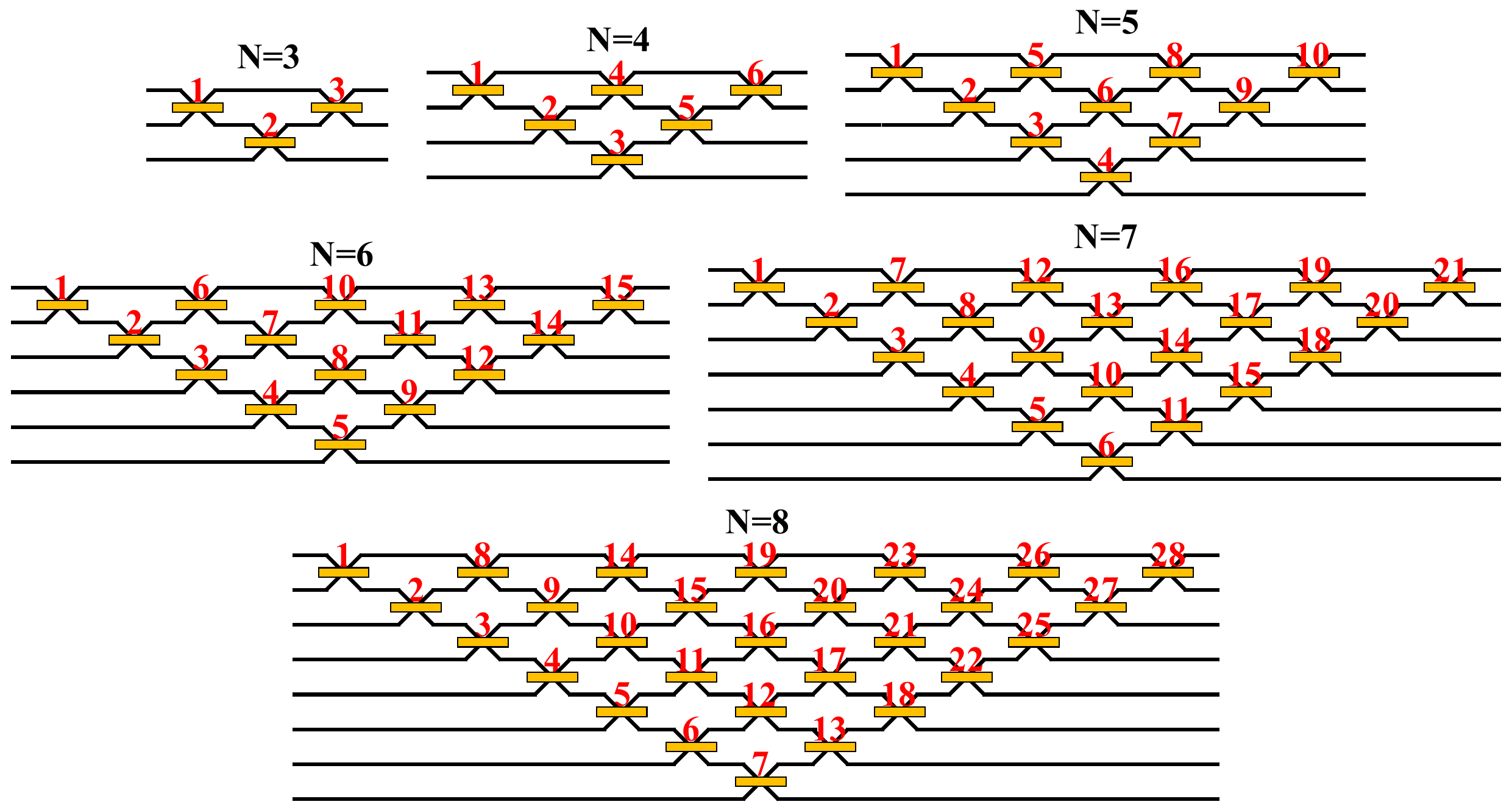}
    \caption{\textbf{Reck scheme for different numbers of modes.}
    The multiport interferometers with Reck mesh for $N=3\ldots8$ number of modes. 
    The MZI is represented by the crossing lines with the yellow rectangle. 
    Above each MZI, the red number specify the MZI position inside the Reck mesh. 
    In the figure the number of the modes are omitted: their labelling start from the upper part from 1 to the $N$.
    Note that the MZI order follows the diagonal slices '\textbackslash' (top-left to bottom-right) of the multiport interferometer.
    }
    \label{fig:reck_app}
\end{figure}

Ref. \cite{reck_experimental_1994} describes an elimination procedure for decomposing a generic $N\times N$ unitary matrix $\mathbf{U}$ into a product of matrices $\mathbf{U}_2$ as follows
\begin{equation}
    \mathbf{U}
    = \mathbf{d}\cdot
    \prod_{k\in\,\Xi_{\rm Reck}} \mathbf{U}^{(k)}_2 \,,
    \label{eq:uNdecu2reck}
\end{equation}
where $\mathbf{d}$ is a $N\times N$ diagonal phase matrix and $\Xi_{\rm Reck}$ specifies the multiplication order in the Reck decomposition. This ordering is reflected in the corresponding MZI mesh: This order multiplication is shown in the MZI mesh of the Reck scheme:  Fig.~\ref{fig:reck_app} illustrates the Reck mesh for dimensions ranging from 3 to 8..

To prove the above decomposition, we repeatedly apply the elimination property given in Eq.~\eqref{eq:eliminationfromright}, for a total of $N(N-1)/2$ steps. 
We begin by right-multiplying $\mathbf{U}$ by $\left(\mathbf{U}_2^{(1)}[\theta_1,\phi_1]\right)^{-1}$  and choosing $(\theta_1,\phi_1)$ so that the $(N,1)$ entry of $\mathbf{U}\cdot\left(\mathbf{U}_2^{(1)}\right)^{-1}$ vanishes. Next, we right-multiply the resulting matrix by  $\left(\mathbf{U}_2^{(2)}[\theta_2,\phi_2]\right)^{-1}$, and choose $(\theta_2,\phi_2)$ so that the $(N,2)$-entry of $\mathbf{U}\cdot\left(\mathbf{U}_2^{(1)}\right)^{-1}\cdot\left(\mathbf{U}_2^{(2)}\right)^{-1}$ vanishes.
We repeat this process until the only non-zero element of the last row or column is present in position $(N,N)$:  
\begin{equation}
\mathbf{U}\cdot\left(\prod_{j=1}^{N-1} \mathbf{U}_2^{(j)}[\theta_j,\phi_j]\right)^{-1} =
    \begin{pmatrix}
    * & \ldots & \ldots & \ldots & * & 0 \\
    \vdots & \ddots & & & \vdots & \vdots \\
    \vdots & & & & \vdots & \vdots \\
    \vdots & & & \ddots & \vdots & \vdots \\
    * & \ldots & \ldots & \ldots & * & 0 \\
    0 & \ldots & \ldots & \ldots & 0 & {\rm e}^{{\rm i} \alpha_N} \\
    \end{pmatrix} \,.
\end{equation}
where the productorial is ordered from right to left, and the phase pairs $\{(\theta_j,\phi_j)\}_{j\in 1\ldots N-1}$ are set using the recipe given in Algorithm \ref{alg:elementelimination} with $c=+1$.
Looking at Fig. \ref{fig:reck_app}, we can see that the product of $\mathbf{U}_2^{(j)}$ with $j\in 1\ldots N-1$ describes exactly the first diagonal layer of MZIs labelled with indexes $j\in 1\ldots N-1$.
At this point, the resulting matrix is block diagonal, with the $(N-1)\times(N-1)$ block composed of a unitary matrix and a $1\times1$ block composed of a phase.
Then, it is enough to apply the same procedure to the $(N-1)\times(N-1)$ block by right-multiplying the resulting matrix of the previous equation by $\left(\mathbf{U}_2^{(j)}\right)^{-1}$ with $j\in 1\ldots N-2$. 
Repeating this process $N-1$ times corresponds  to the generic diagonal layer of the meshes shown in Fig. \ref{fig:reck_app}: at the $i$-th diagonal layer we apply $\left(\mathbf{U}_2^{(j)}\right)^{-1}$ with $j\in 1\ldots N-i$ from the right.
Finally, after step number $N(N-1)/2$ we obtain the phase diagonal matrix:
\begin{equation}
\mathbf{U}\cdot\left(\mathbf{U}_2^{(1)}[\theta_{N(N-1)/2},\phi_{N(N-1)/2}]\right)^{-1}\!\!\!\ldots\left(\prod_{j=1}^{N-2} \mathbf{U}_2^{(j)}[\theta_{N-j+1},\phi_{N-j+1}]\right)^{-1}\!\!\!\!\cdot\left(\prod_{j=1}^{N-1} \mathbf{U}_2^{(j)}[\theta_j,\phi_j]\right)^{-1}\!\!\!\! =
    \mathbf{d} \,,
\end{equation}
which is equivalent to Eq. \eqref{eq:uNdecu2reck}.

The following algorithm summarizes the procedure described above.

\begin{algorithm}[H]
\caption{The algorithm for Reck unitary matrix decomposition}\label{alg:Reck}
\begin{algorithmic}
\Function{reck\_decomposition}{$\mathbf{U}$}
\State \textbf{Output:} $\vec{\theta}, \vec{\phi}, \mathbf{U}_{\rm phases}$
\State $N \gets \text{size}(\mathbf{U})$
\State $U_{work} \gets \mathbf{U}$ \Comment{Create a copy.}
\State $m \gets N(N-1)/2$ \Comment{Number of MZIs, dimension of $\vec\theta$ and $\vec\phi$.}
\State $(\vec\theta,\vec\phi) \gets (\vec 0,\vec 0)$ \Comment{Initialize vectors of dimension $m$ for the phase setting.}
\State $q \gets 1$ \Comment{Index running from 1 to $m$.}

\For{$i = 1$ to $N-1$}
    \For{$j = 1$ to $N-i$}
                
        \State $(\vec\theta(q),\vec\phi(q)) \gets \text{\scriptsize MATRIX\_ELEMENT\_ELIMINATION}(U_{work};N-i+1;j;+1)$
        \State $G\gets \mathbf{U}_2^{(j)}[\vec\theta(q),\vec\phi(q)]$
        \State $U_{work} \gets U_{work} \cdot G^{-1}$ \Comment{Elimination of element $(N-i+1,j)$ from the right.}

        \State $q \gets q + 1$
    \EndFor
\EndFor
\State $\mathbf{U}_{\rm phase} \gets U_{work}$ \Comment{Save the phases of the resulting diagonal matrix.}
\State return $\vec{\theta}, \vec{\phi}, \mathbf{U}_{\rm phases}$
\EndFunction
\end{algorithmic}
\end{algorithm}

Finally, the following algorithm takes as input the dimension of the multiport interferometer and the phase settings, and returns the corresponding unitary matrix in its Reck decomposition.

\begin{algorithm}[H]
\caption{The algorithm to generate a Reck-decomposed unitary matrix}\label{alg:Reckmesh}
\begin{algorithmic}
\Function{reck\_mesh}{$N;\vec\theta;\vec\phi$}
\State $m \gets N(N-1)/2$
\If{ $m \ne \text{size}(\vec\theta) \; \&\; m \ne \text{size}(\vec\phi)$}
    \State Not Compatible inputs
\Else
    \State $\mathbf{U}\gets \mathbf{1}_N$ , $q\gets 1$
    \For{$i = 1$ to $N-1$}
    \For{$j = 1$ to $N-i$} \Comment{The MZI order is given in Fig. \ref{fig:reck_app}.}
        \State $\mathbf{U}\gets \mathbf{U}_2^{(j)}[\vec\theta(q),\vec\phi(q)] \cdot \mathbf{U} $    
        
        \State $q \gets q + 1$
    \EndFor
\EndFor
\EndIf

\State return $\mathbf{U}$
\EndFunction
\end{algorithmic}
\end{algorithm}

\section{Clements scheme}
\label{app:clements}

\begin{figure}[t!]
    \centering
    \includegraphics[width=0.9\textwidth]{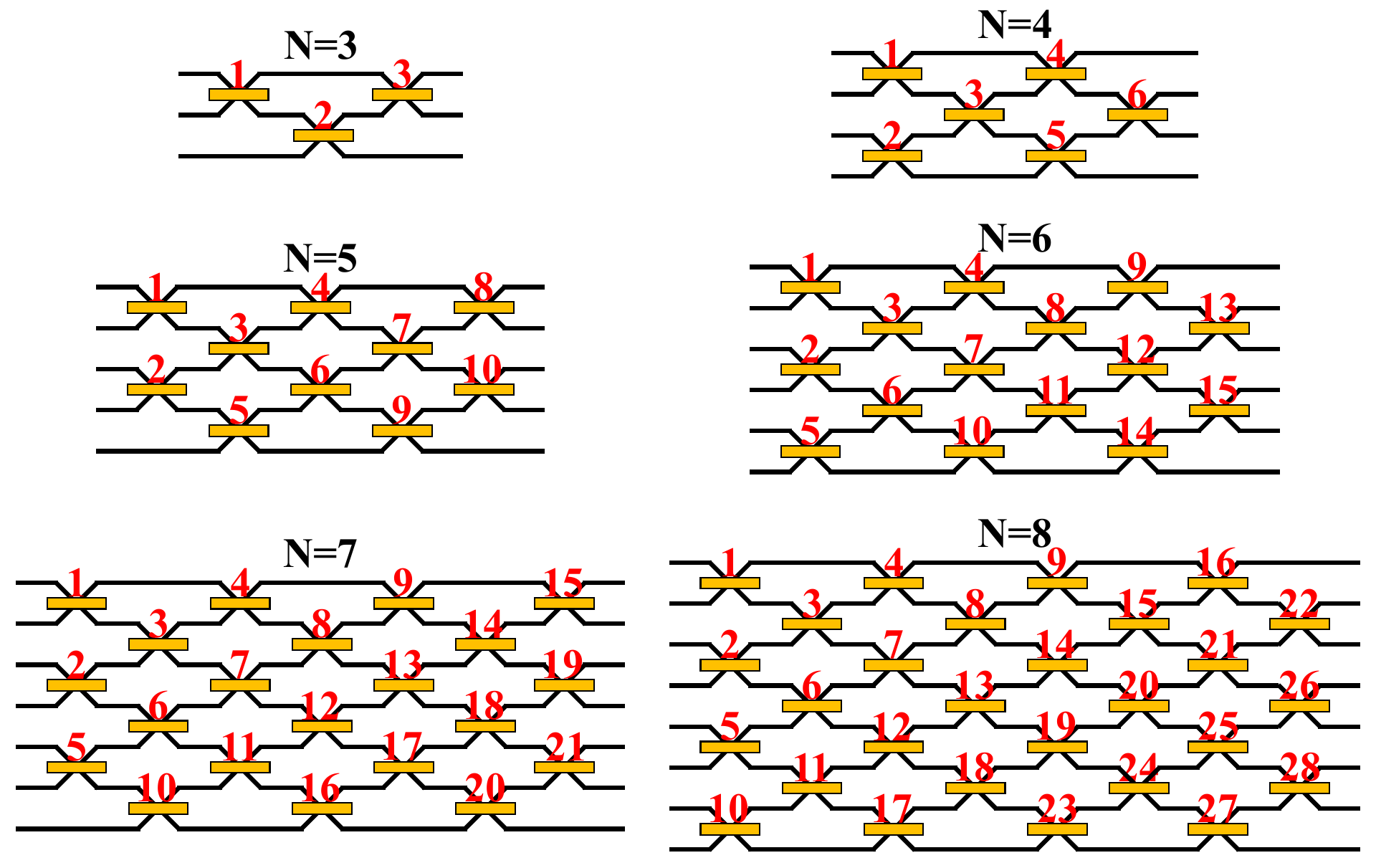}
    \caption{\textbf{Clements scheme for different numbers of modes.}
    The multiport interferometers with Clements mesh for $N=3\ldots8$ number of modes. 
    The MZI is represented by the crossing lines with the yellow rectangle. 
    Above each MZI, the red number specify the MZI position inside the Clements mesh. 
    In the figure the number of the modes are omitted: their labelling start from the upper part from 1 to the $N$.
    Note that the MZI order follows the diagonal slices '/' (bottom-right to top-left) of the multiport interferometer.
    }
    \label{fig:clements_app}
\end{figure}

Ref. \cite{clements_optimal_2016} presents the elimination procedure to decompose a generic unitary matrix $\mathbf{U}$ in terms of matrices $\mathbf{U}_2$ as follows
\begin{equation}
    \mathbf{U}
    = \mathbf{d}\cdot
    \prod_{k\in\,\Xi_{\rm Clements}} \mathbf{U}^{(k)}_2 \,,
    \label{eq:uNdecu2clements}
\end{equation}
where $\mathbf{d}$ is a $N\times N$ diagonal phase matrix and $\Xi_{\rm Clements}$ is the order multiplication of the Clements decomposition. This multiplication order is reflected in the MZI mesh of the Clements scheme. Figure \ref{fig:clements_app} illustrates the mesh for dimensions ranging from 3 to 8.

To prove the above decomposition, we perform $N(N-1)/2$ elimination steps, each consisting of a combination of the two operations described in Eqs.~\eqref{eq:eliminationfromright} and \eqref{eq:eliminationfromleft}.
First, we right-multiply $\mathbf{U}$ by $\left(\mathbf{U}_2^{(1)}[\theta_1,\phi_1]\right)^{-1}$ and  set $(\theta_1,\phi_1)$ so that the  $(N,1)$-entry of $\mathbf{U}\cdot\left(\mathbf{U}_2^{(1)}\right)^{-1}$ vanishes. Second, we left-multiply $\mathbf{U}\cdot\left(\mathbf{U}_2^{(1)}\right)^{-1}$ by $\mathbf{U}_2^{(N-2)}[\theta_2,\phi_2]$  and  set $(\theta_2,\phi_2)$ so that the  $(N-1,1)$-entry of $\mathbf{U}_2^{(N-2)}\cdot\mathbf{U}\cdot\left(\mathbf{U}_2^{(1)}\right)^{-1}$ vanishes.
Third, we left-multiply $\mathbf{U}_2^{(N-2)}\cdot \mathbf{U}\cdot\left(\mathbf{U}_2^{(1)}\right)^{-1}$ by $\mathbf{U}_2^{(N-1)}[\theta_3,\phi_3]$ and  set $(\theta_3,\phi_3)$ so that the  $(N,2)$-entry of $\mathbf{U}_2^{(N-1)}\cdot\mathbf{U}_2^{(N-2)}\cdot\mathbf{U}\cdot\left(\mathbf{U}_2^{(1)}\right)^{-1}$ vanishes.
After this procedures, the result looks as follows:
\begin{equation}
\begin{split}
&\mathbf{U}_2^{(N-1)}[\theta_{3},\phi_{3}]\cdot\,\mathbf{U}_2^{(N-2)}[\theta_{2},\phi_{2}]\cdot\,\mathbf{U}\cdot\left(\mathbf{U}_2^{(1)}[\theta_1,\phi_1]\right)^{-1} 
=
\begin{pmatrix}
* & * & \ldots & \ldots & \ldots & * \\
* & \ddots & & &  & \vdots \\
\vdots & & & &  & \vdots \\
* & \ddots& &  &  & \vdots \\
0 & * & \ddots &  & \ddots & \vdots \\
0 & 0 & * & \ldots & \ldots & * \\
\end{pmatrix} \,,
\end{split}
\label{eq:clementspartial}
\end{equation}
where phase indices do not match the labeling used in Fig. \ref{fig:clements_app}, therefore, the indices must be relabeled at the end of the decomposition. The procedure begins by eliminating the last and second-to-last sub-diagonals: the former is eliminated by right multiplication, whereas the latter is eliminated by two successive left multiplications. The process then proceeds by alternating right and left multiplications to eliminate the remaining sub-diagonals.
After $N(N-1)/2$ multiplication steps, we obtain
\begin{equation}
    \prod_{k\in\,\Xi_{\rm left}} \mathbf{U}^{(k)}_2 \cdot\mathbf{U} \cdot\left(\prod_{k\in\,\Xi_{\rm right}} \mathbf{U}^{(k)}_2\right)^{-1}
    = \tilde{\mathbf{d}}
    \implies 
    \mathbf{U} 
    = \left(\prod_{k\in\,\Xi_{\rm left}} \mathbf{U}^{(k)}_2\right)^{-1} \cdot \tilde{\mathbf{d}} \cdot\prod_{k\in\,\Xi_{\rm right}} \mathbf{U}^{(k)}_2
     \,,
    \label{eq:uNdecu2clementsbis}
\end{equation}
where $\tilde{\mathbf{d}}$ is a $N\times N$ diagonal phase matrix and $\Xi_{\rm right/left}$ is the right/left order multiplication of the algorithm started with Eq. \eqref{eq:clementspartial}. 
Finally, we apply the property in Eq. \eqref{eq:propu2withdiag} to all the factors associated with $\Xi_{\rm left}$,  and perform a final reindexing to match the labeling convention used in Fig.~\ref{fig:clements_app}. This yields the decomposition in the form given by Eq.~\eqref{eq:uNdecu2clements}.

The following algorithm summarizes what we have just explained.

\begin{algorithm}[H]
\caption{The algorithm for Clements unitary matrix decomposition}\label{alg:Clements}
\begin{algorithmic}
\Function{clements\_decomposition}{$\mathbf{U}$}
\State $N \gets \text{size}(\mathbf{U})$
\State $U_{work} \gets \mathbf{U}$ \Comment{Create a copy.}
\State $m \gets N(N-1)/2$ \Comment{Number of MZIs, dimension of $\vec\theta$ and $\vec\phi$.}
\State $(\vec\theta,\vec\phi) \gets (\vec 0,\vec 0)$ \Comment{Initialize vectors of dimension $m$ for the phase setting.}
\State $q \gets 1$ \Comment{Index running from 1 to $m$.}
\State $\vec q_{right} \gets []$ and $\vec n_{right} \gets []$ \Comment{Arrays of indexes for the reordering operation.}
\For{$i = 1$ to $N-1$}
    \If{$i$ odd}
        \For{$j = 0$ to $i-1$}
            
            \State $(\vec\theta(q),\vec\phi(q)) \gets \text{\scriptsize ELEMENT\_ELIMINATION}(U_{work};N-j;i-j;+1)$
            \State $G\gets \mathbf{U}_2^{(i-j)}[\vec\theta(q),\vec\phi(q)]$
            \State $U_{work} \gets U_{work} \cdot G^{-1}$ \Comment{Elimination of element $(N-j,i-j)$ from the right.}
    
            \State $q \gets q + 1$
        \EndFor
    \Else
        \For{$j = 1$ to $i$}
            
            \State $(\vec\theta(q),\vec\phi(q)) \gets \text{\scriptsize ELEMENT\_ELIMINATION}(U_{work};N-i+j;j;-1)$
            \State $G \gets \mathbf{U}_2^{(N-i+j-1)}[\vec\theta(q),\vec\phi(q)]$
            \State $U_{work} \gets G \cdot U_{work} $ \Comment{Elimination of element $(N-i+j,j)$ from the left.}
    
            \State \textbf{insert} $q$ to $\vec q_{right} $ and $(N-i+j-1)$ to $\vec n_{right} $  in the beginning
            \State $q \gets q + 1$
        \EndFor
    \EndIf
    
\EndFor
\State $\mathbf{U}_{\rm phase} \gets U_{work}$ \Comment{Save the phases of the resulting diagonal matrix.}
\State $\vec{\theta}, \vec{\phi}, \mathbf{U}_{\rm phases} \gets \text{\scriptsize CLEMENTS\_REORDERING}(\vec{\theta}; \vec{\phi}; \mathbf{U}_{\rm phases};\vec q_{right};\vec n_{right})$
\State return $\vec{\theta}, \vec{\phi}, \mathbf{U}_{\rm phases}$
\EndFunction
\end{algorithmic}
\end{algorithm}

The following algorithm shows how to do the index re-ordering at the end of the decomposition to match the labelling of Fig. \ref{fig:clements_app}. The reordering process follows the procedure given in the Supplementary Information \cite{Clements2017_supp}.

\begin{algorithm}[H]
\caption{The algorithm for reordering fro the Clement decomposition}\label{alg:clementsreorder}
\begin{algorithmic}
\Function{clements\_reordering}{$\vec{\theta}; \vec{\phi}; \mathbf{U}_{\rm phases};\vec q_{right};\vec n_{right}$}
\For{$p = 1$ to $\text{size}(\vec q_{right})$}
\Comment{Use the property in Eq. \eqref{eq:propu2withdiag}}
    \State $w \gets \vec n_{right}(p)$ and $q \gets \vec q_{right}(p)$ 
    \State $\alpha \gets \mathbf{U}_{\rm phases}(w,w)$ and $\beta \gets \mathbf{U}_{\rm phases}(w+1,w+1)$
    \State $\mathbf{U}_{\rm phases}(w,w) \gets \exp\left({\rm i}(\beta-\vec\phi(q))\right) $, $\vec\theta(q) \gets - \vec\theta(q)$ and $\vec\phi(q) \gets \arg \left( \alpha \right) - \arg \left( \beta \right)$
\EndFor

\State $\vec P_{left} \gets []$ and $\vec P_{right} \gets []$ \Comment{Arrays of indexes for the permutation operation.}
\State $s\gets 1$

\For{$k=1$ to $N-1$} \Comment{The series $\vec P_{left/right}$ are shown in Fig. S1 of Ref. \cite{Clements2017_supp}.}

    \State $b \gets s:1:s+k-1$ 
    \Comment{Create interval from $s$ to $s+k-1$ with unit step.}
    \If{$k$ odd}
        \State insert $b$ to $\vec P_{left}$ in the end
    \Else
        \State flip $b$ and insert $b$ to $\vec P_{right}$ in the beginning
    \EndIf
    \State $s \gets s+k$
\EndFor

\State Concatenate $\vec P_{left}$ and $\vec P_{right}$ to create $\vec P$
\State Permute $\vec\theta$ and $\vec\phi$ with $\vec P$
\Comment{Permutation to recover the MZI labelling of Fig. \ref{fig:clements_app}.}
\State return $\theta, \phi, \mathbf{U}_{\rm phases}$
\EndFunction
\end{algorithmic}
\end{algorithm}

Finally, the following algorithm takes as input the dimension of the multiport interferometer
and the phase settings, and returns the corresponding unitary matrix in its Clements decomposition.

\begin{algorithm}[H]
\caption{The algorithm to generate a Clements-decomposed unitary matrix}\label{alg:Clementsmesh}
\begin{algorithmic}
\Function{clements\_mesh}{$N;\vec\theta;\vec\phi$}
\State $m \gets N(N-1)/2$
\If{ $m \ne \text{size}(\vec\theta) \; \&\; m \ne \text{size}(\vec\phi)$}
    \State Not Compatible inputs
\Else
    \State $\mathbf{U}\gets \mathbf{1}_N$ , $q\gets 1$
    \State $odds \gets 1:2:N-1$ and $evens \gets 2:2:N-1$
    \State Concatenate $odds$ and $evens$ to create $v$
    \For{$i = 1$ to $N-1$}
    \For{$j = 1$ to $v(i)$} \Comment{The MZI order is given in Fig. \ref{fig:clements_app}.}
        \If{$v(\ell)$ odd}
            \State $\mathbf{U}\gets \mathbf{U}_2^{(v(i)-j+1)}[\vec\theta(q),\vec\phi(q)] \cdot \mathbf{U} $         

        \Else
            \State $\mathbf{U}\gets \mathbf{U}_2^{(N-j)}[\vec\theta(q),\vec\phi(q)] \cdot \mathbf{U} $         
            
        \EndIf
        \State $q \gets q + 1$
    \EndFor
\EndFor
\EndIf

\State return $\mathbf{U}$
\EndFunction
\end{algorithmic}
\end{algorithm}

\section{Linear V-shaped scheme}
\label{app:vshape}

Among the linear schemes, we focus on the linear V-shaped scheme, but an analogous procedure can be implemented to obtain the generic routing action on any desired reference output state through the associated linear scheme shape. We adopt the linear V-shaped scheme because it is the most suitable for practical implementations and the most balanced, making it more resilient to MZI losses.
Moreover, between the modes $|\lceil N/2\rceil\rangle$ and $|\lceil N/2\rceil+1\rangle$, , we select the former. Accordingly,  $\mathbf{U}_{k,\lceil N/2\rceil}$ is simply denoted by $\mathbf{U}_{k}$ to simplify the notation.

\begin{figure}[t!]
    \centering
    \includegraphics[width=0.9\textwidth]{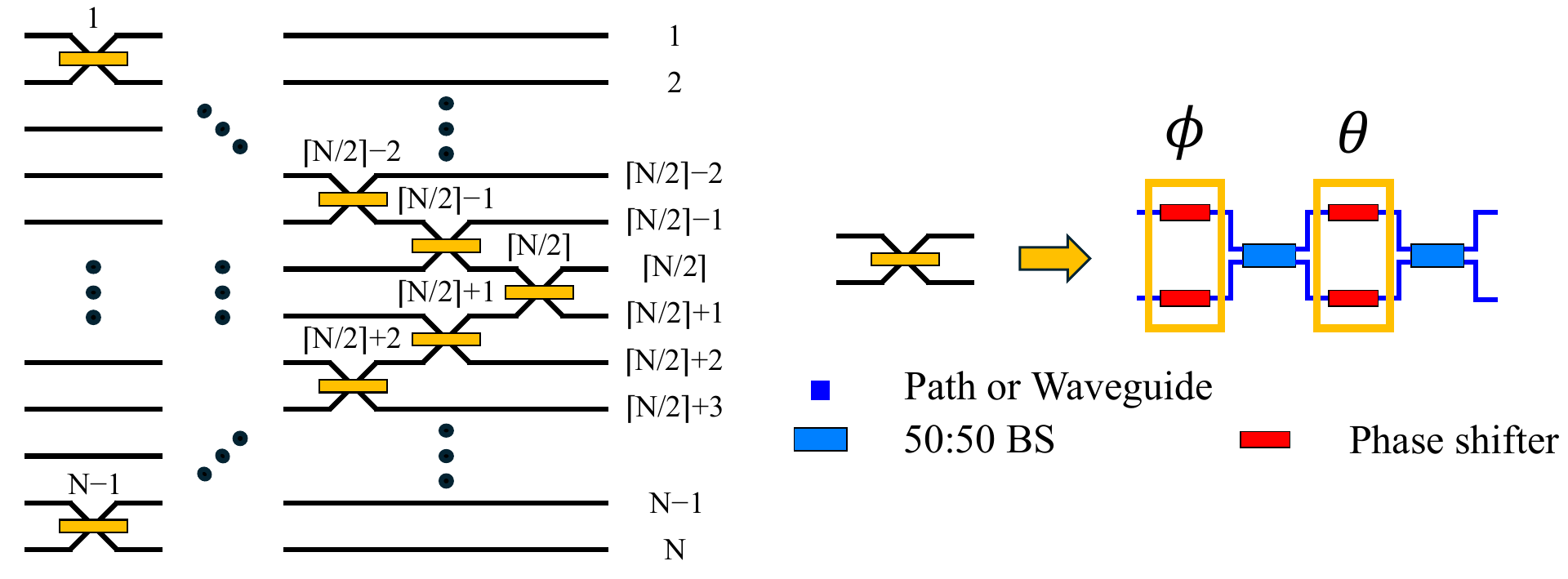}
    \caption{\textbf{Linear V-shaped MZI scheme.}
    On the left, the multiport interferometers with linear V-shaped mesh for generic $N$ number of modes. 
    On the right, the MZI with phase $\psi=0$ representation with the crossing lines with the yellow rectangle, see Fig. \ref{fig:mzi}. 
    Above each MZI, the black numbers specify the MZI position inside the linear V-shaped mesh. 
    The labelling of the mode number starts from the upper part from 1 to the $N$.
    }
    \label{fig:v_shape_gen}
\end{figure}

As mentioned in Section \ref{sec:routers}, vectors $|\psi^{\mathbf{u}}_k\rangle$, Eq. \eqref{eq:psikroutedvector}, can be unitarily transformed to the middle state $|\lceil N/2\rceil\rangle$ through the linear V-shaped scheme.
Looking at Fig. \ref{fig:v_shape_gen}, we can see that every layer contains one or two MZI and there are $\lceil N/2\rceil$ layers. The action of each MZI consists of eliminating an element of $|\psi^{\mathbf{u}}_k\rangle$ by using the property of matrices $\mathbf{u}_2$ in Eq. \eqref{eq:eliminationvector}:
\begin{equation}
\begin{split}
&|\psi^{\mathbf{u}}_k\rangle
=
\begin{pmatrix}
* \\
\vdots \\
\vdots \\
\vdots \\
\vdots \\
* \\
\end{pmatrix}
\to
\begin{pmatrix}
0 \\
* \\
\vdots \\
\vdots \\
\vdots \\
* \\
0 \\
\end{pmatrix}
\to\ldots
\to
\begin{pmatrix}
0 \\
\vdots \\
0 \\
{\rm e}^{{\rm i} \Gamma_k} \\
0 \\
\vdots\\
0
\end{pmatrix}
\,,
\end{split}
\label{eq:linearreduction}
\end{equation}
where ${\rm e}^{{\rm i} \Gamma_k}$ is at position $\lceil N/2\rceil$.
Note that the linear V-shaped routing follows a linear reduction algorithm of the initial vector. At each layer, one or two elements are eliminated.


To prove the general $N$-dimensional routing capability, we repeatedly apply the two-dimensional construction a total of $N-1$ times.
Thus, we start with the first two components of $|\psi_k^{\mathbf{u}}\rangle$, associated with the first two modes $(|1\rangle,|2\rangle)$, and we apply the MZI routing operation, Eq. \eqref{eq:mziasrouter}, to route the state $(\bar{u}_{k,1} , \bar{u}_{k,2} )^{\rm T}$ to the lower output, i.e. $|2\rangle$. This action is described by the following input and output states:
\begin{equation}
    \bar{u}_{k,1} |1\rangle + \bar{u}_{k,2} |2\rangle \to {\rm e}^{{\rm i} \gamma_1} \sqrt{|u_{k,1}|^2+|u_{k,2}|^2} \; |2\rangle \,,
    \label{eq:fiststepvshapealgo}
\end{equation}
where the phase $\gamma_1$ depends on the components $(u_{k,1} , u_{k,2})$. Note that the component of $|\psi_k^{\mathbf{u}}\rangle$ on the first mode has been eliminated, and the vector norm is conserved. At this point, we apply the same operation on the second and the third modes to eliminate the second component of $\mathbf{U}_{\rm MZI}^{(1,2)}\,|\psi_k^{\mathbf{u}}\rangle$. We iterate this procedure until the elimination of the $(\lceil N/2\rceil -1)$-th component of $|\psi_k^{\mathbf{u}}\rangle$, and the resulting state is
\begin{equation}
    |\tilde\psi_k^{\mathbf{u}}\rangle = 
    \prod_{j=1}^{\lceil N/2\rceil-1} \mathbf{U}_{\rm MZI}^{(j,j+1)}
    \,|\psi_k^{\mathbf{u}}\rangle
    =
    {\rm e}^{{\rm i} \gamma_{\lceil N/2\rceil}} \sqrt{\sum_{j=1}^{\lceil N/2\rceil} |u_{k,j}|^2} \; |\lceil N/2\rceil\rangle + \sum_{j=\lceil N/2\rceil+1}^N \bar{u}_{k}\,|j\rangle \,,
\end{equation}
where the phase $\gamma_{\lceil N/2\rceil}$ depends on the components $(u_{k,1},\ldots , u_{k,\lceil N/2\rceil})$ and the product is ordered from right to left.
Then, we consider the $(N-1)$-th and $N$-th components of $|\psi_k^{\mathbf{u}}\rangle$, associated with the last two modes $(|N-1\rangle,|N\rangle)$, and we apply the MZI routing operation, Eq. \eqref{eq:mziasrouter}, to route the state $[\bar{u}_{k,N-1} , \bar{u}_{k,N} ]^{\rm T}$ to the upper output, i.e. $|N-1\rangle$. Analogously to Eq. \eqref{eq:fiststepvshapealgo}, this transformation is described by the following input and output states:
\begin{equation}
    \bar{u}_{k,N-1} |N-1\rangle + \bar{u}_{k,N} |N\rangle \to {\rm e}^{{\rm i} \gamma_N} \sqrt{|u_{k,N-1}|^2+|u_{k,N}|^2} \; |N-1\rangle \,,
\end{equation}
where the phase $\gamma_N$ depends on the components $(u_{k,N-1} , u_{k,N})$. At this point, we apply the same operation on the $(N-2)$-th and the $(N-1)$-th modes to eliminate the $(N-1)$-th component of $\mathbf{U}_{\rm MZI}^{(N-1,N)}\,|\psi_k^{\mathbf{u}}\rangle$. We iterate this procedure until the $(\lceil N/2\rceil +2)$-th component of $|\psi_k^{\mathbf{u}}\rangle$ has been eliminated, and the resulting state is
\begin{equation}
\begin{split}
    &
    \prod_{j=1}^{N-\lceil N/2\rceil-1} \mathbf{U}_{\rm MZI}^{(N-j,N-j+1)} 
    \,|\tilde\psi_k^{\mathbf{u}}\rangle
    =
    \prod_{j=1}^{N-\lceil N/2\rceil-1} \mathbf{U}_{\rm MZI}^{(N-j,N-j+1)} 
    \cdot \prod_{j=1}^{\lceil N/2\rceil-1} \mathbf{U}_{\rm MZI}^{(j,j+1)}
    \,|\psi_k^{\mathbf{u}}\rangle
    \\
    & \hspace{2cm} =
    {\rm e}^{{\rm i} \gamma_{\lceil N/2\rceil}} \sqrt{\sum_{j=1}^{\lceil N/2\rceil} |u_{k,j}|^2} \; |\lceil N/2\rceil\rangle + {\rm e}^{{\rm i} \gamma_{\lceil N/2\rceil+2}} \sqrt{\sum_{j=\lceil N/2\rceil+1}^N |u_{k,j}|^2} \; |\lceil N/2\rceil+1\rangle \,,
    \end{split}
    \label{eq:lastlayervshape}
\end{equation}
where the phase $\gamma_{\lceil N/2\rceil+2}$ depends on the components $(u_{k,\lceil N/2\rceil+1},\ldots , u_{k,N})$ and the productorial is ordered from right to left.
Since the state in Eq. \eqref{eq:lastlayervshape} has only non-zero components for the $\lceil N/2\rceil$-th and $(\lceil N/2\rceil+1)$-th modes, the last step simply consists of the MZI routing action on these modes to route the state in Eq. \eqref{eq:lastlayervshape} to the middle mode $|\lceil N/2\rangle$.
Therefore, the generic unitary router $\mathbf{U}_k$, Eq. \eqref{eq:genericrouter}, is implemented by the sequence of unitary transformations associated with the MZI scheme shown in Fig. \ref{fig:v_shape_gen}:
\begin{equation}
\begin{split}
    \mathbf{U}_{\rm V\,scheme}^{(N)} &\equiv
    \mathbf{U}_{\rm MZI}^{(\lceil N/2\rceil,\lceil N/2\rceil+1)}[\theta_{\lceil N/2\rceil}^{(k)},\phi_{\lceil N/2\rceil}^{(k)}]\cdot \!\!\!\!\!\!\!\!\!
    \prod_{j=1}^{N-\lceil N/2\rceil-1} \!\!\!\!\!\!\!\! \mathbf{U}_{\rm MZI}^{(N-j,N-j+1)}[\theta_j^{(k)}\!\!,\phi_j^{(k)}] \cdot \!\!\!\!\! \prod_{j=1}^{\lceil N/2\rceil-1} \!\!\!\!\mathbf{U}_{\rm MZI}^{(j,j+1)}[\theta_j^{(k)}\!\!,\phi_j^{(k)}] \,, \\
    \mathbf{U}_k &= {\rm e}^{{\rm i} \Gamma_k }\;\mathbf{U}_{\rm V\,scheme}^{(N)} \left[\{(\theta_j^{(k)},\phi_j^{(k)})\}_{j\in 1\ldots (N-1)}\right]
    \,,
\end{split}
\end{equation}
where the phase $\Gamma_{k}$ depends on the components $(u_{k,1},\ldots , u_{k,N})$ and the productorial is ordered from right to left. At the end of the algorithm, the set of phase settings $\{(\theta_j^{(k)},\phi_j^{(k)})\}_{j\in 1\ldots (N-1)}$ required to implement the generic unitary router $\mathbf{U}_k$ is obtained.
The above construction applies to both even and odd values of $N$. In the latter case, the first layer consists of a single MZI acting on the first two modes.

The following algorithm summarizes the decomposition of a unitary matrix in terms of routers with the linear V-shaped mesh.

\begin{algorithm}[H]
\caption{The algorithm for linear V-shaped unitary matrix decomposition}\label{alg:v_shape}
\begin{algorithmic}
\Function{vshape\_decomposition}{$\mathbf{U}$}
\State $N \gets \text{size}(\mathbf{U})$

\State $\{\vec\theta_k,\vec\phi_k\}_{k\in 1\ldots N} \gets \{\vec 0,\vec 0\}_{k\in 1\ldots N}$ \Comment{Initialize $N$ vectors of dimension $N-1$ for MZI phases.}
\State $\vec v_{\rm phases} \gets \vec 0$ \Comment{Initialize vector of dimension $N$ containing the phases for the routers.}

\For{$k = 1$ to $N$}
    \State $\vec{v}_{work} \gets \mathbf{U}(k,:)$ \Comment{Create a copy of each row of $\mathbf{U}$.}
    
    \For{$i = 1$ to $\lceil N/2\rceil-1$}
        
        \State $(\vec\theta_k(i),\vec\phi_k(i)) \gets \text{\scriptsize ELEMENT\_ELIMINATION}(\vec{v}_{work}(i);\vec{v}_{work}(i+1);-1)$
        \State $G\gets \mathbf{U}_2^{(i)}[\vec\theta_k(i),\vec\phi_k(i)]$
        \State $\vec{v}_{work} \gets \vec{v}_{work} \cdot G^{-1}$ \Comment{Route to the lower mode.} 

    \EndFor
    \For{$i = 1$ to $N-\lceil N/2\rceil-1$}
        
        \State $(\vec\theta_k(N-i),\vec\phi_k(N-i)) \gets \text{\scriptsize ELEMENT\_ELIMINATION}(\vec{v}_{work}(N-i);\vec{v}_{work}(N-i+1);+1)$
        \State $G\gets \mathbf{U}_2^{(N-i)}[\vec\theta_k(N-i),\vec\phi_k(N-i)]$
        \State $\vec{v}_{work} \gets \vec{v}_{work} \cdot G^{-1}$ \Comment{Route to the upper mode.}

    \EndFor

    \State $(\vec\theta_k(\lceil N/2\rceil),\vec\phi_k(\lceil N/2\rceil)) \gets \text{\scriptsize ELEMENT\_ELIMINATION}(\vec{v}_{work}(\lceil N/2\rceil);\vec{v}_{work}(\lceil N/2\rceil+1);+1)$ 
        \State $G\gets \mathbf{U}_2^{(\lceil N/2\rceil)}[\vec\theta_k(\lceil N/2\rceil),\vec\phi_k(\lceil N/2\rceil)]$ 
        \State $\vec{v}_{work} \gets \vec{v}_{work} \cdot G^{-1}$ 
    \Comment{Route to the upper mode.}
    
    \State $\vec v_{\rm phase}(k) \gets \vec{v}_{work}(\lceil N/2\rceil)$ \Comment{Save the phase of the resulting phase vector.}
\EndFor

\State return $\{\vec\theta_k,\vec\phi_k\}_{k\in 1\ldots N}, \vec{v}_{\rm phases}$
\EndFunction
\end{algorithmic}
\end{algorithm}

The following algorithm takes as inputs the dimension of multiport interferometer and the phase settings, and returns the corresponding linear V-shaped-decomposed unitary matrix.

\begin{algorithm}[H]
\caption{The algorithm to generate a linear V-shaped-decomposed unitary matrix}\label{alg:vshapesmesh}
\begin{algorithmic}
\Function{vshape\_mesh}{$N;\{\vec\theta_k,\vec\phi_k\}_{k\in 1\ldots N}$}
\If{ $(N-1)\times N \ne \text{size}(\{\vec\theta_k\}_{k\in 1\ldots N}) \; \&\; (N-1)\times N  \ne \text{size}(\{\vec\phi_k\}_{k\in 1\ldots N})$}
    \State Not Compatible inputs
\Else
    \State $\{\mathbf{U}_k\}_{k\in 1\ldots N}\gets \{\mathbf{1}_N\}_{k\in 1\ldots N}$ 
    \For{$k = 1$ to $N$}
    \State $\mathbf{U}_k^{up}\gets \mathbf{1}_N$ and $\mathbf{U}_k^{down}\gets \mathbf{1}_N$
    \For{$i = 1$ to $\lceil N/2\rceil$}
    \Comment{Part '\textbackslash' (top-left to bottom-right).}
        \State $\mathbf{U}_k^{up}\gets \mathbf{U}_2^{(i)}[\vec\theta_k(i),\vec\phi_k(i)] \cdot \mathbf{U}_k^{up} $        
    \EndFor
    \For{$i = 1$ to $N-\lceil N/2\rceil-1$}
    \Comment{Part '/' (top-left to bottom-right).}
        \State $\mathbf{U}_k^{down}\gets \mathbf{U}_2^{(N-i)}[\vec\theta_k(N-i),\vec\phi_k(N-i)] \cdot \mathbf{U}_k^{down} $        
    \EndFor

    \State $\mathbf{U}_k\gets \mathbf{U}_k^{up}\cdot \mathbf{U}_k^{down} $
    \Comment{The MZI order is given in Fig. \ref{fig:v_shape_gen}.}
\EndFor
\EndIf

\State return $\{\mathbf{U}_k\}_{k\in 1\ldots N}$
\EndFunction
\end{algorithmic}
\end{algorithm}

Interestingly, because of the linear elimination procedure, it is possible to find a closed form for the $N$ phase settings of the $N$ unitary routers $\mathbf{U}_k$ associated with the unitary matrix $\mathbf{U}$. 
Indeed, if the number of modes is $N$, the $N$ unitary routers $\mathbf{U}_k$ on mode $|\lceil N/2\rceil\rangle$ implementing $\mathbf{U}$ in $N$ runs are given by the following sets of phases
\begin{equation}
\begin{split}
    \theta_j^{(k)} &= \mbox{atan}\left[\sqrt{\frac{\sum_{\ell=1}^{j}|u_{k,\ell}|^2}{|u_{k,j+1}|^2}}\right]  \quad,\quad
    \mbox{for}\,\,j\in 1\ldots\lceil N/2\rceil-1 \\
    \theta_{\lceil N/2\rceil}^{(k)} &= -\mbox{atan}\left[\sqrt{\frac{\sum_{j=\lceil N/2\rceil+1}^{N}|u_{k,\ell}|^2}{\sum_{j=1}^{\lceil N/2\rceil}|u_{k,\ell}|^2}}\right] \\
    \theta_j^{(k)} &= -\mbox{atan}\left[\sqrt{\frac{\sum_{\ell=j+1}^{N}|u_{k,\ell}|^2}{|u_{k,j}|^2}}\right]  \quad,\quad
    \mbox{for}\,\,j\in \lceil N/2\rceil+1 \ldots N-1 \\
    \phi_j^{(k)} &= \arg\left[u_{k,j}\right]-\arg\left[u_{k,j+1}\right] \quad,\quad
    \mbox{for}\,\,j\in1\ldots \lceil N/2\rceil-1 \\
    \phi_j^{(k)} &= \arg\left[u_{k,j}\right]-\arg\left[u_{k,N}\right] \quad,\quad
    \mbox{for}\,\,j\in \lceil N/2\rceil \ldots N-1
\end{split}
\end{equation}
where the MZI order is given in Fig. \ref{fig:v_shape_gen} and the final phase in Eq. \eqref{eq:linearreduction} $\Gamma_k$ is $\arg\left[ u_{k,\lceil N/2\rceil+1}\right]$.

\section{Tree scheme}
\label{app:triangular}

\begin{figure}[t!]
    \centering
    \includegraphics[width=0.9\textwidth]{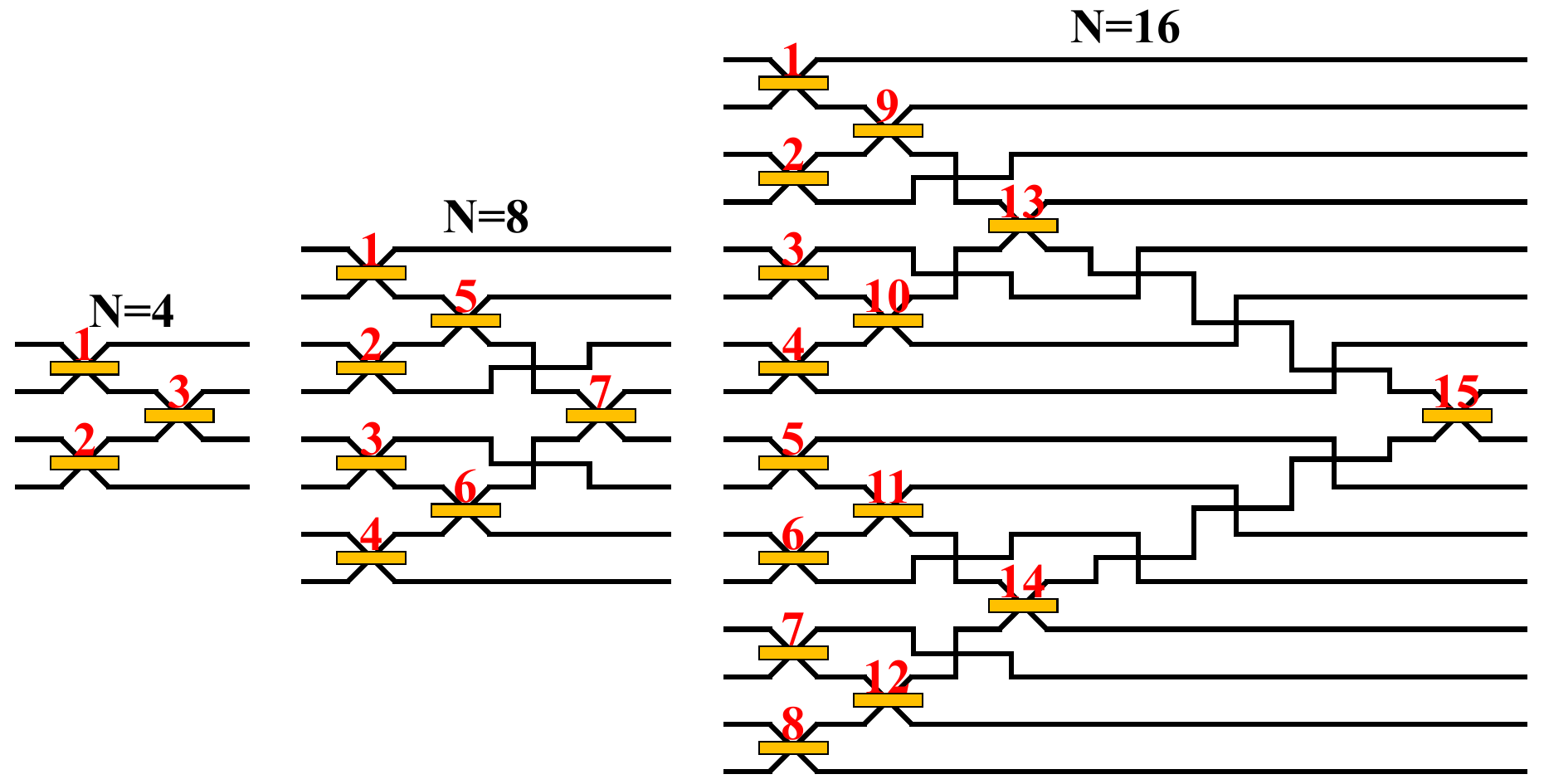}
    \caption{\textbf{Tree scheme for different numbers of modes.}
    The multiport interferometers with tree mesh for $N=[4,8,16]$ number of modes. 
    The MZI is represented by the crossing lines with the yellow rectangle. 
    Above each MZI, the red number specify the MZI position inside the Clements mesh. 
    In the figure the number of the modes are omitted: their labelling start from the upper part from 1 to the $N$.
    As in Fig. \ref{fig:t_shape_gen}, note that the MZI order follows the layer of the multiport interferometer from top to bottom.
    }
    \label{fig:triangular_app}
\end{figure}

As discussed in Section \ref{sec:routers}, the vectors $|\psi^{\mathbf{u}}_k\rangle$, defined in Eq. \eqref{eq:psikroutedvector}, can be transformed unitarily into the middle state $|\lceil N/2\rceil\rangle$ using the tree scheme.
For the tree scheme, it is easier to work with a number of modes that is a power of two. Thus, in the general $N$-dimensional case, we add $2^{\lceil \log_2 N\rceil}-N$ modes in order to complete the tree structure. Then, the MZIs acting on the added modes are used with a phase setting giving an identity evolution.
Note that, between the modes $|\lceil N/2\rceil\rangle$ and $|\lceil N/2\rceil+1\rangle$, we choose the the former. However, this choice is arbitrary and can be changed without difficulty. As in Section \ref{app:vshape}, $\mathbf{U}_{k,\lceil N/2\rceil}$ is simply denoted as $\mathbf{U}_{k}$ in order to simplify the notation.

\begin{figure}[t!]
    \centering
    \includegraphics[width=0.9\textwidth]{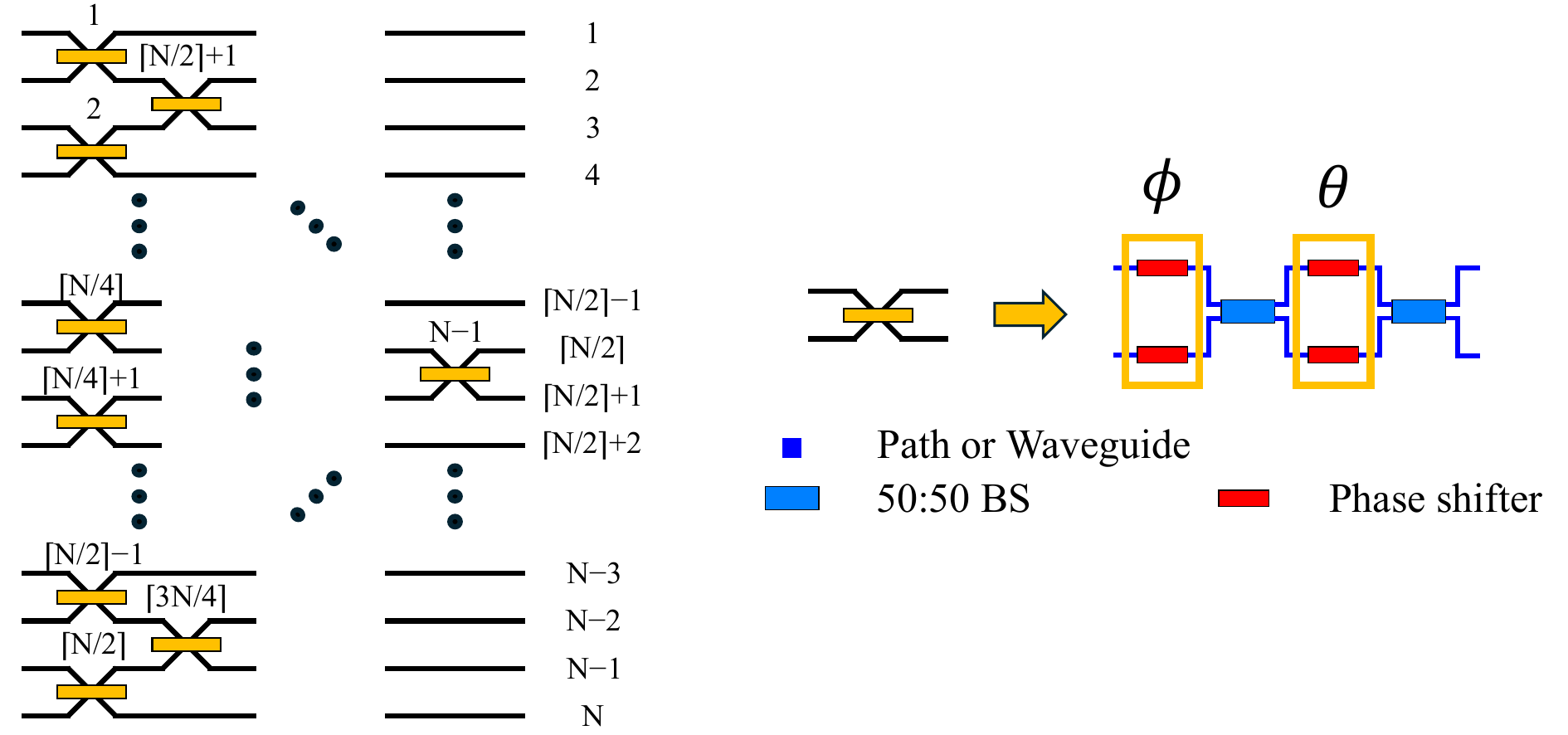}
    \caption{\textbf{Tree shape MZI scheme.}
    On the left, the multiport interferometers with tree mesh for generic $N$ number of modes. 
    On the right, the MZI with phase $\psi=0$ representation with the crossing lines with the yellow rectangle, see Fig. \ref{fig:mzi}. 
    Above each MZI, the black numbers specify the MZI position inside the tree mesh. 
    The labelling of the mode number starts from the upper part from 1 to the $N$.
    }
    \label{fig:t_shape_gen}
\end{figure}

Similarly to the linear V-shaped mesh, the action of each MZI consists of eliminating an element of $|\psi^{\mathbf{u}}_k\rangle$, Eq. \eqref{eq:psikroutedvector}, by using the property of matrices $\mathbf{u}_2$ in Eq. \eqref{eq:eliminationvector}:
\begin{equation}
\begin{split}
&|\psi^{\mathbf{u}}_k\rangle
=
\begin{pmatrix}
* \\
* \\
\vdots \\
* \\
* \\
\vdots \\
* \\
* \\
\end{pmatrix}
\to
\begin{pmatrix}
0 \\
* \\
\vdots \\
* \\
0 \\
\vdots \\
* \\
0 \\
\end{pmatrix}
\to\ldots
\to
\begin{pmatrix}
0 \\
\vdots \\
0 \\
{\rm e}^{{\rm i} \Gamma_k} \\
0 \\
\vdots\\
0
\end{pmatrix}
\,,
\end{split}
\label{eq:binaryreduction}
\end{equation}
where ${\rm e}^{{\rm i} \Gamma_k}$ is at position $\lceil N/2\rceil$. Differently from the linear schemes, the tree routing scheme follows a binary reduction algorithm of the initial vector. At the $n$-th layer, $N/2^n$ elements are eliminated.
Looking at Fig. \ref{fig:t_shape_gen} and Fig. \ref{fig:triangular_app}, we can see that the $n$-th layer contains $N/2^n$ MZIs, and there are $ \log_2 N$ layers.

The tree scheme is based on a binary reduction strategy, in which the number of nonzero components is halved at each layer. In the first layer, the components are grouped into neighboring pairs, and $N/2$ MZIs are used to apply the routing operation defined in Eq.~\eqref{eq:mziasrouter}. Each MZI routes the state associated with its corresponding pair to the lower or upper output, depending on whether the MZI is odd- or even-indexed. 
The resulting state is:

\begin{equation}
    |\tilde\psi_k^{\mathbf{u}}\rangle = 
    \prod_{j=1}^{N/2} \mathbf{U}_{\rm MZI}^{(2j-1,2j)}
    \,|\psi_k^{\mathbf{u}}\rangle
    =
    \sum_{j=1}^{N/2}
    {\rm e}^{{\rm i} \gamma_j} \sqrt{ |u_{k,2j-1}|^2 + |u_{k,2j}|^2} \; | 2j-{\rm mod}_2(j+1) \rangle  \,,
\end{equation}
where $N$ is a power of two, the phases $\gamma_{j}$ depend on the components $(u_{k,2j-1}, u_{k,2j})$ and the product is ordered from right to left.
Then, in the $n$-th layer, we apply $N/2^n$ MZIs routing operation, Eq. \eqref{eq:mziasrouter}, to route the state of each pair to the lower/upper output, for odd/even numbered MZI.
As shown for the eight-dimensional case in Fig. \ref{fig:scheme_8dim}, crossings are needed to route and group non-zero output amplitudes close to the middle mode $|N/2\rangle$. The crossings can be avoided by terminating the paths that are not routed to the middle mode. However, this design choice makes the characterization of the complete multiport interferometer more operationally demanding.
After $\log_2 N$ layers, the input state $|\psi^{\mathbf{u}}_k\rangle$ is routed to the middle mode $|N/2\rangle$. This means that the generic unitary router $\mathbf{U}_k$, Eq. \eqref{eq:genericrouter}, is obtained through the sequence of unitary transformations associated with the MZI scheme shown in Fig. \ref{fig:t_shape_gen}:
\begin{equation}
\begin{split}
    \mathbf{U}_{\rm Tree}^{(N)} &\equiv
    \overbrace{\mathbf{U}_{\rm MZI}^{(N/2, N/2+1)}[\theta_{N-1}^{(k)},\phi_{N-1}^{(k)}]}^{\log_2N-{\rm th} \,\,\,{\rm layer}}\cdot \ldots\cdot\!
    \overbrace{\prod_{j=1}^{N/4} \mathbf{U}_{\rm MZI}^{(f_2(j),f_2(j)+1)} [\theta_j^{(k)}\!\!,\phi_j^{(k)}]}^{2-{\rm nd} \,\,\,{\rm layer}}\cdot \!
    \overbrace{\prod_{j=1}^{N/2} \mathbf{U}_{\rm MZI}^{(f_1(j),f_1(j)+1)} [\theta_j^{(k)}\!\!,\phi_j^{(k)}]}^{1-{\rm st} \,\,\,{\rm layer}} \,, \\
    \mathbf{U}_k &= {\rm e}^{{\rm i} \Gamma_k }\;\mathbf{U}_{\rm Tree}^{(N)} \left[\{(\theta_j^{(k)},\phi_j^{(k)})\}_{j\in 1\ldots (N-1)}\right]
    \,,
\end{split}
\end{equation}
where $f_n(j)=2^{n-1}(2j-1)$ is the function for the index associated with the $j$-th MZI in the $n$-th layer, the crossing maps are present after the second layer, the phase $\Gamma_{k}$ depends on the components $(u_{k,1},\ldots , u_{k,N})$ and the product is ordered from right to left. Also in this case, at the end of the explained procedure, the set of phases $\{(\theta_j^{(k)},\phi_j^{(k)})\}_{j\in 1\ldots (N-1)}$ to implement the generic unitary router $\mathbf{U}_k$ is found.

As we discussed in the main text, the tree scheme needs crossings. Fig. \ref{fig:triangular_app} shows how the number of crossings increases with the mode number. In particular, between the first and the second layer there are zero crossings, between the second and the  third layer there are four crossings, between the third and the fourth layer there are eight crossings, and between $n$-th and $(n+1)$-th layer there are $4(n-1)$ crossings for $n>1$. If a $N$-dimensional tree scheme has $c(N)$ crossings, $2N$-dimensional tree schemes present $c(2N)=2c(N)+N-2$ crossings. Therefore, for the $N$ dimensional tree schemes with $N\ge 4$, there are $2+\frac{N}{2}\left(\log_2N-3\right)$ crossings.

The following algorithm summarizes the decomposition of a unitary matrix in terms of routers with the tree mesh.

\begin{algorithm}[H]
\caption{The algorithm for tree unitary matrix decomposition}\label{alg:t_shape}
\begin{algorithmic}
\Function{tree\_decomposition}{$\mathbf{U}$}
\State $N \gets \text{size}(\mathbf{U})$

\If{$N< 2^{\lceil\log_2N\rceil}$}
    \State $\tilde N \gets 2^{\lceil\log_2N\rceil}$
    \Comment{$\tilde N$ is a power of two.}
    \State Create $\tilde N\times \tilde N$ block diagonal matrix $\tilde{\mathbf{U}}$ with first $N\times N$ block equal to $\mathbf{U}$ and identity for the remaining block.
\EndIf

\State $\{\vec\theta_k,\vec\phi_k\}_{k\in 1\ldots \tilde N} \gets \{\vec 0,\vec 0\}_{k\in 1\ldots \tilde N}$ \Comment{Initialize $\tilde N$ vectors of dimension $\tilde N-1$ for MZI phases.}
\State $\vec v_{\rm phases} \gets \vec 0$ \Comment{Initialize vector of dimension $\tilde N$ containing the phases for the routers.}

\For{$k = 1$ to $\tilde N$}
    \State $\vec{v}_{work} \gets \tilde{\mathbf{U}}(k,:)$ \Comment{Create a copy of each row of $\tilde{\mathbf{U}}$.}
    \State $q \gets 1$
    
    \For{$n = 1$ to $\log_2\tilde N-1$} \Comment{Cycle over the layers of the tree mesh.}
        \State $\vec n_{del} \gets []$
        \Comment{The position of the eliminated components of $\vec{v}_{work}$.}
        \For{$i = 1$ to $\tilde N/2^n$} \Comment{Cycle over the MZIs in each layer.}
            \If{$i$ is odd}
                \State $c\gets-1$ \Comment{Route to the lower mode.} 
                \State insert $2i-1$ in $\vec n_{del}$
            \Else
                \State $c\gets+1$ \Comment{Route to the upper mode.} 
                \State insert $2i$ in $\vec n_{del}$
            \EndIf
            
            \State $(\vec\theta_k(q),\vec\phi_k(q)) \gets \text{\scriptsize ELEMENT\_ELIMINATION}(\vec{v}_{work}(2i-1);\vec{v}_{work}(2i);c)$
            \State $G\gets \tilde{\mathbf{U}}_2^{(i)}[\vec\theta_k(q),\vec\phi_k(q)]$
            \Comment{The dimension of $\tilde{\mathbf{U}}_2^{(i)}$ is $\tilde N/2^{n-1}$.}
            \State $\vec{v}_{work} \gets \vec{v}_{work} \cdot G^{-1}$ 
            \State $q\gets q+1$
        \EndFor
        \State Discard the elements of $\vec{v}_{work}$ with positions given by $\vec n_{del}$
    \EndFor
    
        \State $(\vec\theta_k(\tilde N-1),\vec\phi_k(\tilde N-1)) \gets \text{\scriptsize ELEMENT\_ELIMINATION}(\vec{v}_{work}(1);\vec{v}_{work}(2);+1)$
        \State $G\gets \tilde{\mathbf{U}}_2^{(1)}[\vec\theta_k(\tilde N-1),\vec\phi_k(\tilde N-1)]$ 
        \Comment{The dimension of $\tilde{\mathbf{U}}_2^{(1)}$ is $2$.}
        \State $\vec{v}_{work} \gets \vec{v}_{work} \cdot G^{-1}$ \Comment{Route to the upper mode.}
    
    \State $\vec v_{\rm phase}(k) \gets \vec{v}_{work}(1)$ \Comment{Save the phase of the resulting phase vector.}
\EndFor

\State return $\{\vec\theta_k,\vec\phi_k\}_{k\in 1\ldots \tilde N}, \vec{v}_{\rm phases}$
\EndFunction
\end{algorithmic}
\end{algorithm}

Finally, the following algorithm takes as inputs the dimension of multiport interferometer and the phase setting, and it gives the corresponding tree-decomposed unitary matrix.

\begin{algorithm}[H]
\caption{The algorithm to generate a tree-decomposed unitary matrix}\label{alg:tshapesmesh}
\begin{algorithmic}
\Function{tree\_mesh}{$N;\{\vec\theta_k,\vec\phi_k\}_{k\in 1\ldots N}$}

\If{ $(N-1)\times N \ne \text{size}(\{\vec\theta_k\}_{k\in 1\ldots N}) \; \&\; (N-1)\times N  \ne \text{size}(\{\vec\phi_k\}_{k\in 1\ldots N})$}
    \State Not Compatible inputs
\Else
    \If{$N< 2^{\lceil\log_2N\rceil}$}
        \State $\tilde N \gets 2^{\lceil\log_2N\rceil}$
        \State Add zeros to $\{\vec\theta_k\}_{k\in 1\ldots N})$ and$ \{\vec\phi_k\}_{k\in 1\ldots N}$ to make their size equal to $(\tilde N-1)\times \tilde N$.
    \EndIf
    
    \State $\{\mathbf{U}_k\}_{k\in 1\ldots \tilde N}\gets \{\mathbf{1}_{\tilde N}\}_{k\in 1\ldots \tilde N}$ 
    
    \For{$k = 1$ to $\tilde N$}
    \State $q\gets 1$
    \For{$n = 1$ to $\log_2 \tilde N$} \Comment{The MZI order is given in Fig. \ref{fig:t_shape_gen} and \ref{fig:triangular_app}.}
        
        \State $\vec n_{cross}\gets []$
        \Comment{Indexes for the crossing positions.}
        \For{$i=1$ to $\tilde N/2^{n}$} 
            \State $h\gets2^{n-1}(2i-1)$        
            \State $\mathbf{U}_k\gets \mathbf{U}_2^{(h)}[\vec\theta_k(q),\vec\phi_k(q)] \cdot \mathbf{U}_k $ 
            \State Insert $h$ in the end of $\vec n_{cross}$
            \State $q\gets q+1$
        \EndFor
        \If{$\text{size}(\vec n_{cross})>1$} \Comment{Procedure to insert the crossings.}
            \For{$g=1$ to $\text{size}(\vec n_{cross})$ with two-unit steps}
                \While{$\vec n_{cross}(g+1)-\vec{n}_{cross}(g)>2$}
                    \State $\mathbf{U}_k \gets \mathbf{U}_{cross}^{(\vec n_{cross}(g+1)-1)} \cdot \mathbf{U}_{cross}^{(\vec n_{cross}(g)+1)} \cdot \mathbf{U}_k$
                    \State $\vec n_{cross}(g) \gets \vec n_{cross}(g)+1$ and $\vec n_{cross}(g+1)\gets \vec n_{cross}(g+1)-1$
                \EndWhile
            \EndFor
        \EndIf
    \EndFor
    
\EndFor
\EndIf

\State return $\{\mathbf{U}_k\}_{k\in 1\ldots \tilde N}$
\EndFunction
\end{algorithmic}
\end{algorithm}

\section{Multi-routers and multilinear schemes}
\label{app:multilinear}

In this section, we show how multi-routers and in particular multilinear schemes work. Finally, in subsection \ref{app:multilinear_BS} we demonstrate that the sampling at the $N$ outputs of a generic unitary transformation $\mathbf{U}$ for an input state composed of $m$ photons is equivalent to the sampling at $m$ outputs of the $N!/(m!(N-m)!)$ unitary transformations implementing all multi-routers associated with $\mathbf{U}$.

\begin{figure}[t!]
    \centering
    \includegraphics[width=0.9\textwidth]{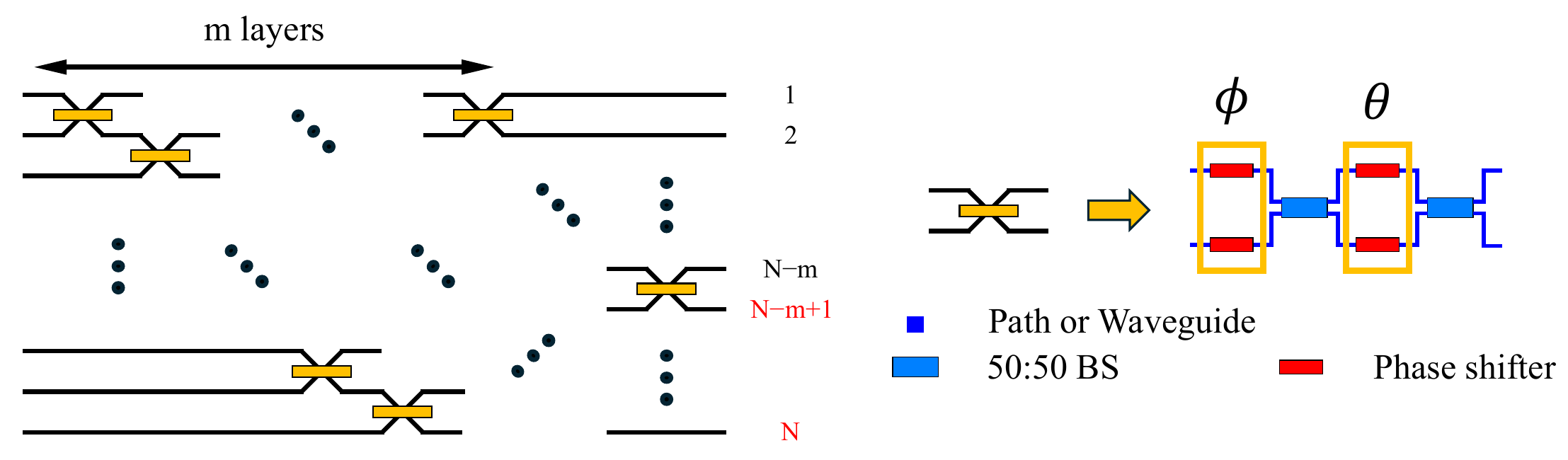}
    \caption{\textbf{Multilinear MZI scheme for $m$ photons.}
    On the left, the multiport interferometers with multilinear mesh for generic $N$ number of modes. 
    On the right, the MZI with phase $\psi=0$ representation with the crossing lines with the yellow rectangle, see Fig. \ref{fig:mzi}. 
    Above each MZI, the black numbers specify the MZI position inside the linear V-shaped mesh. 
    The labelling of the mode number starts from the upper part from 1 to the $N$. The red labels in the outputs give a possible reference output modes' set.
    }
    \label{fig:multilin_gen}
\end{figure}

As said in Section \ref{sec:routers} and in Appendix \ref{app:vshape}, any vectors $|\psi^{\mathbf{u}}_k\rangle$, Eq. \eqref{eq:psikroutedvector}, can be unitarily transformed to the desired reference output mode state $|r\rangle$ through the associated routing scheme. In particular, Appendix \ref{app:vshape} is focused on the linear V-shaped scheme and the reference output mode $|\lceil N/2\rceil\rangle.$
For multi-routing schemes, this technique is repeated for $m$ orthonormal vectors $\{|\psi^{\mathbf{u}}_{k_j}\rangle\}_{j\in 1\ldots m}$ to transform them in the set of reference output modes $\{|r_{j}\rangle\}_{j\in 1\ldots m}$. 
The unitary multi-routing operator $\mathbf{U}_{\mathbf{k},\mathbf{r}}^{(m)}$ can be constructed via routing schemes by means of an iterative procedure. Indeed, $\mathbf{U}_{\mathbf{k},\mathbf{r}}^{(m)}$ can be obtained as a composition of suitable unitaries of the form
\begin{equation}\label{decomposition-scheme}
\mathbf{U}_{\mathbf{k},\mathbf{r}}^{(m)}=\mathbf{U}_{{(k_1,r_1),...,(k_m,r_m)}}\cdot \ldots \cdot\mathbf{U}_{{(k_1,r_1),( k_2,r_2)}}\cdot\mathbf{U}_{{(k_1,r_1)}}\,.
\end{equation}
First of all, $\mathbf{U}_{{(k_1,r_1)}}$ is a unitary operator mapping $|\psi_{(k_1,r_1)}^{\mathbf{u}}\rangle$ into $e^{\gamma_1}|r_1\rangle$. As we have seen in Appendix \ref{app:vshape}, it can be realized by concatenating $N-1$ MZIs in a linear routing scheme.
Note that, by unitarity, $\mathbf{U}_{{(k_1,r_1)}}$ maps the $m-1$ vectors $|\psi_{k_2}^{\mathbf{u}}\rangle, ...,|\psi_{k_m}^{\mathbf{u}}\rangle$ into vectors $|\psi'_{k_j}\rangle\equiv \mathbf{U}_{{(k_1,r_1)}}|\psi_{k_j}^{\mathbf{u}}\rangle$, $j=1,...,m$ that are normalized and orthogonal to $|r_1\rangle$. This property allows to reduce the dimensionality of the problem, since now in the second step we have to implement a unitary operator $\mathbf{U}_{{(k_1,r_1), (k_2,r_2)}}$ acting on the $N-1$-dimensional space generated by the vectors $\{|n\rangle\}_{n\neq r_1}$ and mapping the vector $|\psi'_{k_2}\rangle$ to the reference vector $|r_2\rangle$ up to an arbitrary phase factor. Again, as remarked above, $\mathbf{U}_{{(k_1,r_1), (k_2,r_2)}}$ can be obtained by concatenating $N-2$ MZIs, associated with unitary operators acting only on two modes.
The scheme can be repeated further $(m-2)$ times. Again, the operator $\mathbf{U}_{{(k_1,r_1), (k_2,r_2)}}$ maps the orthonormal vectors $|\psi'_{k_j}\rangle$, $j=3,...,m$, to orthonormal vectors  $|\psi''_{k_j}\rangle\equiv \mathbf{U}_{{(k_1,r_1), (k_2,r_2)}}|\psi'_{k_j}\rangle$ that are also orthogonal to the subspace generated by $\{|r_1\rangle, |r_2\rangle\}$. Hence, the unitary operator $\mathbf{U}_{(k_1,r_1), (k_2,r_2),(k_3,r_3)}$ mapping $|\psi''_{k_j}\rangle$ to $e^{i\gamma_3}|r_3\rangle$ can be obtained via a linear scheme of $(N-3)$ MZIs.
Finally, after further $(m-4)$ steps, the last step amounts to the construction, via a linear scheme consisting of  $(N-m)$ MZIs, of a unitary operator $\mathbf{U}_{{(k_1,r_1),...,(k_m,r_m)}}$ mapping the vector $|\psi^{(m-1)}_{k_j}\rangle\equiv \mathbf{U}_{{(k_1,r_1),...,(k_{m-1},r_{m-1})}}\cdot\ldots \cdot\mathbf{U}_{{(k_1,r_1), (k_2,r_2)}}\cdot\mathbf{U}_{{(k_1,r_1)}}|\psi^{\mathbf{u}}_{k_m}\rangle$ to the final reference vector $|r_m\rangle$.\\
Thus, the action of a multi-router scheme can be summarized as follows:
\begin{equation}
|\psi^{\mathbf{u}}_{k_j}\rangle \to |r_{j}\rangle
\qquad \forall\,j\in 1\ldots m \,.
\end{equation}
This technique can be understood as composed of $m$ steps. At each step, a reduction of one vector $|\psi^{\mathbf{u}}_{k_j}\rangle$ is performed as shown in Eq. \eqref{eq:linearreduction} for the middle mode as reference output mode.
Choosing different linear schemes, Fig. \ref{fig:router_scheme}(a), it is possible to choose any output mode as the reference output mode.
Without loss of generality, if we choose the reference output modes as $\{|N-m+1\rangle,\ldots,|N\rangle\}$, the transformation of the unitary multi-router can be decomposed as follows:
\begin{equation}
\begin{split}
        &|\psi^{\mathbf{u}}_{k_1}\rangle \to {\rm e}^{{\rm i} \Gamma_{1,k_1}}|N\rangle \to {\rm e}^{{\rm i} \Gamma_{1,k_1}}|N\rangle \to \ldots \to {\rm e}^{{\rm i} \Gamma_{1,k_1}}|N\rangle \,,\\
        &|\psi^{\mathbf{u}}_{k_2}\rangle \to \sum_{q=1}^{N-1} c^{\mathbf{u}}_{k_2} |q\rangle \to {\rm e}^{{\rm i} \Gamma_{2,k_2}}|N-1\rangle \to \ldots \to {\rm e}^{{\rm i} \Gamma_{2,k_2}}|N-1\rangle \,,\\
        &\ldots \\
        &|\psi^{\mathbf{u}}_{k_m}\rangle \to \sum_{q=1}^{N-1} c^{\mathbf{u}}_{k_m} |q\rangle \to \sum_{q=1}^{N-2} \tilde{c}^{\mathbf{u}}_{k_m} |q\rangle \to \ldots \to {\rm e}^{{\rm i} \Gamma_{m,k_m}}|N-m+1\rangle \,,
\end{split}
\label{eq:multilinear_action}
\end{equation}
where $\Gamma$ are phases depending on the vectors $\{|\psi^{\mathbf{u}}_{k_j}\rangle\}_{j\in 1\ldots m}$ or, equivalently, on the components $\{(u_{k_1,1},\ldots , u_{k_m,N})\}_{m\in 1\ldots m}$.
At the first step, one vector is mapped to its corresponding reference output mode. As discussed above, the transformation is unitary; therefore, all the remaining vectors are mapped to vectors orthogonal to it and, consequently, have no component along the corresponding reference output mode. This is illustrated by the first arrows in the previous equations. The next transformation then acts on the subspace orthogonal to the first reference mode, and the procedure is iterated $m$ times.
Finally, the above construction can be straightforwardly adapted to any choice of the reference output modes.
In particular, the procedure illustrated in the previous equation matches the scheme shown in Fig. \ref{fig:multilin_gen}.
Specifically, each step of $m$ transformations reported in Eq. \eqref{eq:multilinear_action} is associated with the one diagonal '\textbackslash' layers.
From this perspective, the multilinear scheme can be viewed as a combination of $m$ of progressively decreasing dimension. The first layer has dimension $N$, the second $N-1$, and so on, until the last layer, which has dimension $N-m+1$.\\
Therefore, the generic unitary router $\mathbf{U}_{\mathbf{k},\mathbf{r}}^{(m)}$, Eq. \eqref{eq:unitaryU_as_UV_morephotons}, with $|r_j\rangle=|N-j+1\rangle$ is implemented by the sequence of unitary transformations associated with the MZI scheme shown in Fig. \ref{fig:multilin_gen}:
\begin{equation}
\begin{split}
    \mathbf{U}_{\rm multilin.\,scheme}^{(m,N)} &\equiv
    \prod_{\ell=1}^{m} \,\,\mathbf{U}_{\rm lin.\,scheme}^{(N-\ell+1)}\,, \\
    \mathbf{U}_{\rm lin.\,scheme}^{(N-\ell+1)} &= 
    \prod_{j=1}^{ N-\ell} \,\mathbf{U}_{\rm MZI}^{(j,j+1)}[\theta_j^{(\mathbf{k})}\!\!,\phi_j^{(\mathbf{k})}] \,,\\
    \mathbf{U}_{\mathbf{k},\mathbf{r}}^{(m)} &= \left( {\rm e}^{{\rm i} \Gamma_{\mathbf{k}} } \right)_m\;\mathbf{U}_{\rm multilin.\,scheme}^{(m,N)} \left[\{(\theta_j^{(\mathbf{k})},\phi_j^{(\mathbf{k})})\}_{j\in 1\ldots m \, (N-(m+1)/2)}\right]
    \,,
\end{split}
\end{equation}
where $\mathbf{r}=(N,N-1,\ldots,N-m+1\rangle$, the product is ordered from right to left and $\left( {\rm e}^{{\rm i} \Gamma_{\mathbf{k}} } \right)_m$ is a matrix $N\times N$ with phases $\Gamma$ in the positions of the reference output modes, depending on the components $\{(u_{k_1,1},\ldots , u_{k_m,N})\}_{m\in 1\ldots m}$. At the end of the algorithm, the set of phase settings $\{(\theta_j^{(\mathbf{k})},\phi_j^{(\mathbf{k})})\}_{j\in 1\ldots m \, (N-(m+1)/2)}$ required to implement the generic unitary multi-router $\mathbf{U}_{\mathbf{k},\mathbf{r}}^{(m)}$ is obtained.
This structure is associated with the choice of reference output modes $\{|N-m+1\rangle,\ldots,|N\rangle\}$, but it can easily be changed to any choice of reference output modes.

\subsection{Estimation of boson sampling probabilities via multi-routing schemes}
\label{app:multilinear_BS}
If two photons in the modes $i,j$ are injected in the linear interferometer, that is if the initial state is  $|\psi_{in}\rangle =a^\dag_ia^\dag _j|\Omega\rangle$, $|\Omega\rangle$ being the vacuum state, then
 the output state is given by $|\psi_{out}\rangle =\sum_{n_1,n_2=1}^Nu_{n_1i}u_{n_2j}a^\dag_{n_1}a^\dag _{n_2}|\Omega\rangle$. 
In this case, the probability of detecting one photon in the mode $k_1$ and one photon in the mode $k_2$, with $k_1\neq k_2$, is given by:
\begin{equation}\label{BosonSampling-2photons}
P(k_1,k_2|i,j)=|u_{k_1i}u_{k_2j}+u_{k_1i}u_{k_2j}|^2\,,
\end{equation}
while the probability of coincidences, i.e. of detecting two photons at the mode $k$ is equal to
\begin{equation}\label{BosonSampling-2photons-coinc}
P(k,k|i,j)=2|u_{ki}u_{kj}|^2\,.
\end{equation}
In the general case of $m$, with $m\leq N$, photons injected in the linear interferometer in $m$ different modes $\mathbf{j}\equiv(j_1,...,j_m)$, the input state is \begin{equation}|\psi_{in}\rangle =a^\dag_{j_1}\cdots a^\dag _{j_m}|\Omega\rangle,\end{equation}
while the output state assumes the following form:
\begin{equation}|\psi_{out}\rangle =\sum_{n_1,...,n_m=1}^Nu_{n_1j_1}\cdots u_{n_mj_m}a^\dag_{n_1}\cdots a^\dag _{n_m}|\Omega\rangle\,.\end{equation}
Hence, the probability of detecting at the output of the interferometer  $m$ photons in the $m$ distinct modes $\mathbf{k}\equiv (k_1,...,k_m)$ is given by 
\begin{equation}\label{Boson-sampling-different-modes}P(k_1,...,k_m|j_1,...,j_m)=|\mathrm{Perm}\left(\tilde U_{\mathbf{k},\mathbf{j}}\right)|^2\,,\end{equation}
where $\tilde U_{\mathbf{k},\mathbf{j}}$ is an $m\times m$ matrix made whose rows are labelled by the output modes  $\mathbf{k}$ and the columns are labelled by the input modes $\mathbf{j}$, while $\mathrm{Perm}(\tilde U_{\mathbf{k},\mathbf{j}})$ denotes its permanent, defined for a general $m\times m$ matrix $M$ as:
\begin{equation}\mathrm{Perm}(M)=\sum_{\sigma\in S_m}\prod_{\alpha=1}^mM_{\alpha\sigma(\alpha)}\,,\end{equation}
hence 
\begin{equation}\label{bosonSamplin-mphotons}
    P(k_1,...,k_m|j_1,...,j_m)=|\sum_{\sigma\in S_m}\prod_{\alpha=1}^mu_{k_\alpha j_{\sigma(\alpha)}}|^2\,.
\end{equation}
In order to take into account also the cases where two or more photons are detected in the same mode,  it is convenient to describe the output configuration in terms of the  the occupation number $\mathbf{n}=(n_1,...,n_N)$, with $n_i\geq 0$ denoting the number of photons detected at the mode $i$, fulfilling the constraint $\sum_{i=1}^Nn_i=m$. Similarly, a general  input state  can be described by the input occupation number $\mathbf{s}=(s_1,...,s_N)$, with $\sum_{i=1}^Ns_i=m$ and $s_i\geq 0$ denoting the number of photons injected in the mode $i$. 
The detection probabilities can be computed as
\begin{equation}\label{bosonSamplin-mphotons-coinc}
P(\mathbf{n}|\mathbf{s})=\frac{|\mathrm{Perm}(\tilde U_{\mathbf{n},\mathbf{s}})|^2}{\prod_{i=1}^Ns_i!\prod_{j=1}^Nn_j!}
\end{equation}
where $\tilde U_{\mathbf{n},\mathbf{s}}$ is an $m\times m$ matrix obtained from $\mathbf{U}$ by repeating the rows according to the output occupation numbers and the columns according to the input ones. Clearly, Eq. \eqref{bosonSamplin-mphotons-coinc} reduces to Eq. \eqref{Boson-sampling-different-modes} in the case of no coincidences, i.e. if every occupation number appearing in the vectors $\mathbf{n} $  and $\mathbf{s} $  is either 0 or 1.

Once $m$ distinct reference modes $\mathbf{r}=(r_1,...,r_m)$, each associated with a photon detector at the output of a linear optical interferometer, have been fixed, the boson sampling probabilities in Eq.~\eqref{bosonSamplin-mphotons} can be estimated from the detection probabilities of suitably designed router schemes. To this end, for any possible output sequence $\mathbf{k}=(k_1,...k_m)$ we need to construct a unitary $\mathbf{U}_{\mathbf{k},\mathbf{r}}^{(m)}$ such that for any $j=1,...,m$ and the following holds
\begin{equation}\label{cond-u-km}
    \langle r_j|\mathbf{U}_{\mathbf{k},\mathbf{r}}^{(m)}|\psi\rangle=e^{i\gamma_j}\langle k_j|\textbf{U}|\psi\rangle\,,
\end{equation}
for any one photon state $|\psi\rangle\in \mathbb{C}^N$ and for some phase $\gamma_j$.
Indeed, if \eqref{cond-u-km} holds, then the corresponding boson sampling probabilities associated with the multi-routing scheme
\begin{equation}P_{\mathbf{k}}(r_1,...,r_m|j_1,...,j_m)=|\sum_{\sigma\in S_m}\prod_{\alpha=1}^m\langle r_\alpha|\mathbf{U}_{\mathbf{k},\mathbf{r}}^{(m)}| j_{\sigma(\alpha)}\rangle|^2\,\end{equation}
coincide with the probabilities \eqref{bosonSamplin-mphotons}.
Note that the phases $\gamma_j$ are factorized and do not enter in the previous equation. More generally, the same result holds for the coincidence probabilities in Eq. \eqref{bosonSamplin-mphotons-coinc}. As remarked in the main text, in the case some of the output occupation numbers $n_i$ are strictly greater than 1, there are only $l<m$ output modes $\{k_1,\ldots, k_l\}$ associated with the non-vanishing occupation numbers. Hence, there are  multible experimental settings allowing the estimate of the detection probabilities \eqref{bosonSamplin-mphotons-coinc}, one for each set $\mathbf{k}$ containing  $\{k_1,\ldots, k_l\}$. The redundancy can be leveraged to increase the statistical accuracy of the estimated probabilities. In particular, an unbiased experimental estimator of $P(\mathbf{n}|\mathbf{s})$ is $$\sum_{\mathbf{k}}N_{\mathbf{k},\mathbf{n}}/\sum_{\mathbf{k}}N_{\mathbf{k}},$$ where the sums run over all $m-$tuples $\mathbf{k}$ containing the modes $\{k_1,\ldots, k_l\}$. For each configuration  $\mathbf{k}$, $N_{\mathbf{k}}$  is the total number of experimental runs, whereas $N_{\mathbf{k},\mathbf{n}}$ is the number of runs producing the occupation pattern  $\mathbf{n}$. \\

\section{Tolerance to MZI losses and outputs coupling losses}
\label{app:nonideal}

In this section, we perform additional simulations following the same procedure used for the results reported in Section \ref{sec:disc}. 
The fidelity and the TV distance yield compatible conclusions about the behaviour of the different schemes with respect to the considered non-idealities.

\begin{figure}[H]
    \centering
    \begin{subfigure}[b]{0.45\textwidth}
       \centering
        \includegraphics[width=\textwidth]{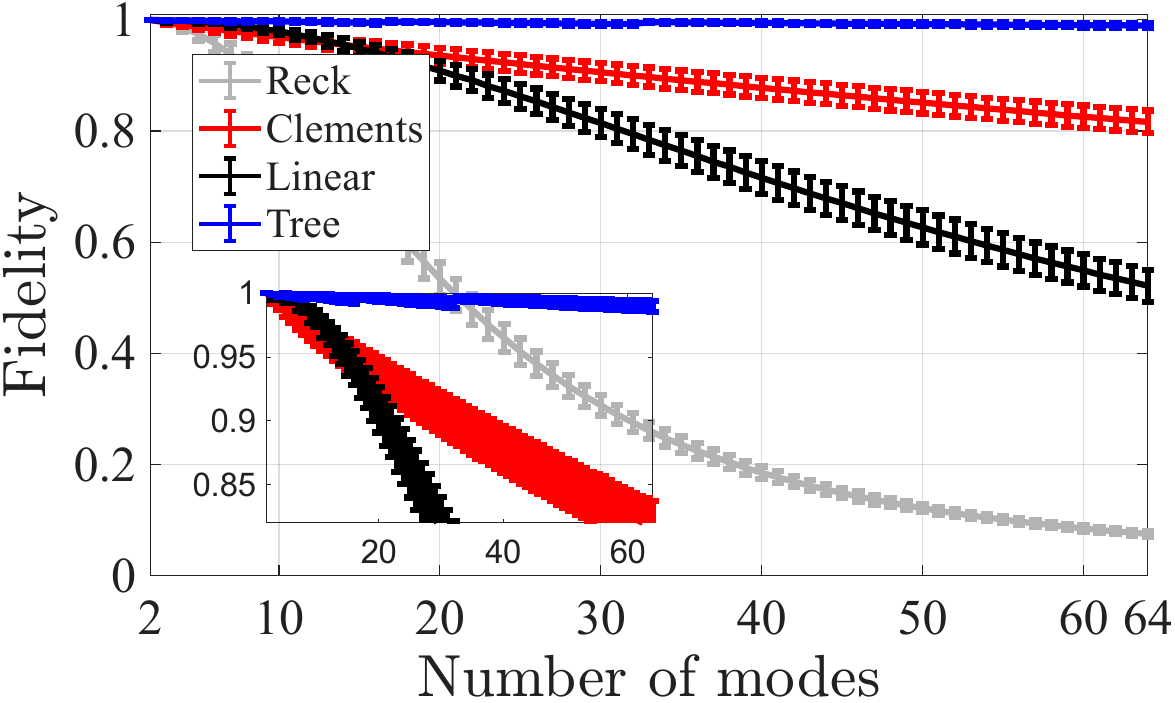}
        \subcaption*{(a)}
    \end{subfigure}
    \hfill
    \begin{subfigure}[b]{0.45\textwidth}
        \centering
        \includegraphics[width=\textwidth]{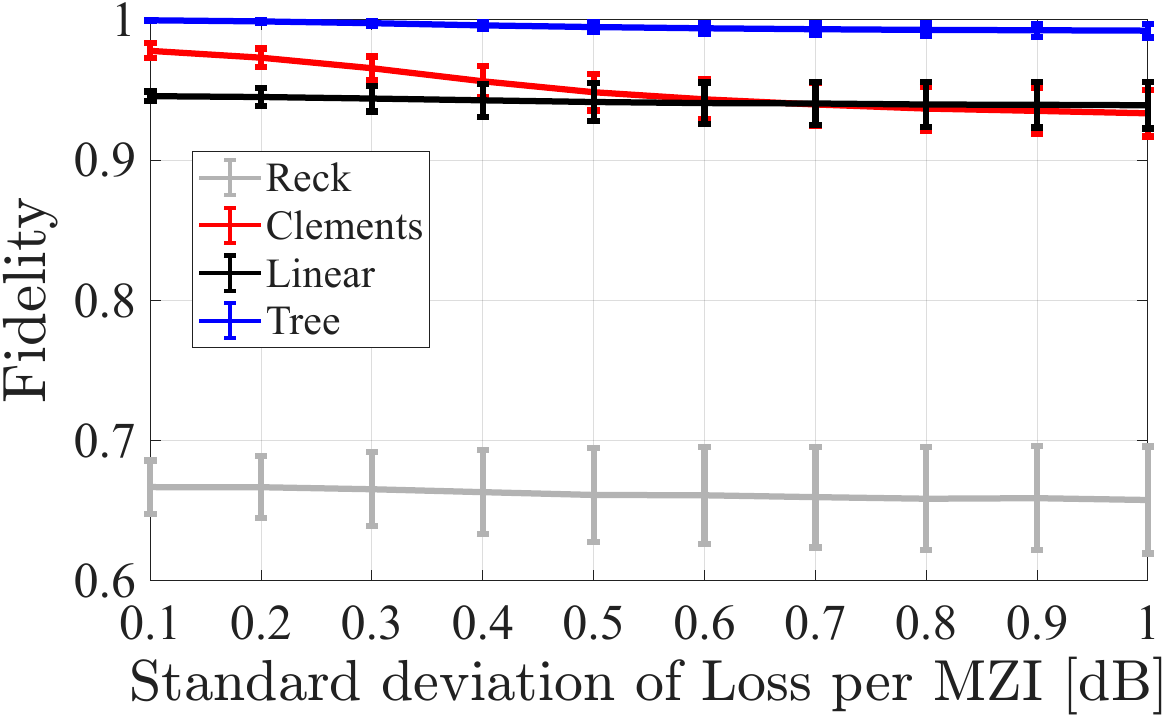}
        \subcaption*{(b)}
    \end{subfigure}
    \begin{subfigure}[b]{0.45\textwidth}
       \centering
        \includegraphics[width=\textwidth]{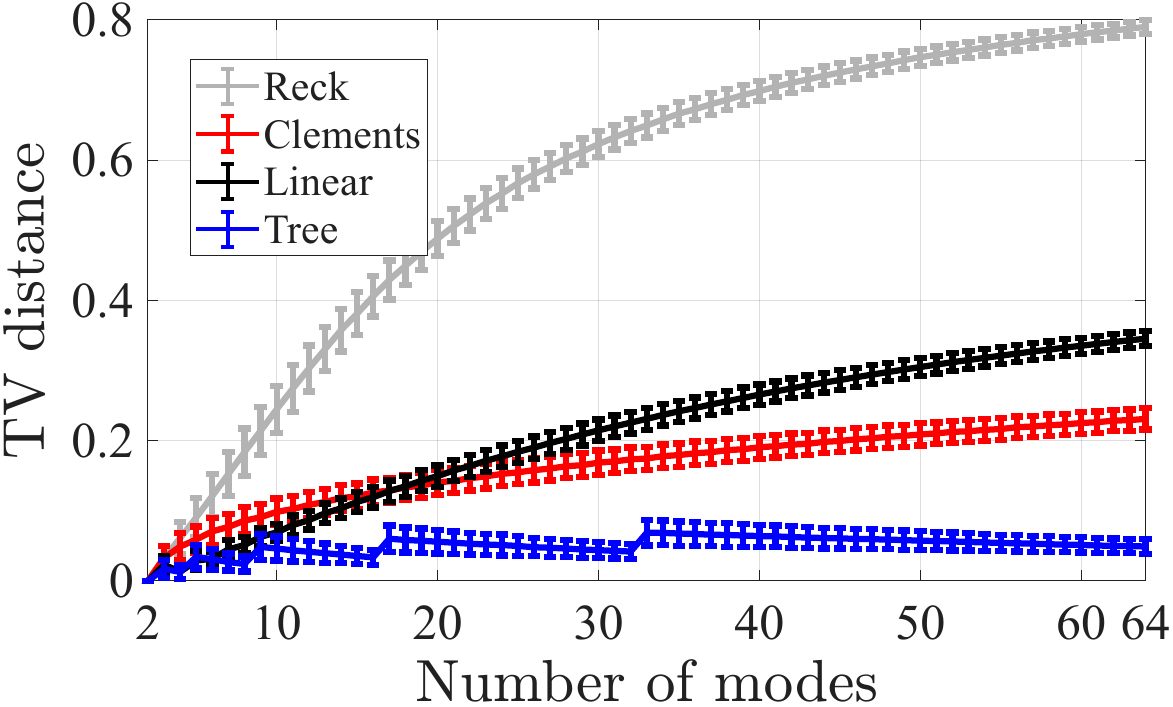}
        \subcaption*{(c)}
    \end{subfigure}
    \hfill
    \begin{subfigure}[b]{0.45\textwidth}
        \centering
        \includegraphics[width=\textwidth]{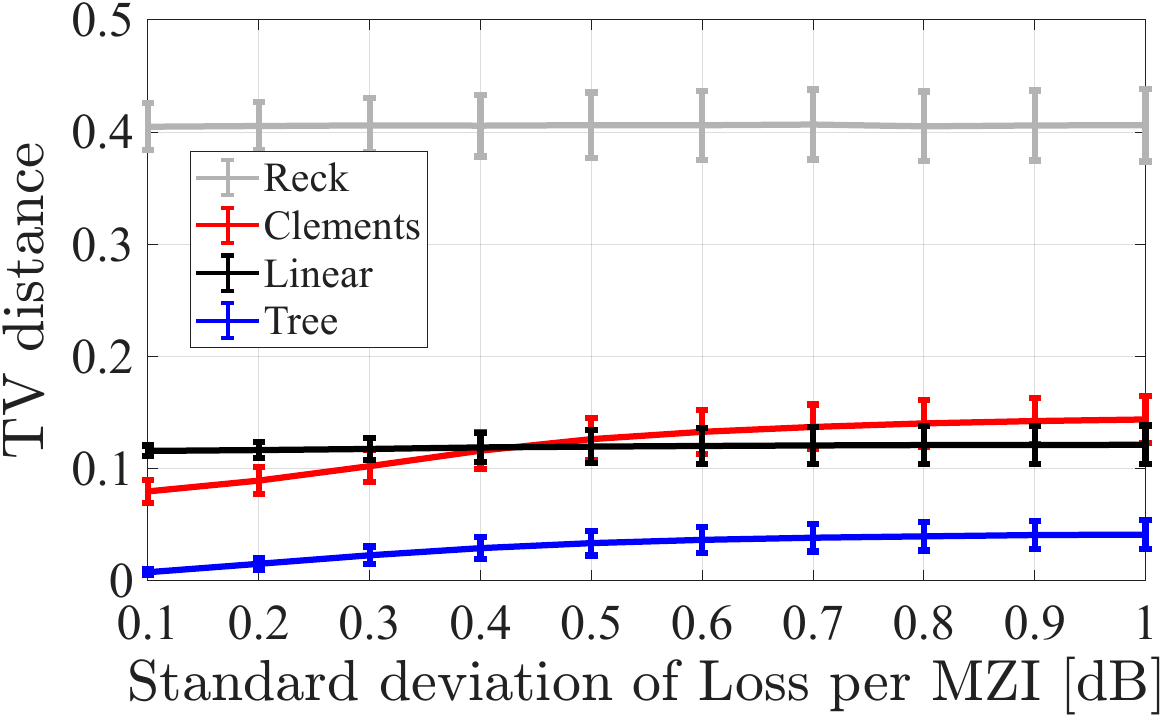}
        \subcaption*{(d)}
    \end{subfigure}
    \caption{ \textbf{Fidelity, Eq. \eqref{eq:fidelity}, and TV distance, Eq. \eqref{KLdivergence}, of the universal schemes and routing schemes for random-lossy MZIs.} 
    Average \textbf{(a)} fidelity and \textbf{(c)} TV distance for different multiport interferometers with Gaussian-distributed MZI loss with mean equal to 1 dB and standard deviation equal to 0.5 dB for interferometers built according to the Reck scheme (grey), Clements scheme (red), linear V-shaped scheme (black) and tree scheme (blue), for different interferometer sizes. Inset: zoom of the fidelity for the Clements, linear V-shaped and the tree schemes.
    The discontinuities for the tree scheme are present for number of modes equal to power of two, and thus with no auxiliary modes with trivial action of the interferometer.
    Average \textbf{(b)} fidelity and \textbf{(d)} TV distance as a function of Gaussian-distributed MZI loss standard deviation for $16\times 16$ multiport interferometers. 
    }
    \label{fig:simu_gaussloss}
\end{figure}

Fig. \ref{fig:simu_gaussloss} shows the average fidelity and TV distance over 1000 random unitary transformations implemented with the different schemes as a function of the mode number with random MZI loss and as a function of constant MZI loss for a fixed mode number.
As for constant loss, the linear V-shaped and tree schemes have higher fidelity/lower TV distance with respect to the Reck scheme, but only the tree scheme has higher fidelity/lower TV distance than Clements. Moreover, also for random MZI loss, if we consider a number of modes below 20, nearly unit fidelity/zero TV distance is reached with Clements scheme or multiple projective measurements with linear V-shaped and tree schemes.

\begin{figure}[H]
    \centering
    \begin{subfigure}[b]{0.45\textwidth}
       \centering
        \includegraphics[width=\textwidth]{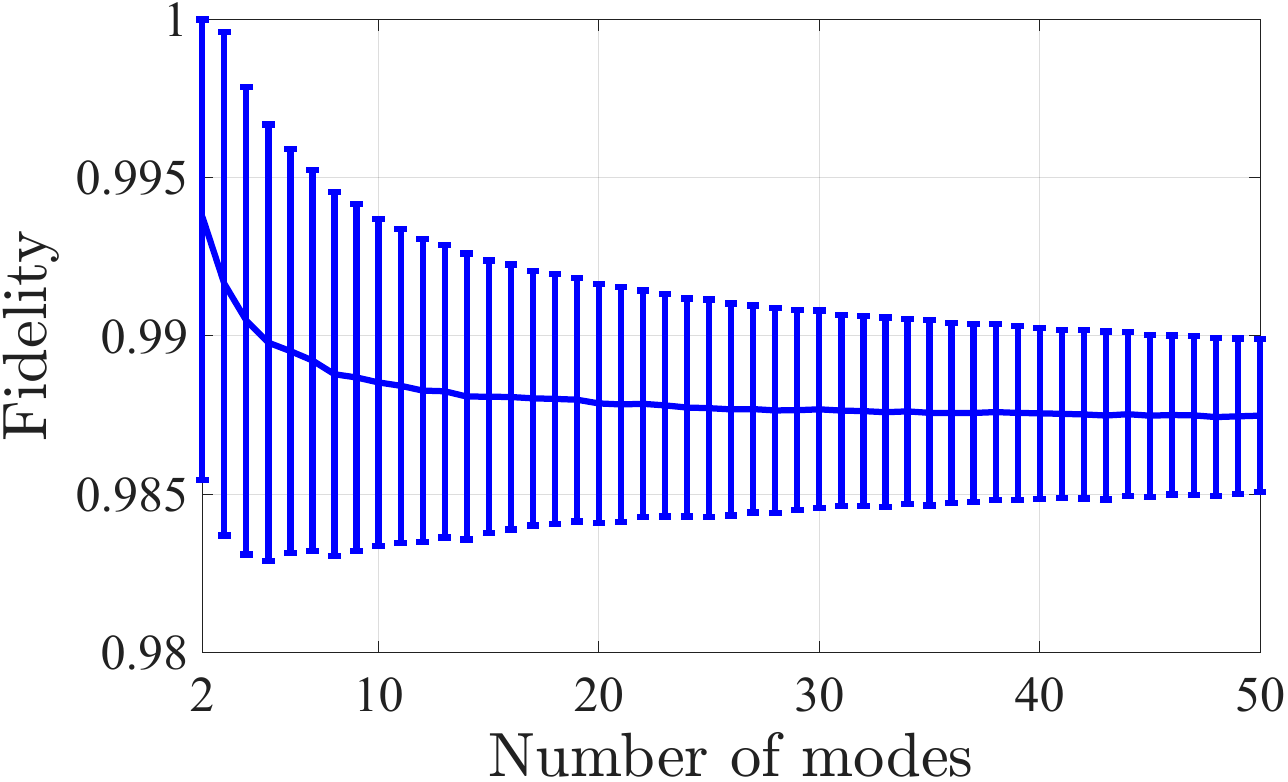}
        \subcaption*{(a)}
    \end{subfigure}
    \hfill
    \begin{subfigure}[b]{0.45\textwidth}
        \centering
        \includegraphics[width=\textwidth]{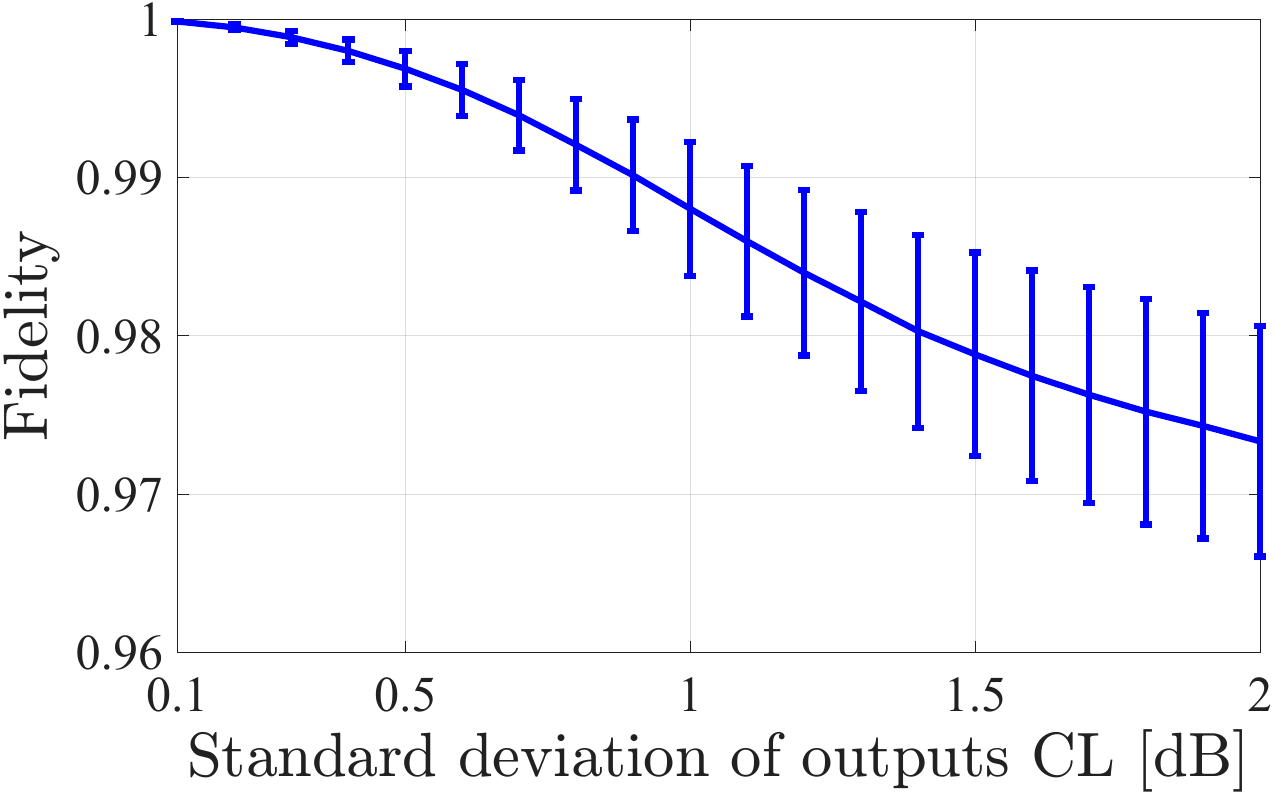}
        \subcaption*{(b)}
    \end{subfigure}
    \begin{subfigure}[b]{0.45\textwidth}
    \includegraphics[width=\textwidth]{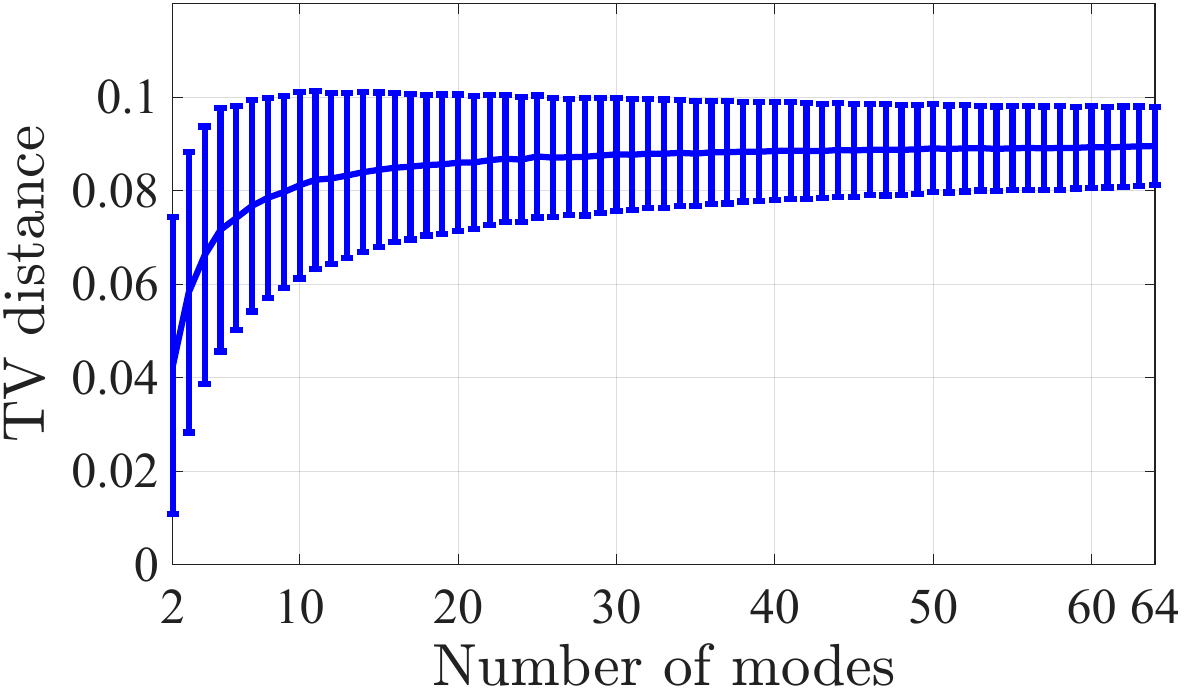}
        \subcaption*{(c)}
    \end{subfigure}
    \hfill
    \begin{subfigure}[b]{0.45\textwidth}
        \centering
        \includegraphics[width=\textwidth]{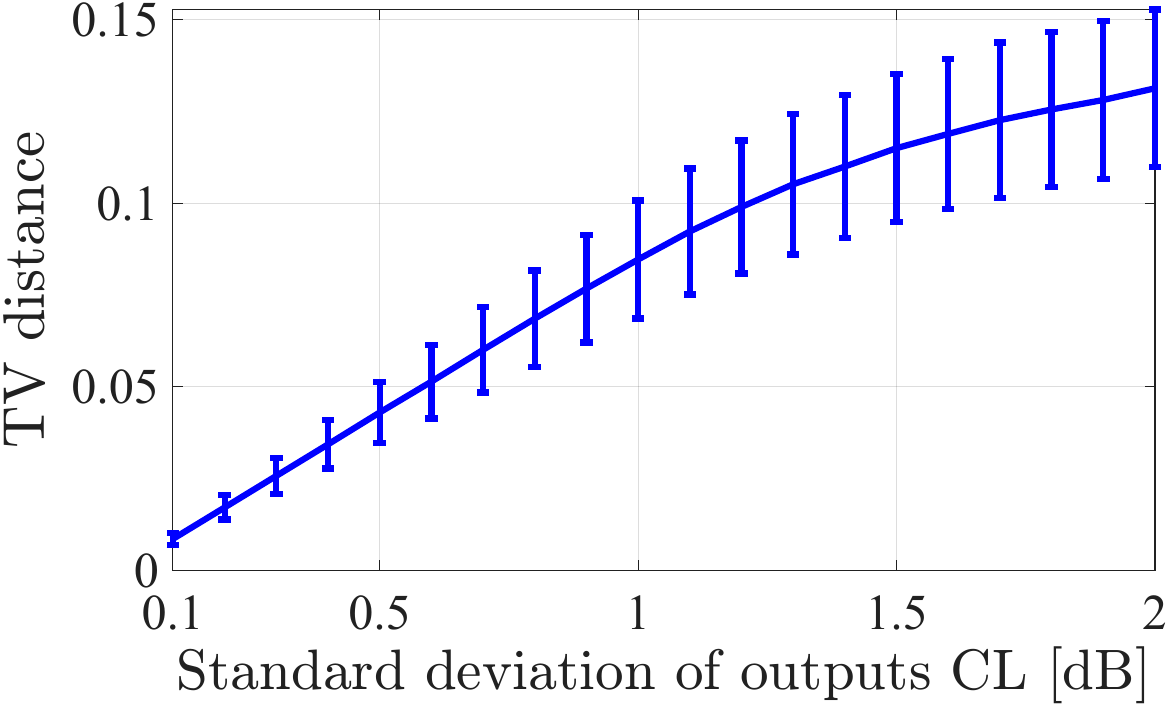}
        \subcaption*{(d)}
    \end{subfigure}
    \caption{
    \textbf{Fidelity, Eq. \eqref{eq:fidelity}, and TV distance, Eq. \eqref{KLdivergence}, of the universal schemes with coupling losses (CL) on the outputs.} 
    Average \textbf{(a)} fidelity and \textbf{(c)} TV distance for different multiport interferometers with Gaussian-distributed coupling losses on the outputs of universal interferometers for different mode numbers and coupling losses standard deviation equal to 1 dB.
    Average \textbf{(b)} fidelity and \textbf{(d)} TV distance as a function of Gaussian-distributed coupling losses standard deviation for $16\times 16$ multiport interferometers. 
    In both simulations, the mean value of the coupling losses on the outputs is 3 dB. 
    }
    \label{fig:simu_CLloss}
\end{figure}

As explained in Section \ref{app:design}, the measurement stage occurs after the unitary manipulation.
Ideally, the coupling losses and the detection efficiency of the different channels are equal, thus they do not affect the fidelity of the transformation.
In reality, the facet of photonic circuits and/or different detection efficiency act as an additional contribution, which can be summarized in a diagonal $N\times N$ matrix $\mathbf{D}$ with elements equal to $10^{-\alpha_k/10}$ with $\alpha_k$ the coupling loss or the detection efficiency of the $k$-th output.
If we insert this effect in Eq. \eqref{eq:fidelity} and assume that $\mathbf{U}^{(\rm exp)}=\mathbf{D}\,\mathbf{U}$, we obtain
\begin{equation}
        F = 
            \frac{\Big|{\rm Tr}\left[ \mathbf{U}^{(\rm exp)} \cdot \mathbf{U}^\dagger\right]\Big|^2}{ N \,{\rm Tr}\left[ \mathbf{U}^{(\rm exp)} \cdot \left(\mathbf{U}^{(\rm exp)}\right)^\dagger\right]} 
            = 
            \frac{\Big|{\rm Tr}\left[ \mathbf{D}\right]\Big|^2}{ N \,{\rm Tr}\left[ \mathbf{D}^{2} \right]} 
            \label{eq:fidelity_CL}
\end{equation}
for universal schemes, while $\mathbf{D}$ do not enter in the fidelity for routers since only one output channel is used. Because of this, there is a cancellation between the loss factor in the numerators and denominators present in Eq. \eqref{eq:fidelity} for routers. Thus, as expected, the fidelity of routing schemes is not affected by unbalanced coupling losses of the outputs.
The same conclusion can be found for the detection probabilities associated with each mode $k$, given by $\tilde p_k$, Eq. \eqref{realistic-probabilities}. Via post-selection, the estimators $\hat p_k$ for the theoretical probabilities $p_k$ are given by $\hat p_k$, Eq. \eqref{estimator-probabilities}. For coupling losses $10^{-\alpha_k/10}$ of the $k$-th output, $\tilde p_k \to 10^{-\alpha_k/10}\,\tilde p_k$ for universal schemes and $\tilde p_k \to 10^{-\alpha_{\lceil N/2\rceil}/10}\,\tilde p_k$ for routing schemes. This means that $\hat p_k$ is affected by different coupling losses for universal schemes, while $\hat p_k$ does not for routing schemes.
Therefore, any figure of merit quantifying the distance between the ideal probability distribution $\{p_k\}_k$ and the estimated one $\{\hat p_k\}_k$ is degraded only for universal schemes.
Fig. \ref{fig:simu_CLloss} shows the effect of unbalanced coupling losses for multiport interferometers with different mode numbers and different values of standard deviation for the coupling losses. 

\end{appendices}

\section*{Declaration statements}

\subsection*{Data Availability}
Data underlying the results presented in this paper are not publicly available, but may be obtained from the authors upon reasonable request.
 
\subsection*{Code Availability}
Not applicable

\subsection*{Acknowledgments}
The work was supported by the Horizon 2020 Framework Programme (899368), Horizon Widera 2023 (101160101) and by the Provincia Autonoma di Trento through the Q@TN joint laboratory.

\subsection*{Competing Interests}
Not applicable

\noindent

\printbibliography

\end{document}